\documentclass[floatfix,a4paper,aps,pra, reprint, amsmath,amsfonts,amssymb,superscriptaddress]{revtex4-2} 

\usepackage{dsfont}
\usepackage{graphicx}
\usepackage{physics}
\usepackage[version=4]{mhchem}
\usepackage{isotope}
\usepackage{bm}
\usepackage[draft=false]{hyperref}
\usepackage{siunitx}
\usepackage[percent]{overpic}
\usepackage{enumitem}
\usepackage{booktabs}
\usepackage[colorinlistoftodos]{todonotes}

\DeclareMathOperator{\re}{\mathrm{Re}}
\DeclareMathOperator{\im}{\mathrm{Im}}

\newcommand{\dyarr}[1]{\overleftrightarrow{#1}}
\newcommand{\useds}[1]{\mathbb{#1}}
\newcommand{\trop}[2]{\hat{\Pi}^{(#1)}_{#2}}
\newcommand{\trfi}[1]{\hat{\Pi}_{#1}}

\newcommand{\co}[1]{#1}

\begin{document}

\title{Waveguide QED with M{\"o}ssbauer Nuclei}

\begin{abstract}
	Thin-film nanostructures with embedded M{\"o}ssbauer nuclei have been successfully used for x-ray quantum optical applications with hard x-rays coupling in grazing incidence. Here we address theoretically a new geometry, in which hard x-rays are coupled in forward incidence (front coupling), setting the stage for waveguide QED with nuclear x-ray resonances. We \co{present in a self-contained manner} a general model based on the Green's function formalism of the field-nucleus interaction in one dimensional waveguides, and show that it combines aspects of both nuclear forward scattering, visible as dynamical beating in the spatio-temporal response, and the resonance structure from grazing incidence, visible in the spectrum of guided modes. The interference of multiple modes is shown to play an important role, resulting in beats with wavelengths on the order of tens of microns, on the scale of practical photolithography. This allows for the design of special sample geometries to explore the resonant response or micro-striped waveguides, opening a new toolbox of geometrical design for hard X-ray quantum optics.
\end{abstract}

\author{Petar Andreji{\'c}}
\email{petar.andrejic@fau.de}
\affiliation{Friedrich-Alexander-Universit\"at Erlangen-N\"urnberg, Staudtstra{\ss}e 7, D-91058 Erlangen, Germany}
\author{Leon Merten Lohse}
\email{llohse@uni-goettingen.de}
\affiliation{Georg-August-Universit{\"a}t G{\"o}ttingen, Friedrich-Hund-Platz 1, D-37077 G{\"o}ttingen, Germany}
\affiliation{Deutches Elektronen-Synchrotron, Notkestraße 85,
D-22607 Hamburg, Germany}
\author{Adriana P{\' a}lffy}
\email{adriana.palffy-buss@physik.uni-wuerzburg.de}
\affiliation{Julius-Maximilians-Universit{\"a}t W{\"u}rzburg, Am Hubland, D-97074 W{\"u}rzburg, Germany}

\date{\today}
\maketitle

\section{Introduction}
The resonant interaction of M{\"o}ssbauer transitions \co{in atomic nuclei} with coherent light from third generation synchrotron and XFEL sources has been shown to be an excellent platform for quantum optics in the X-ray energy scales~\cite{adamsXrayQuantumOptics2013a}. The recoil-free emission due to the M{\"o}ssbauer effect means that scattering is highly elastic, and free from Doppler broadening even at room temperature, while the exceptionally narrow line-width of nuclear transitions means that the temporal response of the nuclei can easily be experimentally resolved \cite{hannonCoherentGrayOptics1999}. 

So far, experiments with these systems have largely been restricted to two scattering geometries: forward scattering, and grazing incidence reflection. In nuclear forward scattering, the target consists of a homogeneous bulk foil containing the resonant nuclei, on the order of \unit{\micro\metre} thickness. The resonant propagation characteristics are that of a homogeneous dielectric medium. The phase difference between the scattered field at the back of the foil compared with the front results in a characteristic spatio-temporal interference pattern known as the `dynamical beat' \cite{hannonCoherentGrayOptics1999}. The analogous visible optical system is the collective emission of a `pencil geometry' of identical atoms, where the dynamical beat is known as a `collective Rabi oscillation'~\cite{svidzinskyNonlocalEffectsSinglephoton2012}. Several interesting quantum optical effects with x-rays have been demonstrated in this geometry, such as magnetic switching \cite{shvydkoStorageNuclearExcitation1996}, coherent pulse shaping~\cite{vagizovCoherentControlWaveforms2014}, electromagnetically induced transparency (EIT)~\cite{coussementControllingAbsorptionGamma2002}, optical control of the nuclear hyperfine spectrum~\cite{kocharovskayaCoherentOpticalControl1999}, pulse shaping~\cite{heegSpectralNarrowingXray2017,heegCoherentXrayOptical2021}, as well as direct observation of the multi-photon dynamics of superradiance~\cite{chumakovSuperradianceEnsembleNuclei2018}.

\co{
In a general X-ray optical context, waveguides have been studied, and demonstrated experimentally to be powerful options for focusing and guiding down to the nanometre scale
\cite{spillerPropagationRaysWaveguides1974, Zwanenburg2000,Pfeiffer2002,Jarre2005,Giewekemeyer_NJoP_2010,Kruger2012,Chen2015,Hoffmann2016,Zhong_XS_2017}. 
Used as optical elements for synchrotron radiation, exploiting their mode-filtering capabilities to provide highly coherent point-like illumination, they have been used for imaging, in particular holographic imaging~\cite{Caro_PRB_2008,Salditt_JoSR_2015}. Tapered waveguides have been used to focus X-rays down to their diffraction limit~\cite{bergemannFocusingXRayBeams2003}. 
In periodically structured waveguides, the photonic bandgap effect has been exploited for single mode propagation~\cite{okamotoXrayWaveguideMode2012}, while longitdunal periodic structures have been demonstrated as efficient mode filters~\cite{Bukreeva_OL_2011,bukreevaPeriodicallyStructuredXray2013}. Due to the relatively large spot sizes of synchrotron beams, direct front coupling of the beam to the waveguide is experimentally challenging, and many experiments are performed using grazing incidence driving. Nevertheless, Fuhse \textit{et al.} have experimentally demonstrated the viability of front coupling using a combination of a Kirkpatrick-Baez mirror and pinhole as a focusing element~\cite{Fuhse_APL_2004}, while Bongaerts \textit{et al.} have used an asymmetric upper and lower cladding at the waveguide entrance to couple the incoming beam via pre-reflection~\cite{bongaertsPropagationPartiallyCoherent2002}.
}

\co{
In the context of M{\"o}ssbauer nuclei, existing experimental work with thin film waveguides has used grazing incidence driving, with layers of M{\"o}ssbauer nuclei embedded as resonant scatterers. Due to an approximate translational symmetry, the system can be analysed in terms of a single Fourier mode, and thus acts analogously to a system of atoms placed between the mirrors of a Fabry-P{\'e}rot resonator. These nanostructures }have proven to be an excellent platform for X-ray quantum optics, with a diverse range of quantum optical phenomena demonstrated, such as superradiance \cite{rohlsbergerCollectiveLambShift2010}, EIT \cite{rohlsbergerElectromagneticallyInducedTransparency2012}, spontaneously generated coherences~\cite{heegVacuumAssistedGenerationControl2013}, Fano resonances and interferometry~\cite{heegInterferometricPhaseDetection2015}, subluminal pulse propagation~\cite{heegTunableSubluminalPropagation2015}, collective strong coupling~\cite{haberCollectiveStrongCoupling2016}, Rabi oscillations between two nuclear ensembles \cite{haberRabiOscillationsXray2017}, \co{as well as a platform for understanding the multiple mode structure of Fabry-P{\'e}rot type resonators~\cite{lentrodtCertifyingMultimodeLightMatter2023}}. 

In this work, \co{we theoretically investigate a different scenario, namely forward incidence, front coupled driving of M{\"o}ssbauer transitions in such thin film nano-structures. Compared to grazing incidence, the broken translational symmetry results in direct driving of a superposition of attenuated guided modes of the waveguide, rather than the grazing incidence picture of a single Fourier mode with conserved wave-vector. Via the spatial structure of the M{\"o}ssbauer scatter layer, this provides direct nuclear control over the interacting waveguide modes, and allows us to engineer the waveguide mediated interaction between sub-ensembles of M{\"o}ssbauer nuclei}. This geometry has been only scarcely investigated so far in the context of resonant nuclear scattering. We are aware only of a recent proposal for embedding M{\"o}ssbauer nuclei in tapered waveguides showing the potential for reaching inversion of the resonant transition~\cite{chenTransientNuclearInversion2022}, while another proposal has considered using slab waveguides with a core filled with density gradients of M{\"o}ssbauer nuclei as a gravitational sensor~\cite{leeGravitationallySensitiveStructured2023}. 

Due to the fact that they are an experimentally well established platform in the grazing incidence geometry, with good control over sample preparation, we will consider slab waveguides as our explicit example. However, the theoretical description we develop is fairly general, and applies to any system with one dimensional propagation, and a resonant nuclear ensemble that is thin in the transverse extent compared to the mode widths. Our theoretical model is based on the Gr{\"u}ner-Welsch quantization of the macroscopic Maxwell's equations~\cite{grunerGreenfunctionApproachRadiationfield1996,dungThreedimensionalQuantizationElectromagnetic1998,dungResonantDipoledipoleInteraction2002}, which allows us to describe the electromagnetic field in a fully quantum way, in terms of the classical dyadic Green's function of the medium.

Our formalism allows us to derive a system of multimode Maxwell-Bloch equations, which in the linear response regime can be rearranged into a matrix differential equation analogous to the equation of motion of ordinary nuclear forward scattering. We are able to obtain an analytic series solution for the spatio-temporal response in the case of two-level nuclei, and demonstrate that this shows the characteristic dynamical beat of forward scattering, but with additional interference beats due to the coupling to multiple guided mode. 

We additionally consider the case of a non-uniform resonant layer; specifically we consider dividing the layer into microscopic sub-ensembles along the propagation direction. In the regime where the ensemble spacing is a similar order of magnitude to the mode interference length, we show that a phenomenon similar to selective sub-radiance occurs, with the waveguide mediated dipole-dipole interaction between sub-ensembles being sensitive to their spacing. In the limit of a half wavelength spacing, we show that the sub-ensembles are at the nodes of the scattered field of their neighbours, and thus the system splits into two non-interacting sub-ensembles, displaying a sensitivity to the even-odd parity of the number of sub-ensembles. This opens an entire new direction of geometrical control of the x-ray scattering, which could be potentially exploited for quantum fluorescence imaging~\cite{trostImagingCorrelationXRay2023}, implementation of mesoscopic models for the investigation of topological edge states~\cite{milicevicEdgeStatesPolariton2015,salaSpinOrbitCouplingPhotons2015,st-jeanLasingTopologicalEdge2017}, as well as the investigation of geometrical radiation phenomena such as selective sub-radiance~\cite{asenjo-garciaExponentialImprovementPhoton2017}.

This paper is structured as follows. Section~\ref{sec:theory} introduces \co{the theoretical formalism we will use to model the waveguide field and its interaction with the resonant nuclei.} This is continued in Section~\ref{sec:eqmot} which gives the solutions of the equations of motion for single as well as multiple modes. The theoretical approach for spatial patterning of the resonant layer is presented in Sec.~\ref{sec:patterning}, while derivations for analytic solutions for structured and unstructured layers of M\"ossbauer nuclei are given in Appendix~\ref{app:dyson} and~\ref{app:temporal}.
Finally, in Section~\ref{sec:numerics} we then give explicit numeric examples, and a detailed qualitative study of these solutions for a realistically implementable two-mode waveguide.

\section{Theoretical Description}\label{sec:theory}
In this section, we introduce the theoretical model for describing the waveguide field and its interaction with the resonant nuclei. \co{We begin with a brief perspective on the macroscopic Maxwell's Green's function approach we will use, and its applications in previous works.}

\co{
The majority of works considering x-ray propagation in the waveguide regime address non-resonant propagation of the X-ray field, i.e.~waveguides in the absence of M{\"o}ssbauer nuclei. 
Previous theoretical studies in the general X-ray context have focused on the mode structure and coherence properties of X-ray waveguides using semi-analytic mode decompositions combined with numeric integration~\cite{bongaertsPropagationPartiallyCoherent2002,bukreevaWaveFieldFormationHollow2006,bukreevaPeriodicallyStructuredXray2013,osterhoffCoherenceFilteringXray2011,lagomarsinoSubmicrometerRayBeam1996}, as well as finite difference~\cite{fuhseFinitedifferenceFieldCalculations2005,Fuhse_OC_2006,pellicciaComputerSimulationsExperimental2007} approaches to the propagation of the guided mode fields. 
In a recent paper, two of us have presented a classical theory for planar X-ray waveguides, discussing the general mode structure of the electromagnetic field and the fields caused by embedded sources, based on classical electromagnetic Green's functions~\cite{Lohse__2023}.
}

\co{
In non-resonant contexts, the refractive indices of the \emph{background} media, the waveguide structure, vary slowly over the bandwidth of interest. In contrast, introducing a nuclear resonance results in a very rapid change in refractive index over the nuclear linewidth compared with the background, and as such it is simplest to treat the nuclear response separately from the background waveguide, which can be taken as constant over the bandwidth of interest. Additionally, with the advent of narrower bandwidth X-ray sources such as XFEL oscillators~\cite{adamsScientificOpportunitiesXray2019}, the saturation of nuclear transitions could become more feasible, and in anticipation of this we wish to develop an \textit{ab initio} microscopic description for the quantum dynamics of both the nuclei and the waveguide fields.
}

\co{
In the interest of a self-contained presentation, we begin with a brief overview of the Gr{\"u}ner-Welsch quantization of the electromagnetic field in terms of the Green's functions of the macroscopic Maxwell's equations.
This approach has been used in the description of collective light-matter interaction, with a wide variety of applications. Asenjo-Garcia \textit{et al}, Chang \textit{et al} and others have used this form to derive an effective dipole-dipole coupling model for atomic lattices, with applications in one dimensional waveguides~\cite{asenjo-garciaAtomlightInteractionsQuasionedimensional2017}, as well as free space lattices~\cite{asenjo-garciaExponentialImprovementPhoton2017,changCavityQEDAtomic2012}. Svidzinsky \textit{et al} have derived an equivalent description \textit{ab initio} in their studies of single photon superradiance, in both isotropic atomic clouds~\cite{svidzinskyCooperativeSpontaneousEmission2008,svidzinskyCooperativeSpontaneousEmission2010}, and one dimensional geometries~\cite{svidzinskyNonlocalEffectsSinglephoton2012}, while Ma and Yelin have used the Green's function as part of a self-consistency approach to study the collective Lamb shift and modified decay rates of atomic clouds~\cite{maCollectiveLambShift2023}. In the non-linear regime, Ruostekoski \textit{et al}~\cite{javanainenExactElectrodynamicsStandard2017,javanainenShiftsResonanceLine2014,cormanTransmissionNearresonantLight2017,jenkinsOpticalResonanceShifts2016} have used this approach to study slab geometries of atomic clouds, developing a hierarchy of equations for the correlation functions of the atomic cloud, coupled via the classical Green's function of the background medium, while Schneider \textit{et al.} have used a Feynman diagram approach to study impurity scattering in waveguides, with the classical waveguide Green's function appearing as the free propagator~\cite{schneider_PRA_2016}. In the X-ray regime, the Gr{\"u}ner-Wlesch formalism has been adapted to describe grazing incidence scattering~\cite{lentrodtInitioQuantumModels2020,kongGreenSfunctionFormalism2020}, which takes advantage of the fact that the Green's function for planar layered media are analytically known. Equivalent expressions appear in the standard treatments of nuclear forward scattering by Kagan \textit{et al}~\cite{kaganExcitationIsomericNuclear1979} and Shvyd'ko~\cite{shvydkoNuclearResonantForward1999}, as well as the general formalism for X-ray resonant scattering of Hannon and Trammel~\cite{hannonMossbauerDiffractionQuantum1968,hannonMossbauerDiffractionII1969,hannonMossbauerDiffractionIII1974,hannonCoherentGrayOptics1999}. 
}
\subsection{Macroscopic QED}
The prototypical example of a M{\"o}ssbauer nucleus is \isotope[57]{Fe}. The metastable internal states of nuclei are characterized primarily by their spin quantum number $I$, and \isotope[57]{Fe} has a relatively low-lying magnetic dipole transition between the $I_g=1/2$ ground states and the $I_e=3/2$ ground states, with an energy of $\hbar \omega_0 = \SI{14.4}{\kilo\electronvolt}$ and an incredibly narrow width of $\hbar\gamma = \SI{4.7}{\nano\electronvolt}$. 

Due to this large energy, the nuclear transition lies well above the largest electronic resonances in the layer materials, and as such the electronic scattering is both weak, and well described as a linear dielectric.
In this regime, to describe the electromagnetic propagation through the medium, we will use the Gr{\"u}ner-Welsch quantization of the macroscopic Maxwell's equations~\cite{grunerGreenfunctionApproachRadiationfield1996}. In this scheme, the polariton-like electromagnetic fields in the medium are quantized via Bosonic noise currents $\hat{f}$, obeying
\begin{align}
	[\hat{f}_{\lambda}(\vec{r},\nu), \hat{f}_{\lambda'}^\dagger(\vec{r}',\nu')] 
	=& \delta_{\lambda \lambda'}\delta^3(\vec{r}- \vec{r}')\delta(\nu-\nu')
	\\
	[\hat{f}_{\lambda}(\vec{r},\nu), \hat{f}_{\lambda'}(\vec{r}',\nu')] 
	=& 0.
\end{align}
Here, $\nu$ is a formal frequency parameter, and $\lambda=e,m$ labels the electric and magnetic polarization of the noise currents respectively. The free field Hamiltonian is then given by
\begin{equation}
	H_F = \sum_{\lambda=e,m} \int_0^\infty \dd{\nu}\int \dd[3]{r} \hbar \nu \hat{f}_\lambda^\dagger(\vec{r},\nu)\hat{f}_\lambda(\vec{r},\nu),
\end{equation}
ensuring that the formal frequency parameter corresponds to a Fourier frequency for the free field,
\begin{equation}
	\partial_t \hat{f}_\lambda(\vec{r},\nu,t) = -\frac{i}{\hbar}[f_\lambda(\vec{r},\nu,t),H_F] = -i\nu \hat{f}_\lambda(\vec{r},\nu,t).
\end{equation}
The electric and magnetic fields are then obtained from the noise currents using the dyadic Green's functions of the macroscopic Maxwell's equations of the material~\cite{buhmannDispersionForces2012},
\begin{align}
	\hat{E}_+(\vec{r}, \nu) &= \sum_{\lambda=e,m}\int \dd[3]{s}\dyarr{\mathcal{\xi}}_\lambda(\vec{r},\vec{s},\nu)\cdot \hat{f}_\lambda(\vec{s},\nu),
	\label{eq:def-e-plus}
	\\
	\hat{E}_-(\vec{r},\nu) &= \hat{E}_+(\vec{r},\nu)^\dagger,
	\\
	\hat{E}(\vec{r}) &= \int_{0}^{\infty}\dd{\nu}\left(\hat{E}_+(\vec{r},\nu) + \hat{E}_-(\vec{r},\nu)\right),
	\\
	\hat{B}_+(\vec{r},\nu) &= \frac{1}{i\nu}\sum_{\lambda=e,m}\int \dd[3]{s}\nabla \times \dyarr{\xi}_\lambda(\vec{r},\vec{s},\nu)\cdot \hat{f}_\lambda(\vec{s}, \nu),
	\\
	\hat{B}_-(\vec{r},\nu) &= \hat{B}_+(\vec{r},\nu)^\dagger,
	\\	
	\hat{B}(\vec{r}) &= \int_{0}^{\infty}\dd{\nu}\left(\hat{B}_-(\vec{r},\nu) + \hat{B}_+(\vec{r},\nu)\right),
	\label{eq:def-total-b}
	\\
	\dyarr{\xi}_e(\vec{r},\vec{r}',\nu) &= i\frac{\nu^2}{c^2}\sqrt{\frac{\hbar}{\pi \varepsilon_0}\Im\varepsilon(\vec{r}',\nu)}\dyarr{G}(\vec{r},\vec{r}',\nu),
	\\
	\dyarr{\xi}_m(\vec{r},\vec{r}',\nu) &= i\frac{\nu}{c}\sqrt{\frac{\hbar}{\pi \varepsilon_0}\frac{\Im\mu(\vec{r}',\nu)}{|\mu(\vec{r}',\nu)|^2}}\dyarr{G}(\vec{r},\vec{r}',\nu)\times \nabla'.
	\label{eq:zeta-m}
\end{align}
In these definitions, $\epsilon_0, \mu_0$ refer to the vacuum permittivity and permeability, $c=\frac{1}{\sqrt{\mu_0 \epsilon_0}}$ the speed of light, while $\varepsilon(\vec{r},\nu),\mu(\vec{r},\nu)$ refer to the dimensionless relative permittivity and permeability of the medium, respectively.
We have introduced the notation $E_{\pm}$ to denote positive and negative frequency field components, and $\dyarr{G}$ denotes the dyadic electric Green's function of the material, obeying
\begin{equation}
	\left(
		\nabla \times \mu^{-1} \nabla \times \quad - \frac{\nu^2}{c^2}\varepsilon 
	\right)
	\dyarr{G}(\vec{r},\vec{r}',\nu) = \dyarr{\delta}(\vec{r}-\vec{r}').
\end{equation}

\subsection{Nuclear Hamiltonian and Lindblad super-operators}
We will model the nuclei using transition operators,
\begin{equation}
	\trop{i}{ab} = \ketbra{a}{b},
\end{equation}
where the bra and ket are implied to act only on the Hilbert space of the $i$th nucleus, and $a,b$ are arbitrary internal states of the nucleus. These obey the commutation relations
\begin{equation}\label{eq:commutator-1}
	[\trop{i}{ab},\trop{j}{cd}] = \delta_{ij}\left(\delta_{bc}\trop{i}{ad} - \delta_{da}\trop{i}{cb}\right).
\end{equation}
For polycrystalline ensembles of nuclei, Bragg scattering is insignificant, and we can model the nuclear layer as a continuum, with number density $\rho(\vec{r})$. We then substitute the transition operators with an operator field,
\begin{equation}
	\trop{i}{ab} \to \trfi{ab}(\vec{r}),
\end{equation}
and \eqref{eq:commutator-1} becomes
\begin{equation}\label{eq:commutator-2}
	[\trfi{ab}(\vec{r}),\trfi{cd}(\vec{r}')] = \frac{1}{\rho(\vec{r})}\delta(\vec{r}-\vec{r}')\left(\delta_{bc}\trfi{ad}(\vec{r}) - \delta_{da}\trfi{cb}(\vec{r})\right).
\end{equation}
The internal nuclear Hamiltonian models the hyperfine interactions of the nucleus, such as the isomer shift, magnetic hyperfine field, and quadrupole splitting~\cite{andrejicSuperradianceAnomalousHyperfine2021}. For the purposes of this article however, it is sufficient to express it in terms of the excited and ground eigenstates,
\begin{multline}
	H_N = \sum_{\mu \in I_e} \int \dd[3]{r}\rho(\vec{r})\hbar(\omega_0 + \Delta_{\mu}) \trfi{\mu\mu}(\vec{r}) 
	\\
	+ \sum_{j\in I_g}  \int \dd[3]{r}\rho(\vec{r}) \hbar\Delta_{j} \trfi{jj}(\vec{r}),
\end{multline}
where we are using Greek indices such as $\mu$ to denote excited eigenstates, and Latin indices such as $j$ to denote ground eigenstates. Here, $\omega_0$ is the reference transition frequency, while $\Delta_\mu, \Delta_j$ are the hyperfine-induced splittings. 

The nuclear excited states decay via both radiative (rad) and electron internal conversion (IC) channels, which can be modelled via Lindblad super-operators,
\begin{widetext}
\begin{align}
	L[\varrho] &= L_{IC}[\varrho] + L_{rad.}[\varrho],
	\\
	L_{IC}[\varrho] &= \sum_{\lambda, l} \Gamma_{IC}(\lambda l, I_e \to I_g) \mathcal{L}_{\lambda l}[\varrho],
	\\
	L_{IC}[\varrho] &= \sum_{\lambda, l} \Gamma_{rad}(\lambda l, I_e \to I_g) \mathcal{L}_{\lambda l}[\varrho],
	\\
	\mathcal{L}_{\lambda l}[\varrho] &=  \int \dd[3]{\vec{r}} \rho(\vec{r}) \sum_{\mu, j} \mathcal{R}(\lambda l, \mu \to j)\left(\trfi{j \mu}(\vec{r})\varrho\trfi{\mu j}(\vec{r})-\frac{1}{2}\{\varrho, \trfi{\mu \mu}(\vec{r})\} \right),
	\\
	\mathcal{L}^H_{\lambda l}[\hat{O}] &=   \int \dd[3]{\vec{r}} \rho(\vec{r})  \sum_{\mu, j} \mathcal{R}(\lambda l,
	\mu \to j)\left(\trfi{\mu j}(\vec{r})\hat{O}\trfi{j \mu}(\vec{r})-\frac{1}{2}\{\hat{O}, \trfi{\mu \mu}(\vec{r})\} \right).
\end{align}
\end{widetext}
Here, $\lambda=\mathcal{E},\mathcal{M}$ denotes the electric or magnetic multipole character of a decay channel, while $l$ denotes the multipole order of the decay. The notation $\mathcal{L}^{H}$ denotes the Heisenberg form of the super-operator, which acts on operators $\hat{O}$ in the Heisenberg picture as opposed to density matrices $\varrho$ in the Schr{\"o}dinger picture.
The sum of the partial rates can be expressed in terms of commonly tabulated quantities,
\begin{align}
	\sum_{\lambda, l} \Gamma_{IC}(\lambda l, I_e \to I_g) =& \frac{\alpha}{1+\alpha} \gamma,
	\\
	\sum_{\lambda, l} \Gamma_{rad}(\lambda l, I_e \to I_g) =& \frac{1}{1+\alpha} \gamma,
\end{align}
where $\gamma$ is the total decay rate, and $\alpha$ the internal conversion coefficient. The rate fractions $\mathcal{R}(\lambda l,\mu \to j)$ can be obtained in terms of the Wigner 3j symbols via~\cite[sec. 5.3]{andrejicControlHighFrequency2023}
\begin{align}\label{eq:rate-fraction}
	\mathcal{R}(\lambda l,\mu \to j) =& \sum_q |C(l q,\mu \to j)|^2,
	\\
	C(kq,\mu\to j) =& \sqrt{2I_e+1} 
	\sum_{m_e,m_g} \bigg[(-1)^{I_e-m_e}
	\label{eq:gen-clebsch}
	\\
	\nonumber&
	\times\braket{\mu}{I_e,m_e}\braket{I_g,m_g}{j}
	\begin{pmatrix}
		I_e & k & I_g \\
		-m_e & q  & m_g
	\end{pmatrix}
	\bigg].
\end{align}
The dominant multipolarity of the resonant transition is $\mathcal{M}1$, i.e. magnetic dipole. The coupling to the field is therefore through the magnetic transition dipole field $\hat{m}(\vec{r})$, which can be expressed in terms of the transition operators as
\begin{align}
	\hat{m}(\vec{r})
	=&
	\hat{m}_+(\vec{r}) + \hat{m}_-(\vec{r}),
	\\
	\hat{m}_+(\vec{r}) =& m_0 \sum_{\mu,j}\vec{d}_{\mu j}^* \trfi{j\mu}(\vec{r})
	\\
	\hat{m}_-(\vec{r}) =& \hat{m}_+(\vec{r})^\dagger.
\end{align}
Here, we have used a generalization of the Wigner-Eckart decomposition;
the prefactor $m_0$ is the usual reduced matrix element of the transition dipole vectors, with magnitude
\begin{equation}
	m_0 = \sqrt{f_{LM} \mathcal{B}(\mathcal{M}1,3/2\to 1/2)},
\end{equation}
where $f_{LM}$ is the Lamb-M{\"o}ssbauer factor, giving the fraction of scattering in the elastic channel, while $ \mathcal{B}(\mathcal{M}1,3/2\to 1/2)$ is the reduced transition probability in Weisskopf units.

The expansion vectors $\vec{d}_{\mu j}$ are dimensionless, and given by
\begin{equation}\label{eq:definition-d-vector}
	\vec{d}_{\mu j} = \sum_{q=-1}^{1}\hat{e}_q C(kq, \mu \to j)
\end{equation}
with $C(kq, \mu \to j)$ as defined in \eqref{eq:gen-clebsch}, and $\hat{e}_q$ the spherical unit vectors,
\begin{align}
	\hat{e}_{-1} &= \frac{1}{\sqrt{2}}(\hat{x} -i \hat{y})
	\\
	\hat{e}_0 &= \hat{z}
	\\
	\hat{e}_1 &= \frac{1}{\sqrt{2}}(\hat{x} +i \hat{y}).
\end{align}

\subsection{Interaction Hamiltonian and Maxwell-Bloch equations}
For the field-nuclei coupling, as we have discussed in the previous section, the dominant multipolarity is magnetic dipole, and we therefore take the interaction Hamiltonian to be
\begin{equation}
	H_I = -\int \dd[3]{r}\rho(r) \hat{B}(\vec{r}) \cdot \hat{m}(\vec{r}).
\end{equation}
We will work in the rotating frame of the nuclei, using the following interaction picture transformation
\begin{align}
	H_T =& H_{T,F} + H_{T,N},
	\\
	H_{T,N} =&
	\hbar \omega_0 \sum_\mu \int \dd[3]{r}\rho(\vec{r})\trfi{\mu\mu}(\vec{r}), 
	\\
	H_{T,F} =& \hbar \omega_0 \sum_{\lambda=e,m}\int_{0}^{\infty}\dd{\nu} \int \dd[3]{r}f_\lambda^\dagger(\vec{r},\nu) f_\lambda(\vec{r},\nu).
\end{align}
The field transformations are then given by
\begin{align}
	\hat{B}_+(\vec{r},\nu) \to& e^{-i\omega_0 t}\hat{B}_+(\vec{r},\nu),
	\\
	\hat{B}(\vec{r}) \to& \hat{B}(\vec{r},t) = \int_0^{\infty}\dd{\nu} e^{-i\omega_0 t}\hat{B}_+(\vec{r},\nu) + \mathrm{h.c.}
	\\
	\trfi{j\mu}(\vec{r}) \to&  e^{-i\omega_0 t} \trfi{j\mu}(\vec{r}),
	\\
	\trfi{\mu \nu}(\vec{r})\to& \trfi{\mu \nu}(\vec{r}),
	\\
	\trfi{jk}(\vec{r})\to& \trfi{jk}(\vec{r}),
	\\
	\hat{m}_+(\vec{r})\to& e^{-i\omega_0 t}\hat{m}_+(\vec{r}),
	\\
	\hat{m}(\vec{r})\to& \hat{m}(\vec{r}, t)= e^{-i\omega_0 t}\hat{m}_+(\vec{r}) + \mathrm{h.c.}
\end{align}
The field equation of motion is then derived via the Heisenberg equations of motion for the field,
\begin{equation}
	\partial_t \hat{B}(\vec{r},t) = -\frac{i}{\hbar}[\hat{B}(\vec{r},t), H_F - H_{F,T} + H_I(t)].
\end{equation}
In Appendix~\ref{app:kramers-kronig}, we show that for the Fourier transformed field,
\begin{equation}
	\hat{B}(\vec{r},\omega) = \int_{-\infty}^{\infty}\dd{t} e^{i \omega t} \hat{B}(\vec{r},t),
\end{equation}
applying the Kramers-Kronig relations leads one to obtain
\begin{multline}\label{eq:maxwell-field}
	\hat{B}(\vec{r},\omega) = \hat{B}_{in}(\vec{r},\omega) - 
	\\
	\mu_0  \int\dd[3]{r'}\rho(r') \dyarr{G}_{mm}(\vec{r},\vec{r}',\omega) \cdot \hat{m}(\vec{r}',\omega),
\end{multline}
where, $\hat{B}_{in}$ is the homogeneous solution in the absence of resonant nuclei, while 
\begin{equation}
	\dyarr{G}_{mm}(\vec{r},\vec{r}',\omega) = \nabla \times \dyarr{G}(\vec{r},\vec{r}',\omega) \times \nabla'
\end{equation}
is the Green's function of the macroscopic Maxwell's equations giving the magnetic field response of a magnetic source.
This therefore demonstrates that the macroscopic Maxwell equations hold in the operator sense for the fully quantized field-nucleus interaction.

We note at this stage that for the interacting system, the Fourier frequency $\omega$ is not the same as the noise current frequency $\nu$ defined in equations \eqref{eq:def-e-plus} through \eqref{eq:def-total-b}. Nevertheless, we may still divide the field into positive and negative frequency components corresponding to annihilation and creation operators of the noise field respectively.

Evaluating the equation of motion of the nuclear transition operators results in the following Bloch equations (see Appendix~\ref{app:optical-bloch} for details),
\begin{widetext}
	\begin{align}
		\label{eq:bloch-1}
		\partial_t \trfi{\mu\nu}(\vec{r},t) &= 
		\left(i(\Delta_\mu - \Delta_k)-\gamma\right)\trfi{\mu	\nu}(\vec{r},t)
		\\
		&\hspace{40pt}\nonumber 
		+\frac{i m_0}{\hbar}\sum_j \left(\trfi{\mu j}(\vec{r},t)\vec{d}_{\nu j}e^{i\omega_0 t}-\trfi{j \nu}(\vec{r},t)\vec{d}_{\mu j}^* e^{-i\omega_0 t}\right)\cdot \hat{B}(\vec{r},t),
		\\
		\label{eq:bloch-2}
		\partial_t\trfi{jk}(\vec{r},t) &=
		i(\Delta_j - \Delta_k)\trfi{j k}(\vec{r},t) + \delta_{jk}\sum_\mu \Gamma(\mu \to j)\trfi{\mu\mu}(\vec{r},t)
		\\
		&\hspace{40pt}\nonumber 
		-\frac{i m_0}{\hbar} \sum_\mu \left(\trfi{\mu k}(\vec{r},t)\vec{d}_{\mu j}e^{i\omega_0 t} - \trfi{j \mu}(\vec{r},t) \vec{d}_{\mu k}^* e^{-i\omega_0 t}\right)\cdot \hat{B}(\vec{r},t),
		\\
		\label{eq:bloch-3}
		\partial_t \trfi{\mu j}(\vec{r},t) &= 
		\left(i(\Delta_\mu - \Delta_j) - \frac{\gamma}{2}\right)\trfi{\mu j}(\vec{r},t)
		\\
		&\hspace{40pt}\nonumber 
		+\frac{i m_0}{\hbar}\left(\sum_{\nu}\trfi{\mu\nu}(\vec{r},t)\vec{d}_{\nu j}^*e^{-i\omega_0 t} -\sum_k \trfi{kj}\vec{d}_{\mu k}^*e^{-i\omega_0 t}\right)\cdot \hat{B}(\vec{r},t),
		\\
		\label{eq:bloch-4}
		\partial_t \trfi{j \mu}(\vec{r},t) &= 
		\left(-i(\Delta_\mu - \Delta_j) - \frac{\gamma}{2}\right)\trfi{j\mu}(\vec{r},t)
		\\
		&\hspace{40pt}\nonumber 
		-\frac{i m_0}{\hbar}\left(\sum_{\nu}\trfi{\nu\mu}(\vec{r},t)\vec{d}_{\nu j}e^{i\omega_0 t} -\sum_k \trfi{jk}\vec{d}_{\mu k}e^{i\omega_0 t}\right)\cdot \hat{B}(\vec{r},t).
	\end{align}
\end{widetext}
Here, $\Gamma(\mu \to j)$ is the total decay rate over all channels from excited state $\mu$ to ground state $j$.
In current experiments, there are few resonant photons per incident pulse, and thus we can consider the linear response for the magnetization, $\trfi{\mu \nu} \approx 0$, $\trfi{jk}\approx \frac{\delta_{jk}}{2I_g + 1}$.
In addition, due to the very large nuclear transition frequency, the rotating wave approximation holds very well, and the positive and negative frequency components of the magnetization can be described with via a linear susceptibility tensor $\dyarr{\chi_m}$ (see Appendix~\ref{app:optical-bloch})
\begin{align}
	\hat{m}_+(\vec{r},\omega) =& \frac{1}{\mu_0}\dyarr{\chi_m}(\omega) \cdot \hat{B}_+(\vec{r},\omega),
	\\
	\dyarr{\chi_m}(\omega) =&
	-\frac{\sigma_{res}}{k_0}
	\dyarr{F}(\omega),
	\label{eq:susceptibility}
	\\
	\sigma_{res} =& \frac{2\pi}{k_0^2}\frac{ f_{LM}}{1+\alpha}
	\frac{2I_e+1}{2I_g+1}.
\end{align}
Here, $\sigma_{res}$ is the cross-section of resonant scattering, $f_{LM}$ is the Lamb-M{\"o}ssbauer factor, $\alpha$ the internal conversion coefficient, and $k_0=\omega_0 c^{-1}$ the overall transition wave-number. We note that $\dyarr{\chi_m}$ has the overall dimension of volume, as we have defined $\hat{m}(\vec{r},\omega)$ via the nuclear transition dipole moments, rather than their density.

The dimensionless response tensor $\dyarr{F}(\omega)$ is given by the sum of the Lorentzian responses of the available transitions
\begin{equation}
	\dyarr{F}(\omega) = \frac{3}{2I_e+1}\sum_{\mu,j}\frac{\gamma / 2}{\omega - \Delta_{\mu} +\Delta_j + i \gamma / 2} \vec{d}_{\mu j}^*\otimes \vec{d}_{\mu j}.
\end{equation} 
In the case of an inhomogeneous ensemble, the response is averaged over the probability distribution of the inhomogeneous hyperfine environment of the nuclei~\cite{andrejicSuperradianceAnomalousHyperfine2021}.

\subsection{Green's function for slab waveguides}
\begin{figure}[ht]
	\centering
	\includegraphics[width=6cm]{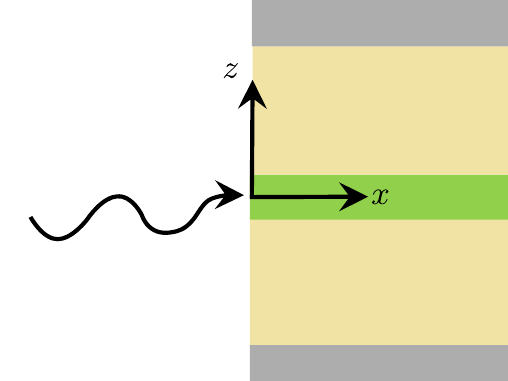}
	\caption{
		Overview of scattering geometry. The waveguide is formed by a stack of dielectric layers in the $z$ axis, while the incident beam propagates along the $x$ axis. The field is sufficiently well collimated that it is uniform along $y$, allowing us to consider a two dimensional problem in the $xz$ plane.
	}
	\label{fig:fc-geometry}
\end{figure}

In Figure~\ref{fig:fc-geometry} we give a schematic view of the scattering geometry used to create a slab waveguide for resonant X-rays. The field propagates in the $x$ direction, with refractive index gradients in the $z$ direction used to create the waveguide structure. The waveguide bulk is translationally symmetric in the $x$ and $y$ directions, and since synchrotron sources are well collimated, we can take the incident field to be uniform in the $y$ direction, making the problem effectively two-dimensional.

For a slab waveguide, the Green's functions are analytically known (see for example \co{\cite{buhmannDispersionForces2012,tomasGreenFunctionMultilayers1995,johanssonElectromagneticGreenFunction2011,hansonOperatorTheoryElectromagnetics2002}}). The Green's functions can be divided into transverse electric and transverse magnetic polarization, with these components in turn being decomposed into a sum of discrete modes, and a continuum of radiative modes \co{\cite{Lohse__2023,hansonOperatorTheoryElectromagnetics2002}},
\begin{multline}\label{eq:greens-function-form}
	\dyarr{G}^s(\vec{r},\vec{r}',\omega) 
	=
	\sum_\lambda \dyarr{g}^s_\lambda(z,z',\omega) e^{i q^s_\lambda(\omega)|x - x'|}
	\\
	+ \dyarr{G}^s_{rad}(z,z',x-x',\omega).
\end{multline}
Here, $s=TE,TM$ labels the polarization while $\lambda$ labels the guided modes, which propagate with complex wave-numbers $q^s_\lambda$, the positive imaginary parts of which give the attenuation of the guided mode. 

In particular, \co{in the hard X-ray regime}, compared to the guided mode contributions, the radiative contribution is small in magnitude and very short range \cite{Lohse__2023}, and due to the correspondingly large bandwidth in momentum space results in an overall Purcell factor and Lamb shift. Thus, we will absorb it into our definition of the transition frequency and decay rate.

Due to the very weak backscattering of X-rays outside the Bragg condition, we can neglect the backward propagating scattered field, with the substitution
\begin{equation}
	e^{iq^s_\lambda|x-x'|}\to \Theta(x-x')e^{iq^s_\lambda(x-x')},
\end{equation}
where $\Theta(x)$ is the Heaviside theta distribution.

Over the resonant bandwidth of the M{\"o}ssbauer nuclei, the envelope of the guided mode components of the Green's function vary very little as functions of frequency, while the wave-numbers have a dispersion of approximately \cite{pellicciaDispersionPropertiesXray2006,Lohse__2023}
\begin{equation}
	\pdv{q_\lambda}{\omega} \approx \frac{1}{c}.
\end{equation}
This linear dispersion can be eliminated by transforming operators with
\begin{equation}
	\hat{O}(x,\omega)\to e^{-i\omega x /c}\hat{O}(x,\omega),
\end{equation}
which has the effect of substituting time in the Fourier inversion with the retarded time,
\begin{equation}
	t \to t_r = t - \frac{x}{c}.
\end{equation}
Thus, we can simply solve for the absence of the linear dispersion, and substitute out ordinary time for the retarded time in our solution.
For \isotope[57]{Fe}, with a lifetime of approximately \SI{142}{\nano\second}, the retardation is on the order of $10^{-5}$ lifetimes per millimetre, and is thus negligible for our purposes, and we will simply use the ordinary time from this point forward.

Within this regime, we can then approximate the Green's function as
\begin{align}
	g^s_\lambda(z,z',\omega) &\approx g^s_\lambda(z,z',\omega_0),
	\\
	q^s_\lambda(\omega) &\approx q^s_\lambda(\omega_0),
\end{align} 
where $\omega_0$ is the mean transition frequency of the nuclei. 

In the geometry and energy scale we have considered, the difference in reflectivity for TE and TM polarizations is negligible, and additionally the longitudinal component of the TM fields are small. Thus, we can approximate the TM components as having the same magnitude but orthogonal polarization dependence to the TE, as well as the same wave-numbers. Therefore, we can express the Green's function in the approximate form
\begin{align}
	\dyarr{G}(\vec{r},\vec{r}',\omega) \approx& (\dyarr{1} - \hat{x}\otimes\hat{x})
	\sum_\lambda  g_\lambda(z,z') e^{i q_\lambda(x-x')},
	\\
	\dyarr{G}_{mm}(\vec{r},\vec{r}',\omega) \approx& k_0^2 \dyarr{G}(\vec{r},\vec{r}',\omega),
\end{align}
where the guided mode envelope $g_\lambda$ is given in terms of the eigenfunctions $u_\lambda$ of the associated Sturm-Liouville problem for TE modes,
\begin{gather}
	g_\lambda(z,z') = \frac{i}{2q_\lambda} u_\lambda(z)u_\lambda(z'),
	\\
	\label{eq:helmholtz-problem}
	\bigg(\mu(z) \partial_z \mu(z)^{-1} \partial_z + k_0^2 n(z)^2 - q_\lambda^2\bigg)u_\lambda(z) = 0,
\end{gather}
where $n(z) = \sqrt{\mu(z)\varepsilon(z)}$ is the refractive index.
The normalizable TE eigenfunctions obey the bi-orthogonality relation
\begin{equation}\label{eq:orthogonality}
	\delta_{\lambda \lambda'} =
	\int_{-\infty}^{\infty} \dd{z} \frac{1}{\mu(z)}u_\lambda(z) u_{\lambda'}(z) 
\end{equation}
A generalization to the non-normalizable leaky modes is also possible \cite{Lohse__2023}, using an analogous regularization method to that of Leung \emph{et al.}~\cite{leungCompletenessTimeindependentPerturbation1996,kristensenGeneralizedEffectiveMode2012}.

For determining the incident field at the air-waveguide interface, the negligible backscattering means that the field normal is approximately equal on either side of the boundary, and therefore we can take the boundary condition to simply be continuity of the field. The incident field at the interface can then be decomposed into the guided mode basis and propagated,
\begin{align}
	\hat{B}_{in}(x,z,\omega) =& \sum_{\lambda}\hat{B}_{in,\lambda}(x,z,\omega),
	\\
	\hat{B}_{in,\lambda}(x,z,\omega) =& u_\lambda(z)e^{iq_\lambda x}\int_{-\infty}^{\infty} \dd{z} \frac{1}{\mu(z)}u_{\lambda}(z)\hat{B}_{in}(0,z,\omega).
\end{align}

For a thin resonant nuclear layer, such that the guided mode envelopes are uniform across the layer coordinate, we can take the nuclear density to be a delta function,
\begin{equation}
	\rho(\vec{r}) = L\delta(z-z_0) \rho_N,
\end{equation}
where $\rho_N$ is the number density of the bulk material, $L$ is the layer thickness, and $z_0$ the $z$ coordinate of the layer centre. The one-dimensional equation of motion then becomes
\begin{multline}\label{eq:eom-full}
	\hat{B}(x,z_0, \omega)
	= 
	\hat{B}_{in}(x,z_0,\omega)
	\\
	-i \frac{\zeta}{2} F(\omega)\int_0^{x}\dd{x'} \sum_\lambda \xi_\lambda e^{iq_\lambda(x-x')} \hat{B}(x',z_0,\omega),
\end{multline}
where
\begin{equation}
	\zeta = \rho_N \sigma_{res},
\end{equation}
is the on-resonance attenuation coefficient for ordinary nuclear forward scattering, while 
\begin{equation}
	\xi_\lambda = k_0 L \frac{u_\lambda(z_0)^2}{q_\lambda}
\end{equation}
is the dimensionless coupling strength for each mode $\lambda$, relative to ordinary nuclear forward scattering.
In particular, we can see that the equation of motion is similar in form to ordinary nuclear forward scattering, with the NFS equation of motion given by~\cite{kaganExcitationIsomericNuclear1979,shvydkoNuclearResonantForward1999}
\begin{multline}\label{eq:eom-nfs}
	\hat{B}(x,\omega) = \hat{B}_{in}(x,\omega) 
	\\
	- i \frac{n \zeta}{2} F(\omega)\int_0^{x}\dd{x'}e^{in k_0(x-x')} \hat{B}(x',\omega).
\end{multline}
Here, we have included both the bulk medium refractive index $n$, and the overall linear dispersion in the bulk medium wave-vector, which are usually neglected in the literature.

\subsection{Matrix form of equations of motion}
For many purposes it is convenient to work with the decomposition of the waveguide field into the guided modes directly. This can be expressed in a matrix-vector notation. To begin with, we define the following vector, comprising the field components of each participating mode, evaluated at the layer position,
\begin{equation}
	\vec{\beta}(x,\omega) = 
	\begin{pmatrix}
		B_1(x,z_0,\omega)
		\\
		\vdots
		\\
		B_{n}(x,z_0,\omega)
	\end{pmatrix}.
\end{equation}
The total field at any $x,z$ coordinate can then be evaluated as 
\begin{equation}
	\hat{B}(x,z,\omega) = \vec{w}(z)^\top \cdot \vec{\beta}(x,\omega),
\end{equation}
where
\begin{equation}
	\vec{w}(z) = \begin{pmatrix}
		\frac{u_1(z)}{u_1(z_0)}
		\\
		\vdots
		\\
		\frac{u_n(z)}{u_n(z_0)}
	\end{pmatrix}.
\end{equation}
In this notation, the equations of motion read
\begin{align}\label{eq:eom-vector1}
	\vec{\beta}(x,\omega) =& \vec{\beta}_{in}(x,\omega)
	- i \frac{\zeta}{2} F(\omega) \times
	\\
	\nonumber
	&\int_0^{x}\dd{x'}\exp(i (Q - \omega /c )(x-x'))\cdot \Lambda \cdot \vec{\beta}(x',\omega),
\end{align}
where $Q$ is the diagonal matrix of wave-numbers,
\begin{equation}
	Q_{\lambda \lambda'} = q_\lambda \delta_{\lambda \lambda'},
\end{equation}
while $\Lambda$ is the dimensionless rank-1 matrix describing the resonant scattering,
\begin{align}
	\Lambda =& \vec{\xi} \otimes \vec{w}(z_0)^\top,
	\\
	\nonumber
	=& \vec{\xi} \otimes 1^\top,
	\\
	\vec{1} =& \begin{pmatrix}
		1
		\\
		\vdots
		\\
		1
	\end{pmatrix},
\end{align}
where $\vec{\xi}$ is the column vector of dimensionless relative coupling strengths for each mode, and we note that $w(z)$ becomes a uniform vector when evaluated at $z_0$. Compared with~\eqref{eq:eom-nfs}, we see that the bulk medium wave-number $n k_0$ is replaced with the mode wave-number matrix $Q$, while the effective coupling strength in the bulk medium $n$ is replaced with the matrix $\Lambda$.
We can then take a spatial derivative to obtain
\begin{equation}\label{eq:eom-vector2}
	\partial_x \vec{\beta}(x,\omega) = i Q \cdot \vec{\beta}(x,\omega) -i\frac{\zeta}{2} F(\omega)\Lambda \cdot \vec{\beta}(x,\omega).
\end{equation}
We note that in transforming to the differential form of the equations of motion, since $\hat{B}_{in}$ is the homogeneous solution, we have
\begin{equation}
	\partial_x \vec{\beta}_{in} - i Q\cdot\vec{\beta}_{in} = 0.
\end{equation}

\section{Solution of the equations of motion}\label{sec:eqmot}
In this section, we will solve the equations of motion~\eqref{eq:eom-vector2}, first for a single mode waveguide, and then for the general case of multiple modes.
\subsection{Single mode solution}
For realistic layer materials, the leaky modes lying above the cut-off wave-number have a very small amplitude in the waveguide core compared with the guided modes, which lie below the cut-off. Thus, we can adjust the waveguide thickness appropriately, such that the desired number of modes are supported, and neglect the rest due to their small amplitudes. In this section, we will consider the simplest system, which consists of a single mode waveguide with a thin layer of resonant nuclei placed in the centre.

For simplicity, we will neglect hyperfine interactions, such that~\cite{andrejicSuperradianceAnomalousHyperfine2021}
\begin{equation}
	\dyarr{F}(\omega) = \dyarr{1}\frac{\gamma / 2}{\omega + i \gamma /2}.
\end{equation}
In this regime, the incident beam polarization is preserved, and we can treat the problem as scalar.
The equation of motion for the single supported mode $\hat{B}_1$ is then given by
	\begin{multline}\label{eq:fws-resonant-mode}
		\hat{B}_1(x,\omega)
		=
		\hat{B}_{in}(x,\omega)
		\\
		-i\xi_1
		\frac{\zeta }{2} F(\omega)
		\int_0^x \dd{x'}e^{iq_1 (x-x')}
		\hat{B}_1(x',\omega),
	\end{multline}
we can see that this is of the same form as the equation for ordinary nuclear forward scattering, \eqref{eq:eom-nfs}, with the attenuation length scaled by $\xi_1$ and the bulk material wave-vector $n k_0$ replaced by the mode wave-vector $q_1$.

As in ordinary nuclear forward scattering, the driving pulse is far shorter in duration than the lifetime of the nuclear transition. We can therefore approximate the driving pulse as
\begin{equation}
	\expval{B_{in}(x=0,t)} \to \frac{B_{0}}{\Gamma_{0}}\delta(t),
\end{equation}
where $\Gamma_{0}$ is the bandwidth of the driving pulse, $\delta(t)$ the Dirac delta distribution, and $B_0$ the peak amplitude of the pulse.

Therefore, \eqref{eq:fws-resonant-mode} can be solved in the same manner as in ordinary nuclear forward scattering. Applying the Kagan Fourier transform method~\cite{kaganExcitationIsomericNuclear1979}, we then obtain
	\begin{align}\label{eq:bessel-response}
		\expval{B_1(x,t)}
		=&
		\frac{B_{0}}{\Gamma_{0}}e^{iq_1 x}
		\left(
			\delta(t)
			- \Theta(t)e^{-\gamma t/ 2}\frac{\gamma \tau_1}{2}
			\frac{J_1(\sqrt{\tau_1 \gamma t})}{\sqrt{\tau_1 \gamma t}}
		\right),
		\\
		\xi_1 =& \frac{k_0 L u_1(z_0)^2 }{q_1},
		\\
		\tau_1 =&   
		\xi_1 \zeta x.
	\end{align}
Here, $\tau_1$ is the effective optical depth, which differs from the bulk material optical depth $\tau = \zeta x$ by the relative coupling strength $\xi_1$. In the limit $q_1 \to n k_0$, $\xi_1 \to n$ we recover the ordinary nuclear forward scattering solution for a bulk material.

\subsection{Multimode solution}
Next, we will consider the case of multiple guided modes. This can be done using the matrix equation \eqref{eq:eom-vector1}.
As a first order vector differential equation, the formal solution is the following matrix exponential,
\begin{equation}
	\vec{\beta}(x,\omega) = \exp(i Q x - i \frac{\zeta}{2} F(\omega)\Lambda x)\cdot \vec{\beta}(0,\omega).
\end{equation}
For analytic Fourier inversion purposes, this solution has the drawback that each term in the series expansion of the matrix exponential is not homogeneous in powers of $F(\omega)$. For these purposes, we will proceed to instead express the solution as a path ordered exponential.

To begin, we eliminate the wave-vectors from the equation of motion by taking the exponential of $Q$, giving the diagonal propagation matrix $S$, that accounts for the mode attenuation and phase as $x$ is varied,
\begin{equation}
	S(x_f-x_i) 
	=
	\exp(i Q(x_f - x_i)).
\end{equation}
We then make the substitution
\begin{equation}
	\vec{\beta}(x,\omega) = S(x)\cdot\tilde{\beta}(x,\omega).
\end{equation}
Note that we have taken the input face of the waveguide to be at the position $x=0$. Under this substitution, the transformed equation of motion is
\begin{equation}\label{eq:transformed-eom}
	\partial_x \tilde{\beta}(x,\omega) = -i\frac{\zeta}{2} F(\omega) \tilde{\Lambda}(x)\cdot \tilde{\beta}(x,\omega),
\end{equation}
where
\begin{equation}
	\tilde{\Lambda}(x) = S^{-1}(x)\cdot \Lambda \cdot S(x),
\end{equation}
which has the matrix elements
\begin{equation}
	\tilde{\Lambda}_{\lambda \lambda'}(x)
	=
	e^{i (q_\lambda - q_{\lambda'}) x} \xi_\lambda.
\end{equation}
The formal solution to \eqref{eq:transformed-eom} is then the path ordered exponential
\begin{equation}
	\tilde{\beta}(x,\omega) = \mathcal{P}\exp\left(-i\frac{\zeta}{2} F(\omega)
	\int_0^{x} \dd{x'}\tilde{\Lambda}(x')
	\right)\tilde{\beta}(0,\omega).
\end{equation}
The full solution can then be obtained via
\begin{equation}
	B(x,\omega) = \vec{1}^\top \cdot S(x) \cdot \mathcal{P}\exp\left(-i\frac{\zeta}{2} F(\omega)
	\int_0^{x} \dd{x'}\tilde{\Lambda}(x')
	\right)\cdot\vec{\beta}(0,\omega),
\end{equation}
where we note that since $S(0)=\mathds{1}$, the initial condition for both the transformed and original mode vector are the same,
\begin{equation}
	\tilde{\beta}(0,\omega) = \vec{\beta}(0,\omega).
\end{equation}
This can be further simplified by defining the geometric factors
\begin{align}
	U(x,x') =&
	\frac{1}{\tr\{\Lambda\}}\tr\{\Lambda 
		\cdot S(x)\cdot S^{-1}(x')
	\}
	\\
	=&
	\frac{1}{\tr\{\Lambda\}}\tr\{\Lambda 
		\cdot \exp(i Q(x-x'))
	\}
	\\
	=& U(x-x'),
\end{align}
where in the last line we have noted that again due to translation symmetry the geometric factor $U$ depends only on the difference of its arguments.
These can be interpreted as the field envelope of the scattered field from position $x'$, evaluated at position $x$, normalized to unit magnitude at $x'$.
The solution can then be expressed as the following Dyson series,
\begin{widetext}
\begin{align}
	B(x,\omega) =& B_{in}(x,\omega)
	\\
	&-i \frac{\zeta}{2} \tr\{\Lambda\} F(\omega) \int_{0}^{x}\dd{x_1}U(x-x_1)B_{in}(x_1,\omega)
	\\
	&-\frac{\zeta^2}{4}\tr\{\Lambda\}^2 F(\omega)^2
	\int_{0}^{x}\dd{x_1}\int_{0}^{x_1}\dd{x_2}U(x-x_1)U(x_1-x_2)B_{in}(x_2,\omega)
	\\ 
	&+ \ldots
\end{align}
\end{widetext}
where
\begin{equation}
	B_{in}(x,\omega) = \vec{1}^\top \cdot \exp(i Q x) \cdot \vec{\beta}(0,\omega)
\end{equation}
is the usual free-field solution in the absence of the resonant nuclei. In this form, the solution's nature as a multiple scattering series becomes transparent; each term is given by the sum of all scattering amplitudes to a given order, with the overall frequency dependence for a given order $m$ simply given by $F(\omega)^m$.

The spatial coefficients can be readily obtained using a recurrence relation and the Laplace transform: writing the series as 
\begin{equation}
	B(x,\omega) = \sum_{n=0}^{\infty}
	\left(-i\frac{\zeta}{2} \tr\{\Lambda\}F(\omega)
	\right)^n t_n(x),
\end{equation}
we have the following recurrence relation,
\begin{align}
	t_n(x) =& \int_0^{x}\dd{x'}U(x-x') t_{n-1}(x'),
	\\
	t_0(x) =& B_{in}(x,\omega).
\end{align}
Applying a Laplace transform gives us 
\begin{equation}
	\tilde{t}_n(s) = \tilde{U}(s) \tilde{t}_{n-1}(s),
\end{equation}
where we have denoted the Laplace transformed variables with a tilde. The solution in Laplace space is therefore simply given by 
\begin{equation}
	\tilde{t}_n(s) = \tilde{U}(s)^n \tilde{B}_{in}(s,\omega).
\end{equation}
In particular, the exact form of $\tilde{U}(s), \tilde{B}_{in}(s)$ are readily evaluated from their definition, and given by
\begin{align}
	\tilde{U}(s) =& \frac{1}{\sum_i \xi_i}\sum_i \frac{\xi_i}{s-iq_i},
	\\
	\tilde{B}_{in}(s,\omega) =& \sum_i \frac{\beta_i(\omega)}{s-iq_i}.
\end{align}
As each Laplace transformed coefficient is a rational function, the inverse transform will be a sum of polynomials multiplied by plane wave envelopes for each mode, with explicit closed-form expressions given in Appendix \ref{app:dyson}.

\section{Spatial patterning}\label{sec:patterning}
So far we have considered only a single uniform resonant layer. While the wavelength of the resonant transition is very small, on the order of angstroms, this largely contributes to an overall phase factor on the order of $e^{ik_0 x}$, which will be uniform across the sample. Any deviation from this overall plane wave phase factor can be expressed as a slowly varying envelope and phase, with length scales on the order of
\begin{equation}
	\delta x \approx \frac{1}{q_\lambda - k_0}
\end{equation}
for any mode $\lambda$. In practice, these can be fairly large, with interference beats on the order of \unit{\um} and attenuation lengths up to \unit{\cm} in scale. As such, this is on a scale at which it is practical to use techniques such as photolithography during sample preparation. Therefore, in this section we will consider layers that are spatially structured on the \unit{\um} scale, and their interaction with the guided modes.

\subsection{Micro-strips}
\begin{figure}[h!]
	\centering
	\includegraphics[width=8cm]{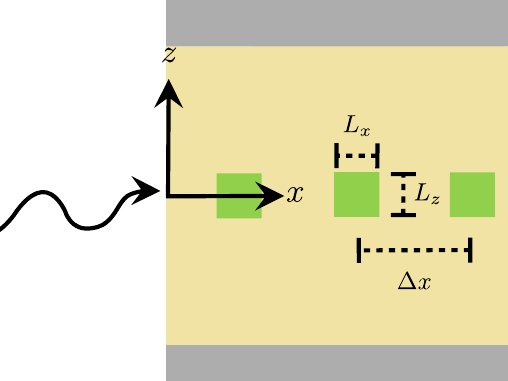}
	\caption{Front coupling geometry, with resonant layer split into micro-strips with extent $L_x, L_z$, and spacing $\Delta x$.}
	\label{fig:fc-split}
\end{figure}

The simplest system to consider is dividing the layer along the propagation direction into micrometre sized strips, Figure~\ref{fig:fc-split}.
If the strip is made sufficiently thin, such that the envelope of the scattered field is uniform across the strip dimension, it will scatter super-radiantly. To begin with, let us consider the response of a single strip. In the uniform envelope regime, the density can be taken to be
\begin{equation}
	\rho(\vec{r}) = \rho_N L_x L_z \delta(z-z_0) \delta(x-x_0),
\end{equation}
where $L_x$ is the strip $x$ extent, $L_z$ its $z$ extent, and $x_0,z_0$ the strip coordinates in the $x,z$ plane.

The equation of motion \eqref{eq:eom-vector1} of a single microstrip then becomes
\begin{equation}\label{eq:eom-micro}
	\vec{\beta}(x,\omega) = \vec{\beta}_{in}(x,\omega) -i \frac{\tau}{2} F(\omega) \exp(i Q (x-x_0))\cdot \Lambda  \cdot \vec{\beta}(x_0, \omega),
\end{equation}
where as before we have
\begin{align}
	\Lambda =& \vec{\xi}\otimes \vec{1}^\top,
	\\
	\xi_\lambda =& k_0 L_z  \frac{u_\lambda(z_0)^2}{q_\lambda},
\end{align}
and $\tau = L_x \zeta$ is the bulk optical depth of the micro-strips $x$ extent.

\subsection{Super-radiance of a single micro-strip}
To solve \eqref{eq:eom-micro}, we must solve for the self-interaction of the field. This is found by evaluating at $x=x_0$, giving
\begin{equation}
	\vec{\beta}(x_0, \omega) = \vec{\beta}_{in}(x_0,\omega)-i \frac{\tau}{2} F(\omega) \Lambda  \cdot \vec{\beta}(x_0, \omega).
\end{equation}
The solution to this equation is given by
\begin{equation}
	\vec{\beta}(x_0, \omega) = \left(\mathds{1} + i\frac{\tau}{2} F(\omega) \Lambda \right)^{-1} \cdot \vec{\beta}_{in}(x_0, \omega).
\end{equation}
This can be further simplified however, by noting that $\Lambda$ is a rank one matrix, and thus we can apply the Sherman-Morrison formula~\cite{weissteinShermanMorrisonFormula} to the inverse to obtain 
\begin{equation}
	\left(\mathds{1} + i \frac{\tau}{2} F(\omega) \Lambda \right)^{-1} = \mathds{1} - \frac{i\frac{\tau}{2} F(\omega) \Lambda}{1 + i \frac{\tau}{2} F(\omega) \tr\{\Lambda\}},
\end{equation}
where we have noted
\begin{equation}
	\tr\{\Lambda\} = \vec{1}^\top \cdot \vec{\xi} = \sum_\lambda \xi_\lambda. 
\end{equation}
Thus, we have
\begin{equation}
	\vec{\beta}(x_0, \omega) = \vec{\beta}_{in}(x_0, \omega) - \frac{i\frac{\tau}{2} F(\omega)\Lambda}{1 + i \frac{\tau}{2} F(\omega) \tr\{\Lambda\}} \cdot \vec{\beta}_{in}(x_0, \omega).
\end{equation}
Evaluating the total field then gives
\begin{equation}
	B(x_0, \omega) = B_{in}(x_0, \omega) - \frac{i\frac{\tau}{2} F(\omega) \tr\{\Lambda\}}{1 + i \frac{\tau}{2} F(\omega) \tr\{\Lambda\}} B_{in}(x_0, \omega).
\end{equation}
This allows us to directly read off the relative susceptibility of
\begin{equation}
	\chi(\omega) = - \frac{i\frac{\tau}{2} F(\omega) \tr\{\Lambda\}}{1 + i\frac{\tau}{2} F(\omega) \tr\{\Lambda\}}.
\end{equation}

\subsection{Multiple scattering}
The transfer matrix of an array of $N$ micro-strips of uniform size, with strip $i$ placed at position $x_i$ is given by
\begin{multline}
	W_{tot}(x_N,x_1,\omega) = 
	\\
	\left(\mathds{1}+ \chi(\omega)\tilde{\Lambda} \right)
	\prod_{i=1}^{N-1}
	\bigg[
	S(x_{i+1},x_i)
	\left(\mathds{1}+ \chi(\omega)\tilde{\Lambda}\right) 
	\bigg]
	,
\end{multline}
where we have defined
\begin{equation}
	\tilde{\Lambda} = \frac{\Lambda}{\tr\{\Lambda\}}.
\end{equation}
The total field can then be obtained via
\begin{equation}
	B(x,\omega) = \vec{1}^{\top}\cdot S(x,x_N)\cdot W_{tot}(x_N,x_1)\cdot S(x_1,0)\cdot \beta_{in}(0,\omega).
\end{equation}
The transmission coefficient is then given by
\begin{equation}
	T(x,\omega) = \frac{B(x,\omega)}{B_{in}(0,\omega)}.
\end{equation}
We will now proceed to expand the transmission coefficient into a multiple scattering series.
At each scattering order, the overall frequency dependence is path independent, given by
\begin{equation}
	\chi(\omega)^m
\end{equation}
for a term corresponding to $m$ scattering events. The expansion coefficient for this order is given by the sum over the geometric factors involving $m$ distinct sites,
\begin{align}
	V_m(x) =& \sum_{i_1 < i_2 \ldots i_{m-1} < i_m} U(x - x_{i_m})
	\\
	&\times\prod_{j=1}^{m-1} \left[U(x_{i_{j+1}} - x_{i_j})\right] \frac{B_{in}(x_{i_1},\omega)}{B_{in}(0,\omega)},
\end{align}
where we note that only sites between $x,x'$ are to be considered. We note at this stage that the transmission is dependent on the spatial profile of the incident field: for each scattering path the resultant amplitude depends on the input field at the first site in the path, which is sensitive to the relative weightings of the two modes in the incident field.

The final transmission coefficient is then given by the sum over all scattering orders,
\begin{equation}
	T(x,\omega) = \sum_{m=0}^{N} \chi(\omega)^m V_m(x).
\end{equation}
We can see that while a given scattering order always gives rise to the same frequency spectrum independent of the geometry of the micro-strips, the superposition of pathways of different scattering order results in interference that is greatly geometrically dependent. In the following section we will examine specific forms of the transmission coefficient for periodic arrangements coupled to two guided modes.

\section{Two mode solution: structured and unstructured layers}\label{sec:two-mode}
In this section, we will investigate in detail the case of two resonant modes, for both solid and micro-patterned resonant layers.

The relevant parameters for such a system are the relative coupling strengths of each mode, $\xi_1,\xi_2$, and the complex wave-numbers $q_1,q_2$. To simplify the analysis, we will divide these into mean and difference, and further decompose the resonant lengths into modulus and phase, and the wave-numbers into real and imaginary parts, as follows,
\begin{align}
	q_1 &= \bar{q} + \delta q + i \bar{\kappa} + i\delta \kappa,
	\\
	q_2 &= \bar{q} - \delta q + i \bar{\kappa} - i\delta \kappa,
	\\
	\bar{q} &= \frac{1}{2} \re(q_1 + q_2),
	\\
	\bar{\kappa} &= \frac{1}{2} \im(q_1 + q_2),
	\\
	\delta q &= \frac{1}{2} \re(q_1 - q_2),
	\\
	\delta \kappa &= \frac{1}{2} \im(q_1 - q_2),
	\\
	\xi_1 &= |\xi_1| e^{i\phi_1} = |\xi_1|e^{i(\bar{\phi} + \delta\phi)},
	\\
	\xi_2 &= |\xi_2| e^{i\phi_2} = |\xi_2|e^{i(\bar{\phi} - \delta\phi)},
	\\
	\bar{\phi} &= \frac{1}{2}(\phi_1 + \phi_2),
	\\
	\delta\phi &= \frac{1}{2}(\phi_1 - \phi_2).
\end{align}
For an ideal lossless waveguide, $\xi_1,\xi_2$ are purely real, and thus $\phi_1,\phi_2=0$. However, for a realistic waveguide they are small but non-vanishing. 

\subsection{Scattered field}
To begin, we will examine the geometric factor for the scattered field, given by
\begin{equation}
	U(x) = \frac{1}{\xi_1+ \xi_2}\left(\xi_1 e^{iq_1 x} + \xi_2 e^{iq_2 x}\right).
\end{equation}
The common phase can be factored out, giving
\begin{widetext}
	\begin{align}
		U(x) =& \frac{1}{
			|\xi_1|e^{i\delta \phi} +|\xi_2|e^{-i\delta \phi}
		}e^{i(\bar{q}+i\bar{\kappa})x + i \bar{\phi}}
		\left(
		|\xi_1|e^{i (\delta q + i \delta \kappa)x +i \delta\phi}
		+
		|\xi_2|e^{-i (\delta q + i \delta \kappa)x - i\delta\phi}
		\right),
		\\
		U(x)|^2 &= \frac{e^{-2\kappa x}}{|\xi_1 + \xi_2|^2}
		\left(
		|\xi_1|^2 e^{2\delta \kappa x} + |\xi_2|^2e^{-2\delta \kappa x} + 2|\xi_1||\xi_2|\re(  e^{i[(\delta q + i \delta \kappa)x + i\delta \phi]})
		\right)
		\label{eq:envelope-beat}
		\\
		\nonumber
		&= 
		\frac{e^{-2\kappa x}}{|\xi_1 + \xi_2|^2}
		\left(
		|\xi_1|^2e^{2\delta \kappa x} + |\xi_2|^2 e^{-2\delta \kappa x} + 2|\xi_1||\xi_2|\cos(\delta q x + \delta \phi)\cosh(\delta\kappa x)
		\right).
	\end{align}
\end{widetext}
In practice, as we shall see in the following section, it is possible to design waveguides such that imaginary parts of $q_1$, $q_2$ are close. We will therefore assume $\delta\kappa\approx 0$, valid for sufficiently short distances. For distances long enough for the mismatch in attenuation to be an issue, the overall attenuation will be strong regardless, so in practice the effect is negligible.

Thus, for negligible attenuation mismatch, \eqref{eq:envelope-beat} reaches an extremum for positions 
\begin{equation}
	\delta q x + \delta \phi = \pi n \quad n\in \useds{Z}.
\end{equation}
Consider a strip placed at $x_0=0$. Let the strips ahead of it be placed at locations
\begin{equation}
	x_n = \frac{\pi n - \delta\phi}{\delta q}, \quad n>0.
\end{equation}
The field reaches its maximum amplitude of 
\begin{equation}
	|U(x_n) / U(0)|^2 = e^{-2\kappa x_n}(|\xi_1| + |\xi_2|)^2.
\end{equation}
However, consider now the scattered field from $x_1$. The path difference is then given by
\begin{equation}
	x_n - x_1 = \frac{\pi (n - 1)}{\delta q},
\end{equation}
which is off target with the anti-nodes of the scattered field from $x_1$ by a distance of $\delta\phi/\delta q$. Therefore, it is impossible to place all the micro-strips to be completely constructive with each other unless $\delta\phi=0$. In practice, as we shall see, for realistic waveguides this effect is small, and over the attenuation length of the cavity modes we can consider all micro-strips to be perfectly constructive.

Let us turn our attention now to destructive interference. This occurs when the beat term is zero,
\begin{equation}
	\delta q x + \delta \phi = \pi (n + \frac{1}{2}) \quad n \in \useds{Z}.
\end{equation}
Thus, the scattered field from a strip at $x=0$ is completely out of phase with strips placed at locations
\begin{equation}
	x_n = \frac{\pi (n+ \frac{1}{2}) - \delta\phi}{\delta q}, n>0.
\end{equation}
At these locations, the unattenuated scattered amplitude reaches its minimum value of
\begin{equation}
	|U(x_n)/U(0)|^2 = e^{-2\kappa x_n}(|\xi_1| - |\xi_2|)^2.
\end{equation}
As we shall see, it is possible in practice to achieve $|\xi_1|, |\xi_2|$ very close to each other, and thus achieve a high level of destructive interference.
However, note it is not possible to get total destructive interference at all positions in a periodic array: consider three micro-strips placed $\pi/2\delta q$ apart. The second strip is transparent to the first, due to the fact that the scattered field of the first strip is completely destructively interfered. The third strip is transparent to the second. However, the third strip is located $\pi/\delta q$ from the first, and thus the first strips field is maximal. Nevertheless, this demonstrates an intriguing sub-radiant phenomenon: a period array of micro-strips at $\pi/2\delta q$ spacing can be divided into two non-interacting ensembles.

\subsection{Transmission coefficients for micro-strips}\label{sec:two-mode-micro}
We will now examine the transmission coefficients for two cases: placing the strips a whole beat and half beat wavelength, which we will refer to as constructively and destructively interfering ensembles, respectively. To understand the qualitative behaviour, we will consider the idealized case of no attenuation mismatch ($\delta\kappa = 0$), and equally coupled modes ($\xi_1 = \xi_2 = \xi$).

We first note that the overall envelope of $e^{i(\bar{q} + i\bar{\kappa})x}$ can be factored out, giving us
\begin{widetext}
	\begin{align}
		V_m(x) =& e^{i(\bar{q}+i\bar{\kappa})x} \bar{V}_m(x),
		\\
		\bar{V}_m(x)
		=&
		\sum_{i_1<i_2\ldots i_{m-1} <i_m} \bar{U}(x- x_{i_m})
		\prod_{j=1}^{m-1}
		[\bar{U}(x_{i_{j+1}} -x_{i_j})]
		\frac{\bar{B}_{in}(x_{i_1},\omega)}{\bar{B}_{in}(0,\omega)},
		\\
		\bar{U}(x) =& \frac{1}{2}(e^{i \delta q x} +e^{-i \delta q x}) = \cos(\delta q x),
		\\
		\bar{B}_{in}(x,\omega) =& \beta_1 e^{i\delta q x} + \beta_2 e^{-i\delta q x}.
	\end{align}
\end{widetext}
In the case of placing the strip locations a beat wavelength apart, $x_n = \frac{(n-1)\pi}{\delta q}$, we have
\begin{equation}
	\bar{U}(x_i - x_j) = \cos((i-j)\pi) = (-1)^{i-j}.
\end{equation}
The geometric factors then evaluate to 
\begin{align}
	\bar{V}_m(x) &= \cos(\delta q x)\sum_{i_1<i_2\ldots i_{m-1} <i_m}
	\nonumber 
	\\
	&=  \cos(\delta q x)\begin{pmatrix}
		N
		\\
		m
	\end{pmatrix},
\end{align}
where we note that all the intermediate phase factors cancel, and the sum simply evaluates to the number of $m$ combinations of the first $N$ natural numbers.
We then simply have
\begin{align}
	T(x,\omega) =& e^{i(\bar{q}+i\bar{\kappa}) x} \cos(\delta q x)\sum_{m=0}^{N} \begin{pmatrix}
		N 
		\\
		m
	\end{pmatrix} \chi(\omega)^m 
	\\
	\nonumber
	=& 
	e^{i(\bar{q}+i\bar{\kappa}) x} \cos(\delta q x) (1 + \chi(\omega))^N.
\end{align}
We note that this is the same as the transmission of $N$ micro-strips interacting with a single mode, with wave-vector $\bar{q}+i\bar{\kappa}$. 

On the other hand, for the destructively interfering strips, we have $x_n = \frac{(n-1)\pi}{2\delta q}$. We have
\begin{equation}
	\forall m \in \mathbb{Z}: U(x_{i + 2m+1}-x_i) = \cos(\left(m+\frac{1}{2}\right)\pi) = 0,
\end{equation}
and therefore any scattering events involving both even and odd positions are vanishing. We can therefore divide the ensemble into even and odd sub-ensembles, with the total transmission given by the independent transmissions of each sub-ensemble,
\begin{equation}
	T(x,\omega) = e^{i(\bar{q}+i\bar{\kappa})x}\left(T_{odd}(x,\omega) + T_{even}(x,\omega)\frac{i(\beta_1 - \beta _2)}{\beta_1 + \beta_2}\right).
\end{equation}
Here, we have used
\begin{align}
	\frac{B_{in}(x_1,\omega)}{B_{in}(0,\omega)}
	=&
	1,
	\\
	\frac{B_{in}(x_2,\omega)}{B_{in}(0,\omega)}
	=&
	\frac{i(\beta_1 - \beta_2)}{\beta_1 + \beta_2}.
\end{align}
The even and odd transmission coefficients are themselves sensitive to whether the chain ends on an even or odd strip, with the even transmission given by
\begin{widetext}
	\begin{equation}\label{eq:even}
		T_{even}(x,\omega) =
		\begin{cases}
				&\sin(\delta q x) (1+\chi(\omega))^{N/2}
				\quad N \textnormal{ is even},
 				\\
				&\sin(\delta q x) (1+\chi(\omega))^{(N-1)/2},
				\quad N \textnormal{ is odd}.
		\end{cases}
	\end{equation}
\end{widetext}

In particular, we note that if the symmetric state is driven, $\beta_1 = \beta_2$, that the even transmission will be completely vanishing, due to the fact that both the incident and scattered field would have their nodes at the even positions.
The odd transmission is given by
\begin{widetext}
	\begin{equation}\label{eq:odd}
		T_{odd}(x,\omega)
		=
		\begin{cases}
			&\cos(\delta q x)(1+\chi(\omega))^{N/2},
				\quad N \textnormal{ is even},
			\\
			&\cos(\delta q x)(1+\chi(\omega))^{(N+1)/2},
			\quad N \textnormal{ is odd}.
		\end{cases}
	\end{equation}
\end{widetext}
The temporal evolution of these solutions can be obtained analytically, and we give the derivation of the necessary response function in Appendix~\ref{app:temporal}. Specifically, in terms of the delayed response of $n$ micro-strips,
\begin{equation}
	R_n(\omega) = (1+\chi(\omega))^n - 1,
\end{equation}
the Fourier inverse of this expression is given by 
\begin{equation}\label{eq:lagurre-response}
	R_n(t) = i\nu_0 e^{-\gamma t/2 +i\nu_0 t}L^{(1)}_{n-1}(-i\nu_0 t),
\end{equation}
where $L^{(1)}_{n-1}$ is a generalized Laguerre polynomial, and 
\begin{equation}
	\nu_0  = i \tau \tr\{\Lambda \}\gamma  /4.
\end{equation}

An intriguing phenomenon is that for the case of even $N$, both the even and odd transmissions have the same number of strips and therefore frequency dependence, and thus the overall frequency dependence is simply that of a single mode waveguide with $N/2$ micro-strips. On the other hand, for odd $N$, the odd sub-ensemble has one more strip than the even sub-ensemble, which will give rise to further interference in the time spectrum due to the superposition of two spectra with different dynamical beats.

\begin{figure}[h]
	\centering
	\includegraphics[width=6cm]{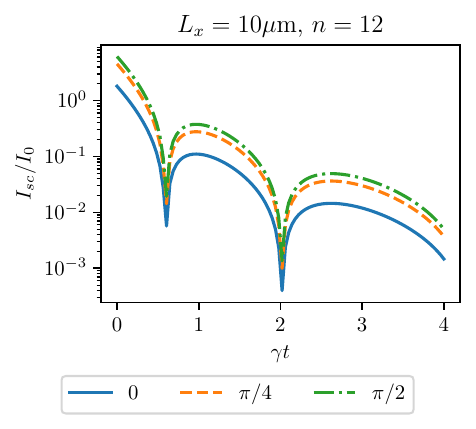}
	\caption{Example of temporal response of an even numbered interfering micro-strips, in a waveguide with parameters considered in Section~\ref{sec:numerics}. The idealized case is considered by neglecting the attentuation mismatch and the mismatch in relative coupling strengths. Three time spectra are compared at different observation points $\delta x$ relative to the last micro-strip position, measured in terms of the interference beat phase $\phi = \pi \delta x \delta q$. Due to the even parity, both sub-ensembles have the same time spectrum, and therefore shifting the observation point only scales the spectrum.}
	\label{fig:interference-even}
\end{figure}

\begin{figure}[h]
	\centering
	\includegraphics[width=6cm]{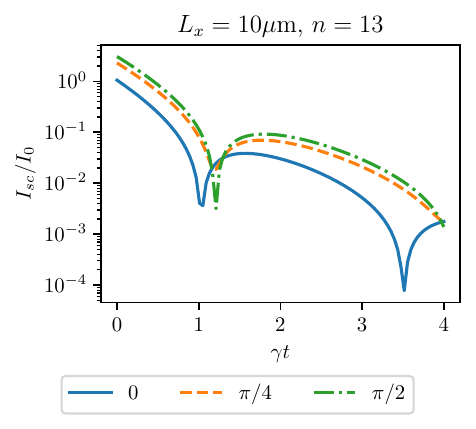}
	\caption{Odd parity case of Figure~\ref{fig:interference-even}. Due to the odd parity, the different sub-ensembles have different beat times, and therefore shifting the observation point results in a noticeable shift of the time spectra.}
	\label{fig:interference-odd}
\end{figure}

As such, the resulting temporal spectrum is sensitive not only to the number of strips in the ensemble, but the parity as well. For $N$ even, both sub-ensembles have the same temporal response, and adjusting the position $x$ at which the spectrum is evaluated results in only an overall re-scaling of the spectrum, Figure~\ref{fig:interference-even}. However, for odd $N$, the odd sub-ensemble has one more strip than the even, and the two spectra have different beat times. Adjusting the position $x$ interpolates between these two spectra, visible as a shift in the beat, Figure~\ref{fig:interference-odd}.

\section{Numerical example: two mode waveguide}\label{sec:numerics}
As a numerical study, we will consider a waveguide with Molybdenum cladding layers, a \SI{1}{\nano\metre} iron layer, and  \SI{15.8}{\nano\metre} of $\ce{B4C}$ filler on either side of the resonant layer. This wave-guide illustrates all the features developed in our model, and thus we will use it as our illustrative example. The numerically obtained parameters for this waveguide are summarized in Table~\ref{tab:summary-coupling}.
\begin{table}[h]
	\centering
	\begin{tabular}{@{}ll@{}}
		\toprule
		Parameter & Value \\
		\midrule
		$\delta q$ & \SI{152.12}{\milli\metre^{-1}}\\
		$\pi / \delta q$ & \SI{20.65}{\micro\metre}\\
		$\delta \kappa$ & \SI{-1.88}{\milli\metre^{-1}}\\
		$Q_{beat} = |\delta q/\delta \kappa|$ & 80.78 \\
		$Q_{atten} = |\delta q/\bar{\kappa}|$ & 45.91 \\
		$\bar{Q} =\sqrt{Q_{beat}Q_{atten}}$ & 60.90\\
		$\bar{q} - k_0$ & \SI{-0.3832}{\micro\metre^{-1}} \\
		$\bar{\kappa}$ & \SI{3.31}{\milli\metre^{-1}}\\
		$\xi_1$ & $\num{2.7311e-04}+\num{2.6046e-06}i$\\
		$\xi_3$ & $\num{2.7425e-04}-\num{4.3168e-06}i$\\
		$B_{1}(0,\omega) / B_{in}$ & $\num{5.4906e+00}-\num{1.9194e-03}i$\\
		$B_{3}(0,\omega) / B_{in}$ & $\num{0.61461}-\num{1.8194e-02}i$\\
		$\delta \phi$ & \SI{0.0252}{\radian}\\
		$|\xi_1/\xi_3|$ & $0.99576$ \\
		$\nu_0/\gamma$ & $\num{6.4056e-04}+\num{0.20477}i$ \\
		\bottomrule
	\end{tabular}
	\caption{Summary of two-mode coupling parameters, for the dominant modes $\lambda=1,3$ in a molybdenum waveguide. All quantities are evaluated at resonant layer centre.
	Waveguide structure is as follows:
	\\
	Mo~($\infty$)/~$\ce{B4C}$~(\SI{15.8}{\nano\metre})/~\isotope[57]{Fe}~(\SI{1}{\nano\metre})/~$\ce{B4C}$~(\SI{15.8}{\nano\metre})/~Mo~($\infty$).
	}
	\label{tab:summary-coupling}
	\end{table}

\subsection{Mode structure}
\begin{figure}[h]
	\centering
	\includegraphics[width=6cm]{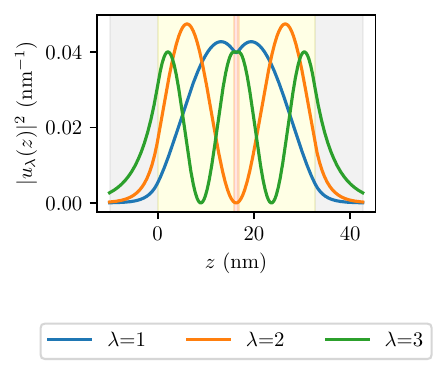}
	\caption{Normalized amplitudes of the guided modes of a molybdenum waveguide. Only the first two even modes, $\lambda=1,3$ couple to the thin nuclear layer, giving us a two mode geometry. The layer widths have been optimized for the two modes to couple almost exactly equally to the resonant layer, giving a strong interference beat in their collective radiation field.}
	\label{fig:guided-modes}
\end{figure}

\begin{figure}[h]
	\centering
	\includegraphics[width=6cm]{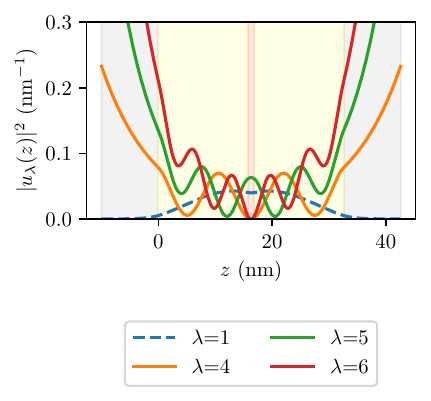}
	\caption{Normalized amplitudes of the first few leaky modes of a molybdenum waveguide, which correspond to resonances of the radiative modes. Superimposed, and dashed, is the amplitude of the first guided mode, $\lambda=1$. The exponential divergence of the leaky modes is clearly visible, demonstrating their nature as an asymptotic expansion for the near field. Although the leaky modes have amplitudes of similar magnitude to the guided modes at the resonant layer (red shading), Figures \ref{fig:modes-moly} and \ref{fig:moly-greens-function} demonstrate how the overall coupling strength is suppressed by their large attenuation.}
	\label{fig:leaky-modes}
\end{figure}

\begin{figure}[h]
	\centering
	\includegraphics[width=6cm]{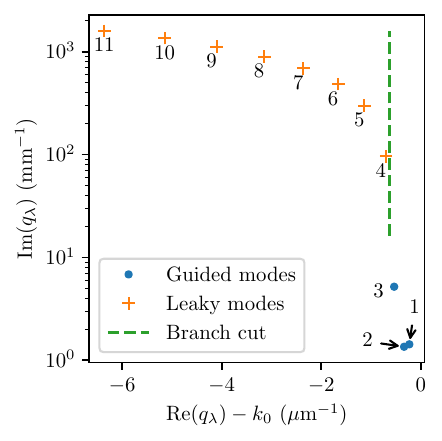}
	\caption{Relative mode wave-numbers and radiative mode branch cut for molybdenum waveguide. One can clearly see that leaky modes and guided modes are separated by the branch cut. The leaky modes are significantly attenuated compared to the guided modes, and as such are only relevant at very close range, on the order of \SI{1}{\micro\metre}.}
	\label{fig:modes-moly}
\end{figure}

\begin{figure}[h]
	\centering
	\includegraphics[width=6cm]{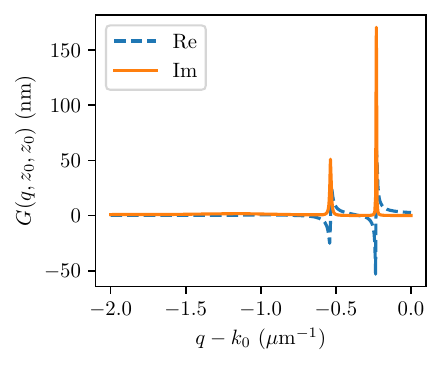}
	\caption{Fourier transformed Green's function of a molybdenum clad waveguide, evaluated at resonant layer position. One can clearly see that only the two guided modes couple with any appreciable amplitude to the nuclei, with the leaky modes heavily suppressed by attenuation.}
	\label{fig:moly-greens-function}
\end{figure}

First, we illustrate the guided and leaky mode profiles in Figures \ref{fig:guided-modes} and \ref{fig:leaky-modes}, as a function of layer depth. This waveguide supports three guided modes, but only the even modes, i.e. those that are symmetric upon reflections about $z_0$, have appreciable magnitude when evaluated at the nuclear layer. The leaky modes have similar magnitudes to the guided modes, however their attenuation is far larger, which can be observed in Figure \ref{fig:modes-moly}. This Figure illustrates the location of the guided modes, leaky modes and branch cut in the complex $q$ plane. Due to the larger attenuation of the leaky modes, their corresponding residues are suppressed by a proportional factor. To illustrate this, in Figure \ref{fig:moly-greens-function} we present the Fourier transformed Green's function along the real $q$ axis. The dominant contribution by far is that of the two even guided modes, $\lambda=1,3$, and the rest can be treated as a constant background, renormalizing the single particle decay rate.

\begin{figure}[h]
	\centering
	\includegraphics[width=6cm]{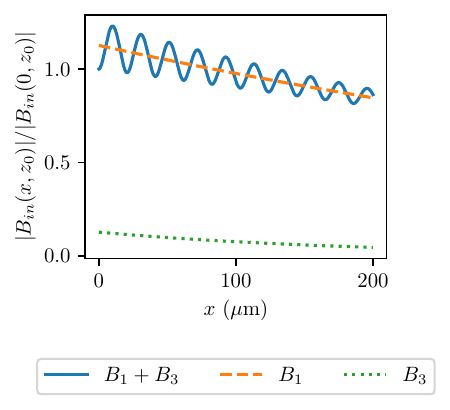}
	\caption{Amplitudes of input fields, evaluated at the layer depth $z_0$, as they propagate through the waveguide. Note the long attenuation lengths. The interference between the two modes is visible as a beat pattern with a wavelength of approximately \SI{20}{\micro\metre}. Field is normalized by total field at beginning of resonant layer; as the input fields are initially out of phase, peak values with this choice of normalization are greater than one.}
	\label{fig:moly-input}
\end{figure}

\begin{figure}[h]
	\centering
	\includegraphics[width=6cm]{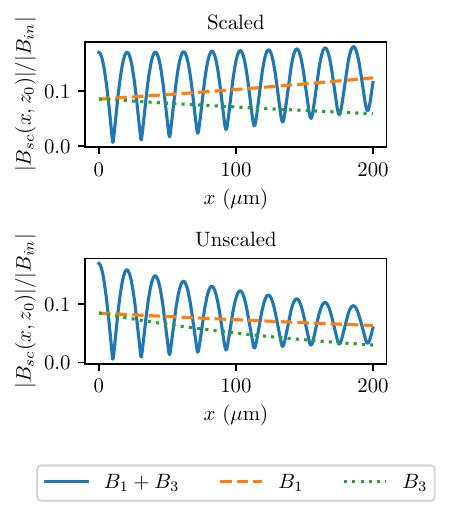}
	\caption{On-resonance scattered field for a single micro-strip, both scaled to remove the overall attenuation (top), and unscaled (bottom). One can observe the interference beat of the two participating modes. As the collective mode is the symmetric superposition of the two participating modes, a mismatch in the attenuation lengths causes the scattered field to gradually drift out of the collective mode, clearly visible as the reduced visibility of the interference beats.}
	\label{fig:moly-scatter}
\end{figure}

To evaluate the expansion coefficients for the input field of each mode, we assume a broadband, collimated input, with the free space field given by
\begin{equation}
	B_{free}(x,z,t) = \frac{B_0}{\Gamma_0} \delta\left(t - \frac{x}{c}\right).
\end{equation}
The Fourier transformed input field at the interface $x=0$ is then given by
\begin{equation}
	B_{in}(0,z,\omega) = \frac{B_0}{\Gamma_0},
\end{equation}
with the initial conditions for the mode expansions simply given by
\begin{equation}
	B_\lambda(0,\omega) = \frac{B_0}{\Gamma_0}\int_{-\infty}^{\infty}\dd{z}\frac{1}{\mu(z)}u_\lambda(z).
\end{equation}
The resultant input field intensity evaluated at the resonant layer is illustrated in Figure~\ref{fig:moly-input}. Clearly visible is the beat pattern resulting from the interference of the two modes. The first guided mode has a larger relative amplitude due to the fact that it oscillates less within the waveguide core, and as such has a larger component in the uniform input profile.
From Table~\ref{tab:summary-coupling}, we can see that the wavelength of the interference beat between the two guided modes is approximately \SI{20}{\micro\metre}. On the other hand, the attenuation lengths are much smaller, on the millimetre scale. This motivates the definition of two Q factors for the system. The first is the `beat Q factor',
\begin{equation}
	Q_{beat} = \frac{\delta q}{\delta \kappa}.
\end{equation}
This is to be qualitatively interpreted as the number of beats that occur before the attenuation mismatch causes visibility to diminish significantly. For this waveguide, it has a value of approximately 81. The second is the `attenuation Q factor',
\begin{equation}
	Q_{atten} = \frac{\delta q}{\bar{\kappa}},
\end{equation}
which measures the number of beats that occur before overall attenuation dissipates the field. For this waveguide, it is lower than $Q_{beat}$, with a value of approximately 46. We take the overall Q factor for the collective mode to be the geometric mean of these two Q factors, as both the overall attenuation and attenuation mismatch should be minimized to optimize the cavity for long range sustained collective interference. For this waveguide, the geometric mean gives an overall Q factor of approximately 61. The overall effect of attenuation is clearly illustrated in Figure~\ref{fig:moly-scatter}, which illustrates how the attenuation mismatch causes the relative strengths of the constituent fields to diverge throughout the waveguide, and thus reduces the visibility of the interference beat.

To evaluate the effect of the phase mismatch between the guided modes, which evaluates to approximately $\delta \phi = \SI{0.0252}{\radian}$, we consider the difference between perfect constructive interference, and one that is slightly off target by $\delta \phi$. This gives
\begin{equation}
	1 - \cos(0.0252) \approx 0.03\%.
\end{equation}
As such, this is negligible, especially compared with the effects of attenuation mismatch.

\subsection{Bulk layer}
\begin{figure}[h]
	\centering
	\includegraphics[width=8cm]{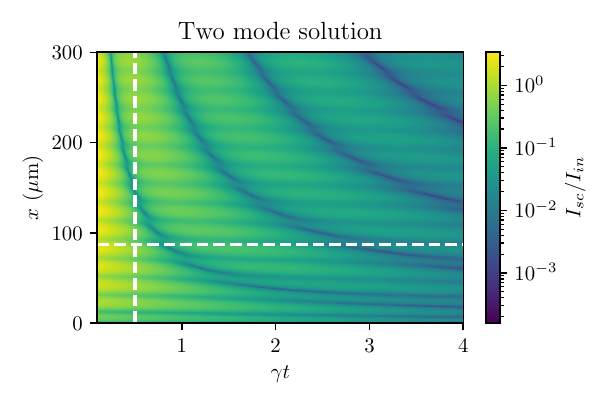}
	\caption{Scattered intensity as a function of both propagation coordinate $x$ and time $t$. Clearly visible are the approximately horizontal minimum contours, corresponding to the interference beats of the symmetric superposition of the two guided modes. For longer times, the minima are shifted closer together. White dotted horizontal and vertical lines illustrate the particular spatial and temporal slice considered in Figures \ref{fig:x-plot} and \ref{fig:t-plot} respectively.}
	\label{fig:2d-plot}
\end{figure}

\begin{figure}[h]
	\centering
	\includegraphics[width=8cm]{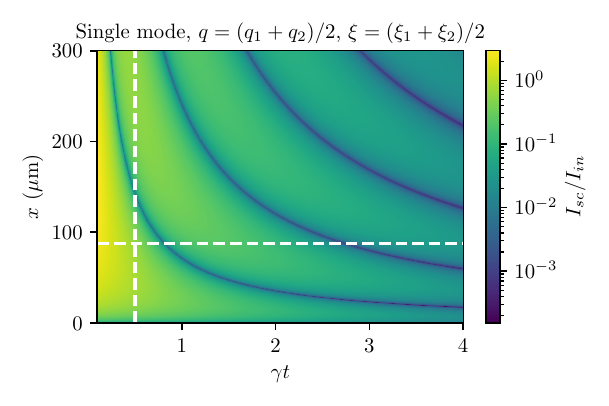}
	\caption{Scattered intensity for the mean of the two guided modes wave-numbers and optical depths, as a function of both propagation coordinate $x$ and time $t$. This gives the overall envelope of Figure~\ref{fig:2d-plot}, without the modulation of the interference beat.}
	\label{fig:2d-plot-single}
\end{figure}

First, we will examine the scattered response of a bulk layer. Figure~\ref{fig:2d-plot} gives the intensity of the scattered field as a function of propagation coordinate $x$, as well as time. For comparison, in Figure~\ref{fig:2d-plot-single} we show that overall scattered intensity resembles that of a single mode with wave-number $(q_1 + q_2)/2$ and optical depth $(\xi_1 + \xi_2)\tau/2$, where $\tau$ is the bulk material optical depth. The resemblance indicates that the resonant scattering largely occurs in the symmetric mode. 

\begin{figure}[h]
	\centering
	\includegraphics[width=6cm]{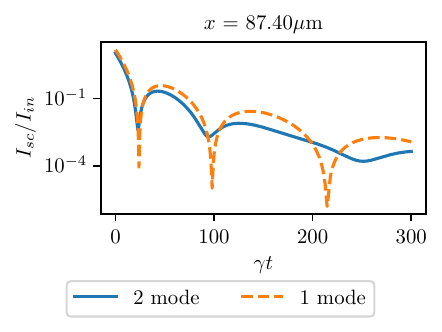}
	\caption{Scattered intensity as a function of time, for a fixed spatial extend of waveguide (blue, solid). The overall envelope somewhat resembles that of the symmetric superposition of the two modes (orange, dashed), however the interference results in a reduced amplitude and shift for the third interference beat.}
	\label{fig:x-plot}
\end{figure}

\begin{figure}[h]
	\centering
	\includegraphics[width=6cm]{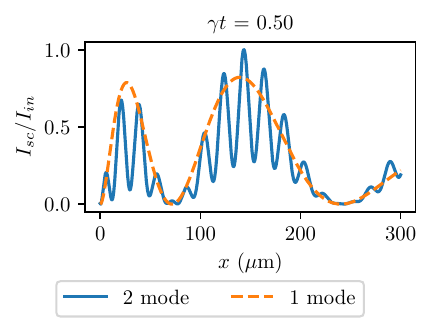}
	\caption{Scattered intensity as a function of propagation coordinate, for a fixed scattering time (blue, solid). The overall envelope strongly resembles that of the symmetric superposition of the two modes (orange, dashed), however the interference of the two modes is visible in a rapid modulation of the amplitude.}
	\label{fig:t-plot}
\end{figure}

The interference of the scattered field is visible in the periodic, approximately horizontal minima, which disrupt the dynamical beat of the symmetric mode. This affects both the temporal and spatial responses in different ways, with Figure~\ref{fig:x-plot} demonstrating that the temporal response is affected in the form missing beats. In contrast, Figure~\ref{fig:t-plot} demonstrates the scattered intensity as a function of the propagation coordinate, at a fixed time slice. Visible are the interplay of two, almost periodic oscillations, the shorter wavelength corresponding to the interference beats, with the larger wavelength corresponding to the spatial pattern of the dynamical beats.

\subsection{Microstrips}
\begin{figure}[h]
	\centering
	\includegraphics[width=6cm]{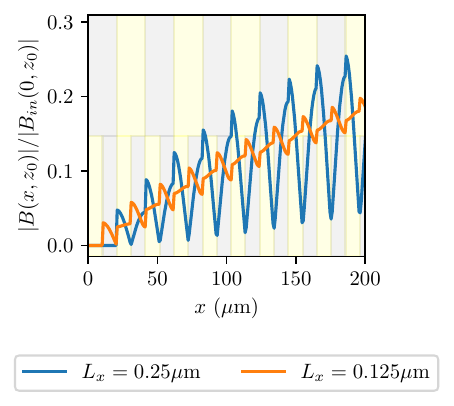}
	\caption{Comparison of on-resonance scattered intensity for super-radiant (blue) and sub-radiant (orange) geometries, with identical combined strip thickness. The super-radiant state reaches a higher peak scattered intensity, but displays the pronounced beat of the collective interference. The sub-radiant geometry displays a suppressed beat, due to the out of phase emission of the two sub-ensembles. Shading displays strip locations for constructive (top) and destructive (bottom) geometries.}
	\label{fig:moly-saturation-2}
\end{figure}

Let us now compare the constructive and destructive scattering ensembles, for an equivalent total combined strip thickness. To begin with, in Figure~\ref{fig:moly-saturation-2} we illustrate the on-resonance scattered intensity along the propagation axis, in a realistic, non-ideal waveguide. One can clearly see that the constructive ensemble reaches a larger maximum, while the destructive ensemble has a greatly suppressed interference beat due to the out of phase emission of the two sub-ensembles. However, due to the attenuation mismatch, the effect is not perfect, and the contrast in peak field strength between the two ensembles is not as high as the ideal case. 

Due to the narrow strip width, and the relatively large wavelength of the interference beat, the field envelope is very uniform over the strip's longitudinal extent. For a \SI{1}{\micro\metre} strip, the change in amplitude is approximately
\begin{equation}\label{eq:width-tradeoff}
	1 - \cos(1/20) \approx 0.12\%.
\end{equation}
Thus, we can consider the strip to follow Dicke model dynamics. This can easily be seen by the susceptibility of a single strip,
\begin{equation}
	\chi(\omega) = - \frac{i\frac{\tau}{2} F(\omega) \tr\{\Lambda\}}{1 + i\frac{\tau}{2} F(\omega) \tr\{\Lambda\}}
	=
	-\frac{\nu_0}{\omega + i\gamma/2 + \nu_0}
	,
\end{equation}
where $\nu_0 = i \gamma \tau \tr\{\Lambda\}/4$.
This is identical in form to the collective response of a grazing incidence Dicke mode~\cite{heegXrayQuantumOptics2013,lentrodtInitioQuantumModels2020,kongGreenSfunctionFormalism2020}. Compared to the response of a single nucleus, this results in an additional overall collective Lamb shift and broadening, however the effect is small, approximately $0.2\gamma$ for the broadening, and negligible Lamb shift.

\begin{figure}[h]
	\centering
	\includegraphics[width=8cm]{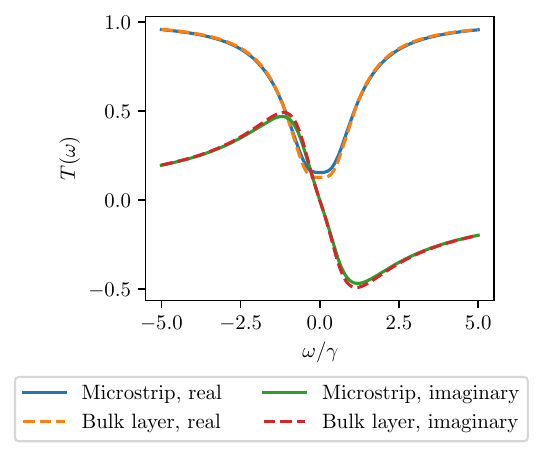}
	\caption{Comparison of absorption spectra for 30 constructively interfering micro-strips of \SI{1}{\micro\metre} width, with a single mode forward scattering spectrum with the same effective optical depth. One can see that both are qualitatively very similar.}
	\label{fig:absoprtion}
\end{figure}
\begin{figure}[h]
	\centering
	\includegraphics[width=6cm]{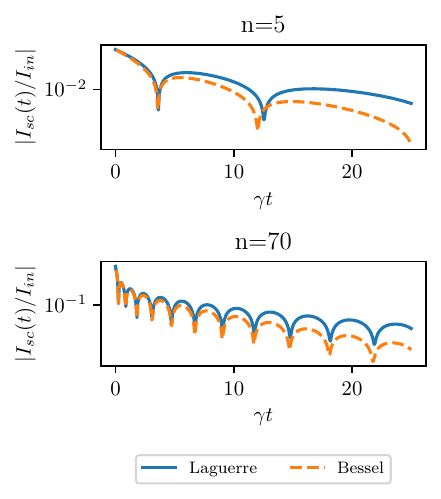}
	\caption{Comparison of scattered intensity in time domain for the microstrip Laguerre polynomial solution (solid) and solid layer Bessel function limit (dashed), for equivalent total resonant length. The responses match for short times. For larger numbers of strips, the dynamical beat of the Bessel function response matches the Laguerre response qualitatively for longer durations, however the decay of a Bessel response is more rapid.}
	\label{fig:beats}
\end{figure}

Qualitatively, the transmission spectra resemble those of nuclear forward scattering for an equivalent optical depth, as illustrated in Figure~\ref{fig:absoprtion}. As we saw in equations \eqref{eq:lagurre-response}, \eqref{eq:bessel-response}, the nuclear forward scattering spectrum is reached as the limit of large strip number. This is illustrated in Figure~\ref{fig:beats}, which compares the Laguerre polynomial response of a finite number of strips, to the large $N$ Bessel function limit. One can see that for larger strip numbers the Bessel function limit and the Laguerre response match for longer times.

\begin{figure}[h]
	\centering
	\includegraphics[width=6cm]{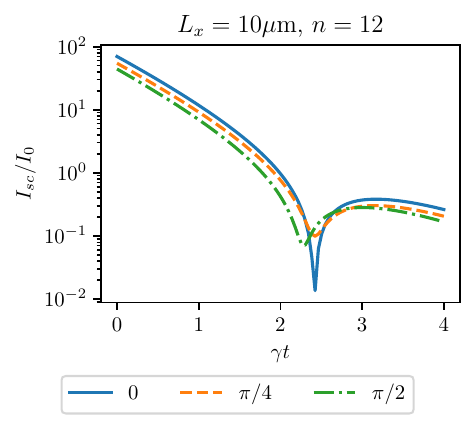}
	\caption{Example of temporal response of an even numbered interfering micro-strips, for realistic parameters considered in Section~\ref{sec:numerics}. Three time spectra are compared at different observation points $\delta x$ relative to the last micro-strip position, measured in terms of the interference beat phase $\phi = \pi \delta x \delta q$. Because of the attenuation and coupling strength mismatch of the two modes, the sub-ensembles are not completely non-interacting, and a small shift in the beat is observed (compare with Figure~\ref{fig:interference-even}).}
	\label{fig:interference-actual-even}
\end{figure}

\begin{figure}[h]
	\centering
	\includegraphics[width=6cm]{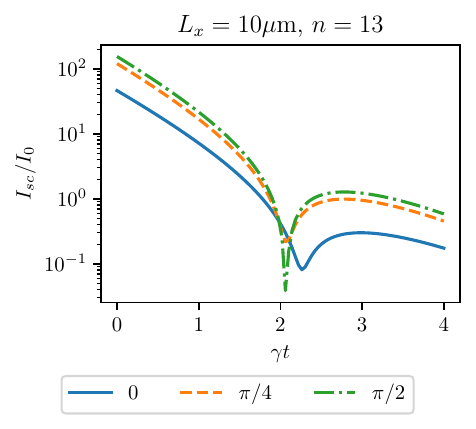}
	\caption{Odd parity case of Figure~\ref{fig:interference-actual-even}, equivalent to Figure~\ref{fig:interference-odd} with the full consideration of attenutation mismatch. Due to the odd parity, the different sub-ensembles have noticeably different beat times, and therefore shifting the observation point results in a larger shift of the time spectra.}
	\label{fig:interference-actual-odd}
\end{figure}

Due to both the attenuation mismatch and the mismatch of the relative coupling strengths of the two modes, the even-odd interference phenomenon seen in Equations \eqref{eq:even} and \eqref{eq:odd} are somewhat suppressed. This is illustrated in Figures~\ref{fig:interference-actual-even} and~\ref{fig:interference-actual-odd}, which show the temporal response of the destructively interfering ensemble for the case of 12 and 13 strips respectively. Compared to the idealized case considered in Figures~\ref{fig:interference-even} and~\ref{fig:interference-odd}, the attenuation mismatch causes a small shift in the beat time even for the case of an even number of strips.


\section{Conclusion}
We have shown that by changing the boundary conditions to forward incidence, thin film nano-structures can act as X-ray waveguides with embedded M{\"o}ssbauer nuclei. In contrast to the grazing incidence boundary condition, in the forward incidence regime the explicitly broken translational symmetry results in propagation characteristics analogous to forward scattering. As a result, dynamical beats are observed, in contrast to the single wave-vector response of grazing incidence. We demonstrated that the interaction of multiple modes with a thin resonant layer results in interference phenomena over a significantly larger length scale than the wavelength of the nuclear transition, opening a new toolbox of geometrical design for hard X-ray quantum optics. 

As a particular example of the kinds of geometric effects possible, we considered patterned micro-strips, and demonstrated novel phenomena such as a temporal response that is sensitive to the even-odd parity of the ensemble number, with a reduced optical depth compared with the bulk layer. The possible geometric designs are not limited to one dimension however, and we wish to examine two-dimensional patterned ensembles in future works. In particular, ensembles that couple in a direction transverse to the propagation direction of the incident pulse do so via a transverse wave-number that is far smaller than $k_0$. Thus, backscattering in these transverse directions is far more significant, and we hope that this could be used to implement bi-directionally coupled models that were otherwise unfeasible with ordinary forward scattering.

While in this work we have considered only slab X-ray waveguides explicitly, our approach applies to any waveguide where the propagation is unidirectional, and the waveguide has negligible dispersion across the resonant bandwidth of the scatterer. In general, in this case the guided modes will propagate with some wave-vector with respect to this coordinate system, and the Green's function will have a similar form to the expression given in~\eqref{eq:greens-function-form}, with the substitution of the $z$ coordinate with the appropriate guided mode coordinate. As such, our findings have general applicability, and could also be applied to analogous systems, such as atomic gases in hollow core fibres.

The linear nuclear response described via the linear susceptibility Equation~\eqref{eq:susceptibility} is completely justified for experiments at current generation synchrotron sources, where only a few resonant photons per shot are available. However, with XFEL sources the available bandwidths are already orders of magnitudes narrower than synchrotron sources, and with the advent of seeded XFEL sources this is set to improve even further. As such, we can expect that nonlinearity could play a larger role in future experiments. In this regime, the macroscopic Maxwell's equations for the field, Equation~\eqref{eq:maxwell-field}, will still hold at the operator level, as long as the waveguide is cooled sufficiently such that the electronic scattering remains linear. However, the magnetization field will no longer be described by a linear susceptibility, and the full Maxwell-Bloch equations for the nucleus-field interaction will have to be considered. 

\begin{acknowledgments}
P.A acknowledges funding by the Deutsches Forschungsgemeinschaft (DFG) through Projects No. 429529648 (TRR 306 QuCoLiMa) (“Quantum Cooperativity of Light and Matter”). 
L.M.L. acknowledges funding by the DFG through Projects No. 432680300 (SFB
1456-C03) and Max Planck School of Photonics. 
A.P. acknowledges funding from the DFG through the Heisenberg Program (Project PA 2508/3-1) and the Cluster of Excellence on Complexity and Topology in Quantum Matter - ct.qmat (EXC 2147, project.id 390858490).
\end{acknowledgments}

\bibliography{article.bib,revision2.bib}

\begin{thebibliography}{82}%
\makeatletter
\providecommand \@ifxundefined [1]{%
 \@ifx{#1\undefined}
}%
\providecommand \@ifnum [1]{%
 \ifnum #1\expandafter \@firstoftwo
 \else \expandafter \@secondoftwo
 \fi
}%
\providecommand \@ifx [1]{%
 \ifx #1\expandafter \@firstoftwo
 \else \expandafter \@secondoftwo
 \fi
}%
\providecommand \natexlab [1]{#1}%
\providecommand \enquote  [1]{``#1''}%
\providecommand \bibnamefont  [1]{#1}%
\providecommand \bibfnamefont [1]{#1}%
\providecommand \citenamefont [1]{#1}%
\providecommand \href@noop [0]{\@secondoftwo}%
\providecommand \href [0]{\begingroup \@sanitize@url \@href}%
\providecommand \@href[1]{\@@startlink{#1}\@@href}%
\providecommand \@@href[1]{\endgroup#1\@@endlink}%
\providecommand \@sanitize@url [0]{\catcode `\\12\catcode `\$12\catcode
  `\&12\catcode `\#12\catcode `\^12\catcode `\_12\catcode `\%12\relax}%
\providecommand \@@startlink[1]{}%
\providecommand \@@endlink[0]{}%
\providecommand \url  [0]{\begingroup\@sanitize@url \@url }%
\providecommand \@url [1]{\endgroup\@href {#1}{\urlprefix }}%
\providecommand \urlprefix  [0]{URL }%
\providecommand \Eprint [0]{\href }%
\providecommand \doibase [0]{https://doi.org/}%
\providecommand \selectlanguage [0]{\@gobble}%
\providecommand \bibinfo  [0]{\@secondoftwo}%
\providecommand \bibfield  [0]{\@secondoftwo}%
\providecommand \translation [1]{[#1]}%
\providecommand \BibitemOpen [0]{}%
\providecommand \bibitemStop [0]{}%
\providecommand \bibitemNoStop [0]{.\EOS\space}%
\providecommand \EOS [0]{\spacefactor3000\relax}%
\providecommand \BibitemShut  [1]{\csname bibitem#1\endcsname}%
\let\auto@bib@innerbib\@empty
\bibitem [{\citenamefont {Adams}\ \emph {et~al.}(2013)\citenamefont {Adams},
  \citenamefont {Buth}, \citenamefont {Cavaletto}, \citenamefont {Evers},
  \citenamefont {Harman}, \citenamefont {Keitel}, \citenamefont {P{\'a}lffy},
  \citenamefont {Pic{\'o}n}, \citenamefont {R{\"o}hlsberger}, \citenamefont
  {Rostovtsev},\ and\ \citenamefont {Tamasaku}}]{adamsXrayQuantumOptics2013a}%
  \BibitemOpen
  \bibfield  {author} {\bibinfo {author} {\bibfnamefont {B.~W.}\ \bibnamefont
  {Adams}}, \bibinfo {author} {\bibfnamefont {C.}~\bibnamefont {Buth}},
  \bibinfo {author} {\bibfnamefont {S.~M.}\ \bibnamefont {Cavaletto}}, \bibinfo
  {author} {\bibfnamefont {J.}~\bibnamefont {Evers}}, \bibinfo {author}
  {\bibfnamefont {Z.}~\bibnamefont {Harman}}, \bibinfo {author} {\bibfnamefont
  {C.~H.}\ \bibnamefont {Keitel}}, \bibinfo {author} {\bibfnamefont
  {A.}~\bibnamefont {P{\'a}lffy}}, \bibinfo {author} {\bibfnamefont
  {A.}~\bibnamefont {Pic{\'o}n}}, \bibinfo {author} {\bibfnamefont
  {R.}~\bibnamefont {R{\"o}hlsberger}}, \bibinfo {author} {\bibfnamefont
  {Y.}~\bibnamefont {Rostovtsev}},\ and\ \bibinfo {author} {\bibfnamefont
  {K.}~\bibnamefont {Tamasaku}},\ }\bibfield  {title} {\bibinfo {title} {X-ray
  quantum optics},\ }\href {https://doi.org/10.1080/09500340.2012.752113}
  {\bibfield  {journal} {\bibinfo  {journal} {Journal of Modern Optics}\
  }\textbf {\bibinfo {volume} {60}},\ \bibinfo {pages} {2} (\bibinfo {year}
  {2013})}\BibitemShut {NoStop}%
\bibitem [{\citenamefont {Hannon}\ and\ \citenamefont
  {Trammell}(1999)}]{hannonCoherentGrayOptics1999}%
  \BibitemOpen
  \bibfield  {author} {\bibinfo {author} {\bibfnamefont {J.}~\bibnamefont
  {Hannon}}\ and\ \bibinfo {author} {\bibfnamefont {G.}~\bibnamefont
  {Trammell}},\ }\bibfield  {title} {\bibinfo {title} {Coherent
  {{$\gamma$}}-{{Ray}} optics},\ }\href
  {https://doi.org/10.1023/A:1017011621007} {\bibfield  {journal} {\bibinfo
  {journal} {Hyperfine Interact.}\ }\textbf {\bibinfo {volume} {123}},\
  \bibinfo {pages} {127} (\bibinfo {year} {1999})}\BibitemShut {NoStop}%
\bibitem [{\citenamefont
  {Svidzinsky}(2012)}]{svidzinskyNonlocalEffectsSinglephoton2012}%
  \BibitemOpen
  \bibfield  {author} {\bibinfo {author} {\bibfnamefont {A.~A.}\ \bibnamefont
  {Svidzinsky}},\ }\bibfield  {title} {\bibinfo {title} {Nonlocal effects in
  single-photon superradiance},\ }\href
  {https://doi.org/10.1103/PhysRevA.85.013821} {\bibfield  {journal} {\bibinfo
  {journal} {Phys. Rev. A}\ }\textbf {\bibinfo {volume} {85}},\ \bibinfo
  {pages} {013821} (\bibinfo {year} {2012})}\BibitemShut {NoStop}%
\bibitem [{\citenamefont {Shvyd'ko}\ \emph {et~al.}(1996)\citenamefont
  {Shvyd'ko}, \citenamefont {Hertrich}, \citenamefont {{van B{\"u}rck}},
  \citenamefont {Gerdau}, \citenamefont {Leupold}, \citenamefont {Metge},
  \citenamefont {R{\"u}ter}, \citenamefont {Schwendy}, \citenamefont {Smirnov},
  \citenamefont {Potzel},\ and\ \citenamefont
  {Schindelmann}}]{shvydkoStorageNuclearExcitation1996}%
  \BibitemOpen
  \bibfield  {author} {\bibinfo {author} {\bibfnamefont {{\relax Yu}.~V.}\
  \bibnamefont {Shvyd'ko}}, \bibinfo {author} {\bibfnamefont {T.}~\bibnamefont
  {Hertrich}}, \bibinfo {author} {\bibfnamefont {U.}~\bibnamefont {{van
  B{\"u}rck}}}, \bibinfo {author} {\bibfnamefont {E.}~\bibnamefont {Gerdau}},
  \bibinfo {author} {\bibfnamefont {O.}~\bibnamefont {Leupold}}, \bibinfo
  {author} {\bibfnamefont {J.}~\bibnamefont {Metge}}, \bibinfo {author}
  {\bibfnamefont {H.~D.}\ \bibnamefont {R{\"u}ter}}, \bibinfo {author}
  {\bibfnamefont {S.}~\bibnamefont {Schwendy}}, \bibinfo {author}
  {\bibfnamefont {G.~V.}\ \bibnamefont {Smirnov}}, \bibinfo {author}
  {\bibfnamefont {W.}~\bibnamefont {Potzel}},\ and\ \bibinfo {author}
  {\bibfnamefont {P.}~\bibnamefont {Schindelmann}},\ }\bibfield  {title}
  {\bibinfo {title} {Storage of {{Nuclear Excitation Energy}} through
  {{Magnetic Switching}}},\ }\href
  {https://doi.org/10.1103/PhysRevLett.77.3232} {\bibfield  {journal} {\bibinfo
   {journal} {Phys. Rev. Lett.}\ }\textbf {\bibinfo {volume} {77}},\ \bibinfo
  {pages} {3232} (\bibinfo {year} {1996})}\BibitemShut {NoStop}%
\bibitem [{\citenamefont {Vagizov}\ \emph {et~al.}(2014)\citenamefont
  {Vagizov}, \citenamefont {Antonov}, \citenamefont {Radeonychev},
  \citenamefont {Shakhmuratov},\ and\ \citenamefont
  {Kocharovskaya}}]{vagizovCoherentControlWaveforms2014}%
  \BibitemOpen
  \bibfield  {author} {\bibinfo {author} {\bibfnamefont {F.}~\bibnamefont
  {Vagizov}}, \bibinfo {author} {\bibfnamefont {V.}~\bibnamefont {Antonov}},
  \bibinfo {author} {\bibfnamefont {Y.~V.}\ \bibnamefont {Radeonychev}},
  \bibinfo {author} {\bibfnamefont {R.~N.}\ \bibnamefont {Shakhmuratov}},\ and\
  \bibinfo {author} {\bibfnamefont {O.}~\bibnamefont {Kocharovskaya}},\
  }\bibfield  {title} {\bibinfo {title} {Coherent control of the waveforms of
  recoilless {{$\gamma$}}-{{Ray}} photons},\ }\href
  {https://doi.org/10.1038/nature13018} {\bibfield  {journal} {\bibinfo
  {journal} {Nature}\ }\textbf {\bibinfo {volume} {508}},\ \bibinfo {pages}
  {80} (\bibinfo {year} {2014})}\BibitemShut {NoStop}%
\bibitem [{\citenamefont {Coussement}\ \emph {et~al.}(2002)\citenamefont
  {Coussement}, \citenamefont {Rostovtsev}, \citenamefont {Odeurs},
  \citenamefont {Neyens}, \citenamefont {Muramatsu}, \citenamefont {Gheysen},
  \citenamefont {Callens}, \citenamefont {Vyvey}, \citenamefont {Kozyreff},
  \citenamefont {Mandel}, \citenamefont {Shakhmuratov},\ and\ \citenamefont
  {Kocharovskaya}}]{coussementControllingAbsorptionGamma2002}%
  \BibitemOpen
  \bibfield  {author} {\bibinfo {author} {\bibfnamefont {R.}~\bibnamefont
  {Coussement}}, \bibinfo {author} {\bibfnamefont {Y.}~\bibnamefont
  {Rostovtsev}}, \bibinfo {author} {\bibfnamefont {J.}~\bibnamefont {Odeurs}},
  \bibinfo {author} {\bibfnamefont {G.}~\bibnamefont {Neyens}}, \bibinfo
  {author} {\bibfnamefont {H.}~\bibnamefont {Muramatsu}}, \bibinfo {author}
  {\bibfnamefont {S.}~\bibnamefont {Gheysen}}, \bibinfo {author} {\bibfnamefont
  {R.}~\bibnamefont {Callens}}, \bibinfo {author} {\bibfnamefont
  {K.}~\bibnamefont {Vyvey}}, \bibinfo {author} {\bibfnamefont
  {G.}~\bibnamefont {Kozyreff}}, \bibinfo {author} {\bibfnamefont
  {P.}~\bibnamefont {Mandel}}, \bibinfo {author} {\bibfnamefont
  {R.}~\bibnamefont {Shakhmuratov}},\ and\ \bibinfo {author} {\bibfnamefont
  {O.}~\bibnamefont {Kocharovskaya}},\ }\bibfield  {title} {\bibinfo {title}
  {Controlling {{Absorption}} of {{Gamma Radiation}} via {{Nuclear Level
  Anticrossing}}},\ }\href {https://doi.org/10.1103/PhysRevLett.89.107601}
  {\bibfield  {journal} {\bibinfo  {journal} {Phys. Rev. Lett.}\ }\textbf
  {\bibinfo {volume} {89}},\ \bibinfo {pages} {107601} (\bibinfo {year}
  {2002})}\BibitemShut {NoStop}%
\bibitem [{\citenamefont {Kocharovskaya}\ \emph {et~al.}(1999)\citenamefont
  {Kocharovskaya}, \citenamefont {Kolesov},\ and\ \citenamefont
  {Rostovtsev}}]{kocharovskayaCoherentOpticalControl1999}%
  \BibitemOpen
  \bibfield  {author} {\bibinfo {author} {\bibfnamefont {O.}~\bibnamefont
  {Kocharovskaya}}, \bibinfo {author} {\bibfnamefont {R.}~\bibnamefont
  {Kolesov}},\ and\ \bibinfo {author} {\bibfnamefont {Y.}~\bibnamefont
  {Rostovtsev}},\ }\bibfield  {title} {\bibinfo {title} {Coherent optical
  control of {{M\"ossbauer}} spectra},\ }\href
  {https://doi.org/10.1103/PhysRevLett.82.3593} {\bibfield  {journal} {\bibinfo
   {journal} {Phys. Rev. Lett.}\ }\textbf {\bibinfo {volume} {82}},\ \bibinfo
  {pages} {3593} (\bibinfo {year} {1999})}\BibitemShut {NoStop}%
\bibitem [{\citenamefont {Heeg}\ \emph {et~al.}(2017)\citenamefont {Heeg},
  \citenamefont {Kaldun}, \citenamefont {Strohm}, \citenamefont {Reiser},
  \citenamefont {Ott}, \citenamefont {Subramanian}, \citenamefont {Lentrodt},
  \citenamefont {Haber}, \citenamefont {Wille}, \citenamefont {Goerttler},
  \citenamefont {R{\"u}ffer}, \citenamefont {Keitel}, \citenamefont
  {R{\"o}hlsberger}, \citenamefont {Pfeifer},\ and\ \citenamefont
  {Evers}}]{heegSpectralNarrowingXray2017}%
  \BibitemOpen
  \bibfield  {author} {\bibinfo {author} {\bibfnamefont {K.~P.}\ \bibnamefont
  {Heeg}}, \bibinfo {author} {\bibfnamefont {A.}~\bibnamefont {Kaldun}},
  \bibinfo {author} {\bibfnamefont {C.}~\bibnamefont {Strohm}}, \bibinfo
  {author} {\bibfnamefont {P.}~\bibnamefont {Reiser}}, \bibinfo {author}
  {\bibfnamefont {C.}~\bibnamefont {Ott}}, \bibinfo {author} {\bibfnamefont
  {R.}~\bibnamefont {Subramanian}}, \bibinfo {author} {\bibfnamefont
  {D.}~\bibnamefont {Lentrodt}}, \bibinfo {author} {\bibfnamefont
  {J.}~\bibnamefont {Haber}}, \bibinfo {author} {\bibfnamefont {H.-C.}\
  \bibnamefont {Wille}}, \bibinfo {author} {\bibfnamefont {S.}~\bibnamefont
  {Goerttler}}, \bibinfo {author} {\bibfnamefont {R.}~\bibnamefont
  {R{\"u}ffer}}, \bibinfo {author} {\bibfnamefont {C.~H.}\ \bibnamefont
  {Keitel}}, \bibinfo {author} {\bibfnamefont {R.}~\bibnamefont
  {R{\"o}hlsberger}}, \bibinfo {author} {\bibfnamefont {T.}~\bibnamefont
  {Pfeifer}},\ and\ \bibinfo {author} {\bibfnamefont {J.}~\bibnamefont
  {Evers}},\ }\bibfield  {title} {\bibinfo {title} {Spectral narrowing of
  {{X-ray}} pulses for precision spectroscopy with nuclear resonances},\ }\href
  {https://doi.org/10.1126/science.aan3512} {\bibfield  {journal} {\bibinfo
  {journal} {Science}\ }\textbf {\bibinfo {volume} {357}},\ \bibinfo {pages}
  {375} (\bibinfo {year} {2017})},\ \Eprint {https://arxiv.org/abs/28751603}
  {28751603} \BibitemShut {NoStop}%
\bibitem [{\citenamefont {Heeg}\ \emph {et~al.}(2021)\citenamefont {Heeg},
  \citenamefont {Kaldun}, \citenamefont {Strohm}, \citenamefont {Ott},
  \citenamefont {Subramanian}, \citenamefont {Lentrodt}, \citenamefont {Haber},
  \citenamefont {Wille}, \citenamefont {Goerttler}, \citenamefont {R{\"u}ffer},
  \citenamefont {Keitel}, \citenamefont {R{\"o}hlsberger}, \citenamefont
  {Pfeifer},\ and\ \citenamefont {Evers}}]{heegCoherentXrayOptical2021}%
  \BibitemOpen
  \bibfield  {author} {\bibinfo {author} {\bibfnamefont {K.~P.}\ \bibnamefont
  {Heeg}}, \bibinfo {author} {\bibfnamefont {A.}~\bibnamefont {Kaldun}},
  \bibinfo {author} {\bibfnamefont {C.}~\bibnamefont {Strohm}}, \bibinfo
  {author} {\bibfnamefont {C.}~\bibnamefont {Ott}}, \bibinfo {author}
  {\bibfnamefont {R.}~\bibnamefont {Subramanian}}, \bibinfo {author}
  {\bibfnamefont {D.}~\bibnamefont {Lentrodt}}, \bibinfo {author}
  {\bibfnamefont {J.}~\bibnamefont {Haber}}, \bibinfo {author} {\bibfnamefont
  {H.-C.}\ \bibnamefont {Wille}}, \bibinfo {author} {\bibfnamefont
  {S.}~\bibnamefont {Goerttler}}, \bibinfo {author} {\bibfnamefont
  {R.}~\bibnamefont {R{\"u}ffer}}, \bibinfo {author} {\bibfnamefont {C.~H.}\
  \bibnamefont {Keitel}}, \bibinfo {author} {\bibfnamefont {R.}~\bibnamefont
  {R{\"o}hlsberger}}, \bibinfo {author} {\bibfnamefont {T.}~\bibnamefont
  {Pfeifer}},\ and\ \bibinfo {author} {\bibfnamefont {J.}~\bibnamefont
  {Evers}},\ }\bibfield  {title} {\bibinfo {title} {Coherent {{X-ray-optical}}
  control of nuclear excitons},\ }\href
  {https://doi.org/10.1038/s41586-021-03276-x} {\bibfield  {journal} {\bibinfo
  {journal} {Nature}\ }\textbf {\bibinfo {volume} {590}},\ \bibinfo {pages}
  {401} (\bibinfo {year} {2021})}\BibitemShut {NoStop}%
\bibitem [{\citenamefont {Chumakov}\ \emph {et~al.}(2018)\citenamefont
  {Chumakov}, \citenamefont {Baron}, \citenamefont {Sergueev}, \citenamefont
  {Strohm}, \citenamefont {Leupold}, \citenamefont {Shvyd'ko}, \citenamefont
  {Smirnov}, \citenamefont {R{\"u}ffer}, \citenamefont {Inubushi},
  \citenamefont {Yabashi}, \citenamefont {Tono}, \citenamefont {Kudo},\ and\
  \citenamefont {Ishikawa}}]{chumakovSuperradianceEnsembleNuclei2018}%
  \BibitemOpen
  \bibfield  {author} {\bibinfo {author} {\bibfnamefont {A.~I.}\ \bibnamefont
  {Chumakov}}, \bibinfo {author} {\bibfnamefont {A.~Q.~R.}\ \bibnamefont
  {Baron}}, \bibinfo {author} {\bibfnamefont {I.}~\bibnamefont {Sergueev}},
  \bibinfo {author} {\bibfnamefont {C.}~\bibnamefont {Strohm}}, \bibinfo
  {author} {\bibfnamefont {O.}~\bibnamefont {Leupold}}, \bibinfo {author}
  {\bibfnamefont {Y.}~\bibnamefont {Shvyd'ko}}, \bibinfo {author}
  {\bibfnamefont {G.~V.}\ \bibnamefont {Smirnov}}, \bibinfo {author}
  {\bibfnamefont {R.}~\bibnamefont {R{\"u}ffer}}, \bibinfo {author}
  {\bibfnamefont {Y.}~\bibnamefont {Inubushi}}, \bibinfo {author}
  {\bibfnamefont {M.}~\bibnamefont {Yabashi}}, \bibinfo {author} {\bibfnamefont
  {K.}~\bibnamefont {Tono}}, \bibinfo {author} {\bibfnamefont {T.}~\bibnamefont
  {Kudo}},\ and\ \bibinfo {author} {\bibfnamefont {T.}~\bibnamefont
  {Ishikawa}},\ }\bibfield  {title} {\bibinfo {title} {Superradiance of an
  ensemble of nuclei excited by a free electron laser},\ }\href
  {https://doi.org/10.1038/s41567-017-0001-z} {\bibfield  {journal} {\bibinfo
  {journal} {Nat. Phys.}\ }\textbf {\bibinfo {volume} {14}},\ \bibinfo {pages}
  {261} (\bibinfo {year} {2018})}\BibitemShut {NoStop}%
\bibitem [{\citenamefont {Spiller}\ and\ \citenamefont
  {Segm{\"u}ller}(1974)}]{spillerPropagationRaysWaveguides1974}%
  \BibitemOpen
  \bibfield  {author} {\bibinfo {author} {\bibfnamefont {E.}~\bibnamefont
  {Spiller}}\ and\ \bibinfo {author} {\bibfnamefont {A.}~\bibnamefont
  {Segm{\"u}ller}},\ }\bibfield  {title} {\bibinfo {title} {Propagation of x
  rays in waveguides},\ }\href {https://doi.org/10.1063/1.1655093} {\bibfield
  {journal} {\bibinfo  {journal} {Applied Physics Letters}\ }\textbf {\bibinfo
  {volume} {24}},\ \bibinfo {pages} {60} (\bibinfo {year} {1974})}\BibitemShut
  {NoStop}%
\bibitem [{\citenamefont {Zwanenburg}\ \emph {et~al.}(2000)\citenamefont
  {Zwanenburg}, \citenamefont {Bongaerts}, \citenamefont {Peters},
  \citenamefont {Riese},\ and\ \citenamefont {van {der
  Veen}}}]{Zwanenburg2000}%
  \BibitemOpen
  \bibfield  {author} {\bibinfo {author} {\bibfnamefont {M.}~\bibnamefont
  {Zwanenburg}}, \bibinfo {author} {\bibfnamefont {J.}~\bibnamefont
  {Bongaerts}}, \bibinfo {author} {\bibfnamefont {J.}~\bibnamefont {Peters}},
  \bibinfo {author} {\bibfnamefont {D.}~\bibnamefont {Riese}},\ and\ \bibinfo
  {author} {\bibfnamefont {J.}~\bibnamefont {van {der Veen}}},\ }\bibfield
  {title} {\bibinfo {title} {Focusing of coherent {{X-rays}} in a tapered
  planar waveguide},\ }\href {https://doi.org/10.1016/S0921-4526(99)02003-7}
  {\bibfield  {journal} {\bibinfo  {journal} {Physica B: Condensed Matter}\
  }\textbf {\bibinfo {volume} {283}},\ \bibinfo {pages} {285} (\bibinfo {year}
  {2000})}\BibitemShut {NoStop}%
\bibitem [{\citenamefont {Pfeiffer}\ \emph {et~al.}(2002)\citenamefont
  {Pfeiffer}, \citenamefont {David}, \citenamefont {Burghammer}, \citenamefont
  {Riekel},\ and\ \citenamefont {{T. Salditt}}}]{Pfeiffer2002}%
  \BibitemOpen
  \bibfield  {author} {\bibinfo {author} {\bibfnamefont {F.}~\bibnamefont
  {Pfeiffer}}, \bibinfo {author} {\bibfnamefont {C.}~\bibnamefont {David}},
  \bibinfo {author} {\bibfnamefont {M.}~\bibnamefont {Burghammer}}, \bibinfo
  {author} {\bibfnamefont {C.}~\bibnamefont {Riekel}},\ and\ \bibinfo {author}
  {\bibnamefont {{T. Salditt}}},\ }\bibfield  {title} {\bibinfo {title}
  {Two-dimensional {{X-ray}} waveguides and point sources},\ }\href
  {https://doi.org/10.1126/science.1071994} {\bibfield  {journal} {\bibinfo
  {journal} {Science}\ }\textbf {\bibinfo {volume} {297}},\ \bibinfo {pages}
  {230} (\bibinfo {year} {2002})},\ \Eprint
  {https://arxiv.org/abs/https://www.science.org/doi/pdf/10.1126/science.1071994}
  {https://www.science.org/doi/pdf/10.1126/science.1071994} \BibitemShut
  {NoStop}%
\bibitem [{\citenamefont {Jarre}\ \emph {et~al.}(2005)\citenamefont {Jarre},
  \citenamefont {Fuhse}, \citenamefont {Ollinger}, \citenamefont {Seeger},
  \citenamefont {Tucoulou},\ and\ \citenamefont {Salditt}}]{Jarre2005}%
  \BibitemOpen
  \bibfield  {author} {\bibinfo {author} {\bibfnamefont {A.}~\bibnamefont
  {Jarre}}, \bibinfo {author} {\bibfnamefont {C.}~\bibnamefont {Fuhse}},
  \bibinfo {author} {\bibfnamefont {C.}~\bibnamefont {Ollinger}}, \bibinfo
  {author} {\bibfnamefont {J.}~\bibnamefont {Seeger}}, \bibinfo {author}
  {\bibfnamefont {R.}~\bibnamefont {Tucoulou}},\ and\ \bibinfo {author}
  {\bibfnamefont {T.}~\bibnamefont {Salditt}},\ }\bibfield  {title} {\bibinfo
  {title} {Two-dimensional hard {{X-ray}} beam compression by combined focusing
  and waveguide optics},\ }\href
  {https://doi.org/10.1103/PhysRevLett.94.074801} {\bibfield  {journal}
  {\bibinfo  {journal} {Phys. Rev. Lett.}\ }\textbf {\bibinfo {volume} {94}},\
  \bibinfo {pages} {074801} (\bibinfo {year} {2005})}\BibitemShut {NoStop}%
\bibitem [{\citenamefont {Giewekemeyer}\ \emph {et~al.}(2010)\citenamefont
  {Giewekemeyer}, \citenamefont {Neubauer}, \citenamefont {Kalbfleisch},
  \citenamefont {Kr{\"u}ger},\ and\ \citenamefont
  {Salditt}}]{Giewekemeyer_NJoP_2010}%
  \BibitemOpen
  \bibfield  {author} {\bibinfo {author} {\bibfnamefont {K.}~\bibnamefont
  {Giewekemeyer}}, \bibinfo {author} {\bibfnamefont {H.}~\bibnamefont
  {Neubauer}}, \bibinfo {author} {\bibfnamefont {S.}~\bibnamefont
  {Kalbfleisch}}, \bibinfo {author} {\bibfnamefont {S.~P.}\ \bibnamefont
  {Kr{\"u}ger}},\ and\ \bibinfo {author} {\bibfnamefont {T.}~\bibnamefont
  {Salditt}},\ }\bibfield  {title} {\bibinfo {title} {Holographic and
  diffractive x-ray imaging using waveguides as quasi-point sources},\ }\href
  {https://doi.org/10.1088/1367-2630/12/3/035008} {\bibfield  {journal}
  {\bibinfo  {journal} {New Journal of Physics}\ }\textbf {\bibinfo {volume}
  {12}},\ \bibinfo {pages} {035008} (\bibinfo {year} {2010})}\BibitemShut
  {NoStop}%
\bibitem [{\citenamefont {Kr{\"u}ger}\ \emph {et~al.}(2012)\citenamefont
  {Kr{\"u}ger}, \citenamefont {Neubauer}, \citenamefont {Bartels},
  \citenamefont {Kalbfleisch}, \citenamefont {Giewekemeyer}, \citenamefont
  {Wilbrandt}, \citenamefont {Sprung},\ and\ \citenamefont
  {Salditt}}]{Kruger2012}%
  \BibitemOpen
  \bibfield  {author} {\bibinfo {author} {\bibfnamefont {S.~P.}\ \bibnamefont
  {Kr{\"u}ger}}, \bibinfo {author} {\bibfnamefont {H.}~\bibnamefont
  {Neubauer}}, \bibinfo {author} {\bibfnamefont {M.}~\bibnamefont {Bartels}},
  \bibinfo {author} {\bibfnamefont {S.}~\bibnamefont {Kalbfleisch}}, \bibinfo
  {author} {\bibfnamefont {K.}~\bibnamefont {Giewekemeyer}}, \bibinfo {author}
  {\bibfnamefont {P.~J.}\ \bibnamefont {Wilbrandt}}, \bibinfo {author}
  {\bibfnamefont {M.}~\bibnamefont {Sprung}},\ and\ \bibinfo {author}
  {\bibfnamefont {T.}~\bibnamefont {Salditt}},\ }\bibfield  {title} {\bibinfo
  {title} {Sub-10nm beam confinement by {{X-ray}} waveguides: Design,
  fabrication and characterization of optical properties},\ }\href
  {https://doi.org/10.1107/S0909049511051983} {\bibfield  {journal} {\bibinfo
  {journal} {Journal of Synchrotron Radiation}\ }\textbf {\bibinfo {volume}
  {19}},\ \bibinfo {pages} {227} (\bibinfo {year} {2012})}\BibitemShut
  {NoStop}%
\bibitem [{\citenamefont {Chen}\ \emph {et~al.}(2015)\citenamefont {Chen},
  \citenamefont {Hoffmann},\ and\ \citenamefont {Salditt}}]{Chen2015}%
  \BibitemOpen
  \bibfield  {author} {\bibinfo {author} {\bibfnamefont {H.-Y.}\ \bibnamefont
  {Chen}}, \bibinfo {author} {\bibfnamefont {S.}~\bibnamefont {Hoffmann}},\
  and\ \bibinfo {author} {\bibfnamefont {T.}~\bibnamefont {Salditt}},\
  }\bibfield  {title} {\bibinfo {title} {X-ray beam compression by tapered
  waveguides},\ }\href {https://doi.org/10.1063/1.4921095} {\bibfield
  {journal} {\bibinfo  {journal} {Applied Physics Letters}\ }\textbf {\bibinfo
  {volume} {106}},\ \bibinfo {pages} {194105} (\bibinfo {year} {2015})},\
  \Eprint {https://arxiv.org/abs/https://doi.org/10.1063/1.4921095}
  {https://doi.org/10.1063/1.4921095} \BibitemShut {NoStop}%
\bibitem [{\citenamefont {{Hoffmann-Urlaub}}\ \emph {et~al.}(2016)\citenamefont
  {{Hoffmann-Urlaub}}, \citenamefont {H{\"o}hne}, \citenamefont {Kanbach},\
  and\ \citenamefont {Salditt}}]{Hoffmann2016}%
  \BibitemOpen
  \bibfield  {author} {\bibinfo {author} {\bibfnamefont {S.}~\bibnamefont
  {{Hoffmann-Urlaub}}}, \bibinfo {author} {\bibfnamefont {P.}~\bibnamefont
  {H{\"o}hne}}, \bibinfo {author} {\bibfnamefont {M.}~\bibnamefont {Kanbach}},\
  and\ \bibinfo {author} {\bibfnamefont {T.}~\bibnamefont {Salditt}},\
  }\bibfield  {title} {\bibinfo {title} {Advances in fabrication of {{X-ray}}
  waveguides},\ }\href {https://doi.org/10.1016/j.mee.2016.07.010} {\bibfield
  {journal} {\bibinfo  {journal} {Microelectronic Engineering}\ }\textbf
  {\bibinfo {volume} {164}},\ \bibinfo {pages} {135} (\bibinfo {year}
  {2016})}\BibitemShut {NoStop}%
\bibitem [{\citenamefont {Zhong}\ \emph {et~al.}(2017)\citenamefont {Zhong},
  \citenamefont {Osterhoff}, \citenamefont {Wen}, \citenamefont {Wang},\ and\
  \citenamefont {Salditt}}]{Zhong_XS_2017}%
  \BibitemOpen
  \bibfield  {author} {\bibinfo {author} {\bibfnamefont {Q.}~\bibnamefont
  {Zhong}}, \bibinfo {author} {\bibfnamefont {M.}~\bibnamefont {Osterhoff}},
  \bibinfo {author} {\bibfnamefont {M.~W.}\ \bibnamefont {Wen}}, \bibinfo
  {author} {\bibfnamefont {Z.~S.}\ \bibnamefont {Wang}},\ and\ \bibinfo
  {author} {\bibfnamefont {T.}~\bibnamefont {Salditt}},\ }\bibfield  {title}
  {\bibinfo {title} {X-ray waveguide arrays: tailored near fields by multi-beam
  interference},\ }\href {https://doi.org/10.1002/xrs.2740} {\bibfield
  {journal} {\bibinfo  {journal} {X-Ray Spectrometry}\ }\textbf {\bibinfo
  {volume} {46}},\ \bibinfo {pages} {107} (\bibinfo {year} {2017})}\BibitemShut
  {NoStop}%
\bibitem [{\citenamefont {Caro}\ \emph {et~al.}(2008)\citenamefont {Caro},
  \citenamefont {Giannini}, \citenamefont {Pelliccia}, \citenamefont {Mocuta},
  \citenamefont {Metzger}, \citenamefont {Guagliardi}, \citenamefont {Cedola},
  \citenamefont {Burkeeva},\ and\ \citenamefont {Lagomarsino}}]{Caro_PRB_2008}%
  \BibitemOpen
  \bibfield  {author} {\bibinfo {author} {\bibfnamefont {L.~D.}\ \bibnamefont
  {Caro}}, \bibinfo {author} {\bibfnamefont {C.}~\bibnamefont {Giannini}},
  \bibinfo {author} {\bibfnamefont {D.}~\bibnamefont {Pelliccia}}, \bibinfo
  {author} {\bibfnamefont {C.}~\bibnamefont {Mocuta}}, \bibinfo {author}
  {\bibfnamefont {T.~H.}\ \bibnamefont {Metzger}}, \bibinfo {author}
  {\bibfnamefont {A.}~\bibnamefont {Guagliardi}}, \bibinfo {author}
  {\bibfnamefont {A.}~\bibnamefont {Cedola}}, \bibinfo {author} {\bibfnamefont
  {I.}~\bibnamefont {Burkeeva}},\ and\ \bibinfo {author} {\bibfnamefont
  {S.}~\bibnamefont {Lagomarsino}},\ }\bibfield  {title} {\bibinfo {title}
  {In-line holography and coherent diffractive imaging with x-ray waveguides},\
  }\href {https://doi.org/10.1103/physrevb.77.081408} {\bibfield  {journal}
  {\bibinfo  {journal} {Physical Review B}\ }\textbf {\bibinfo {volume} {77}},\
  \bibinfo {pages} {081408} (\bibinfo {year} {2008})}\BibitemShut {NoStop}%
\bibitem [{\citenamefont {Salditt}\ \emph {et~al.}(2015)\citenamefont
  {Salditt}, \citenamefont {Osterhoff}, \citenamefont {Krenkel}, \citenamefont
  {Wilke}, \citenamefont {Priebe}, \citenamefont {Bartels}, \citenamefont
  {Kalbfleisch},\ and\ \citenamefont {Sprung}}]{Salditt_JoSR_2015}%
  \BibitemOpen
  \bibfield  {author} {\bibinfo {author} {\bibfnamefont {T.}~\bibnamefont
  {Salditt}}, \bibinfo {author} {\bibfnamefont {M.}~\bibnamefont {Osterhoff}},
  \bibinfo {author} {\bibfnamefont {M.}~\bibnamefont {Krenkel}}, \bibinfo
  {author} {\bibfnamefont {R.~N.}\ \bibnamefont {Wilke}}, \bibinfo {author}
  {\bibfnamefont {M.}~\bibnamefont {Priebe}}, \bibinfo {author} {\bibfnamefont
  {M.}~\bibnamefont {Bartels}}, \bibinfo {author} {\bibfnamefont
  {S.}~\bibnamefont {Kalbfleisch}},\ and\ \bibinfo {author} {\bibfnamefont
  {M.}~\bibnamefont {Sprung}},\ }\bibfield  {title} {\bibinfo {title} {Compound
  focusing mirror and x-ray waveguide optics for coherent imaging and
  nano-diffraction},\ }\href {https://doi.org/10.1107/s1600577515007742}
  {\bibfield  {journal} {\bibinfo  {journal} {Journal of Synchrotron
  Radiation}\ }\textbf {\bibinfo {volume} {22}},\ \bibinfo {pages} {867}
  (\bibinfo {year} {2015})}\BibitemShut {NoStop}%
\bibitem [{\citenamefont {Bergemann}\ \emph {et~al.}(2003)\citenamefont
  {Bergemann}, \citenamefont {Keymeulen},\ and\ \citenamefont {{van der
  Veen}}}]{bergemannFocusingXRayBeams2003}%
  \BibitemOpen
  \bibfield  {author} {\bibinfo {author} {\bibfnamefont {C.}~\bibnamefont
  {Bergemann}}, \bibinfo {author} {\bibfnamefont {H.}~\bibnamefont
  {Keymeulen}},\ and\ \bibinfo {author} {\bibfnamefont {J.~F.}\ \bibnamefont
  {{van der Veen}}},\ }\bibfield  {title} {\bibinfo {title} {Focusing {{X-Ray
  Beams}} to {{Nanometer Dimensions}}},\ }\href
  {https://doi.org/10.1103/PhysRevLett.91.204801} {\bibfield  {journal}
  {\bibinfo  {journal} {Phys. Rev. Lett.}\ }\textbf {\bibinfo {volume} {91}},\
  \bibinfo {pages} {204801} (\bibinfo {year} {2003})}\BibitemShut {NoStop}%
\bibitem [{\citenamefont {Okamoto}\ \emph {et~al.}(2012)\citenamefont
  {Okamoto}, \citenamefont {Noma}, \citenamefont {Komoto}, \citenamefont
  {Kubo}, \citenamefont {Takahashi}, \citenamefont {Iida},\ and\ \citenamefont
  {Miyata}}]{okamotoXrayWaveguideMode2012}%
  \BibitemOpen
  \bibfield  {author} {\bibinfo {author} {\bibfnamefont {K.}~\bibnamefont
  {Okamoto}}, \bibinfo {author} {\bibfnamefont {T.}~\bibnamefont {Noma}},
  \bibinfo {author} {\bibfnamefont {A.}~\bibnamefont {Komoto}}, \bibinfo
  {author} {\bibfnamefont {W.}~\bibnamefont {Kubo}}, \bibinfo {author}
  {\bibfnamefont {M.}~\bibnamefont {Takahashi}}, \bibinfo {author}
  {\bibfnamefont {A.}~\bibnamefont {Iida}},\ and\ \bibinfo {author}
  {\bibfnamefont {H.}~\bibnamefont {Miyata}},\ }\bibfield  {title} {\bibinfo
  {title} {X-ray {{Waveguide Mode}} in {{Resonance}} with a {{Periodic
  Structure}}},\ }\href {https://doi.org/10.1103/PhysRevLett.109.233907}
  {\bibfield  {journal} {\bibinfo  {journal} {Phys. Rev. Lett.}\ }\textbf
  {\bibinfo {volume} {109}},\ \bibinfo {pages} {233907} (\bibinfo {year}
  {2012})}\BibitemShut {NoStop}%
\bibitem [{\citenamefont {Bukreeva}\ \emph {et~al.}(2011)\citenamefont
  {Bukreeva}, \citenamefont {Cedola}, \citenamefont {Sorrentino}, \citenamefont
  {Pelliccia}, \citenamefont {Asadchikov},\ and\ \citenamefont
  {Lagomarsino}}]{Bukreeva_OL_2011}%
  \BibitemOpen
  \bibfield  {author} {\bibinfo {author} {\bibfnamefont {I.}~\bibnamefont
  {Bukreeva}}, \bibinfo {author} {\bibfnamefont {A.}~\bibnamefont {Cedola}},
  \bibinfo {author} {\bibfnamefont {A.}~\bibnamefont {Sorrentino}}, \bibinfo
  {author} {\bibfnamefont {D.}~\bibnamefont {Pelliccia}}, \bibinfo {author}
  {\bibfnamefont {V.}~\bibnamefont {Asadchikov}},\ and\ \bibinfo {author}
  {\bibfnamefont {S.}~\bibnamefont {Lagomarsino}},\ }\bibfield  {title}
  {\bibinfo {title} {Resonance modes filtering in structured x-ray
  waveguides},\ }\href {https://doi.org/10.1364/ol.36.002602} {\bibfield
  {journal} {\bibinfo  {journal} {Optics Letters}\ }\textbf {\bibinfo {volume}
  {36}},\ \bibinfo {pages} {2602} (\bibinfo {year} {2011})}\BibitemShut
  {NoStop}%
\bibitem [{\citenamefont {Bukreeva}\ \emph {et~al.}(2013)\citenamefont
  {Bukreeva}, \citenamefont {Sorrentino}, \citenamefont {Cedola}, \citenamefont
  {Giovine}, \citenamefont {Diaz}, \citenamefont {Scarinci}, \citenamefont
  {Jark}, \citenamefont {Ognev},\ and\ \citenamefont
  {Lagomarsino}}]{bukreevaPeriodicallyStructuredXray2013}%
  \BibitemOpen
  \bibfield  {author} {\bibinfo {author} {\bibfnamefont {I.}~\bibnamefont
  {Bukreeva}}, \bibinfo {author} {\bibfnamefont {A.}~\bibnamefont
  {Sorrentino}}, \bibinfo {author} {\bibfnamefont {A.}~\bibnamefont {Cedola}},
  \bibinfo {author} {\bibfnamefont {E.}~\bibnamefont {Giovine}}, \bibinfo
  {author} {\bibfnamefont {A.}~\bibnamefont {Diaz}}, \bibinfo {author}
  {\bibfnamefont {F.}~\bibnamefont {Scarinci}}, \bibinfo {author}
  {\bibfnamefont {W.}~\bibnamefont {Jark}}, \bibinfo {author} {\bibfnamefont
  {L.}~\bibnamefont {Ognev}},\ and\ \bibinfo {author} {\bibfnamefont
  {S.}~\bibnamefont {Lagomarsino}},\ }\bibfield  {title} {\bibinfo {title}
  {Periodically structured {{X-ray}} waveguides},\ }\href
  {https://doi.org/10.1107/S0909049513018657} {\bibfield  {journal} {\bibinfo
  {journal} {J Synchrotron Rad}\ }\textbf {\bibinfo {volume} {20}},\ \bibinfo
  {pages} {691} (\bibinfo {year} {2013})}\BibitemShut {NoStop}%
\bibitem [{\citenamefont {Fuhse}\ \emph {et~al.}(2004)\citenamefont {Fuhse},
  \citenamefont {Jarre}, \citenamefont {Ollinger}, \citenamefont {Seeger},
  \citenamefont {Salditt},\ and\ \citenamefont {Tucoulou}}]{Fuhse_APL_2004}%
  \BibitemOpen
  \bibfield  {author} {\bibinfo {author} {\bibfnamefont {C.}~\bibnamefont
  {Fuhse}}, \bibinfo {author} {\bibfnamefont {A.}~\bibnamefont {Jarre}},
  \bibinfo {author} {\bibfnamefont {C.}~\bibnamefont {Ollinger}}, \bibinfo
  {author} {\bibfnamefont {J.}~\bibnamefont {Seeger}}, \bibinfo {author}
  {\bibfnamefont {T.}~\bibnamefont {Salditt}},\ and\ \bibinfo {author}
  {\bibfnamefont {R.}~\bibnamefont {Tucoulou}},\ }\bibfield  {title} {\bibinfo
  {title} {Front-coupling of a prefocused x-ray beam into a monomodal planar
  waveguide},\ }\href {https://doi.org/10.1063/1.1791736} {\bibfield  {journal}
  {\bibinfo  {journal} {Applied Physics Letters}\ }\textbf {\bibinfo {volume}
  {85}},\ \bibinfo {pages} {1907} (\bibinfo {year} {2004})}\BibitemShut
  {NoStop}%
\bibitem [{\citenamefont {Bongaerts}\ \emph {et~al.}(2002)\citenamefont
  {Bongaerts}, \citenamefont {David}, \citenamefont {Drakopoulos},
  \citenamefont {Zwanenburg}, \citenamefont {Wegdam}, \citenamefont {Lackner},
  \citenamefont {Keymeulen},\ and\ \citenamefont {Van
  Der~Veen}}]{bongaertsPropagationPartiallyCoherent2002}%
  \BibitemOpen
  \bibfield  {author} {\bibinfo {author} {\bibfnamefont {J.~H.~H.}\
  \bibnamefont {Bongaerts}}, \bibinfo {author} {\bibfnamefont {C.}~\bibnamefont
  {David}}, \bibinfo {author} {\bibfnamefont {M.}~\bibnamefont {Drakopoulos}},
  \bibinfo {author} {\bibfnamefont {M.~J.}\ \bibnamefont {Zwanenburg}},
  \bibinfo {author} {\bibfnamefont {G.~H.}\ \bibnamefont {Wegdam}}, \bibinfo
  {author} {\bibfnamefont {T.}~\bibnamefont {Lackner}}, \bibinfo {author}
  {\bibfnamefont {H.}~\bibnamefont {Keymeulen}},\ and\ \bibinfo {author}
  {\bibfnamefont {J.~F.}\ \bibnamefont {Van Der~Veen}},\ }\bibfield  {title}
  {\bibinfo {title} {Propagation of a partially coherent focused {{X-ray}} beam
  within a planar {{X-ray}} waveguide},\ }\href
  {https://doi.org/10.1107/S0909049502016308} {\bibfield  {journal} {\bibinfo
  {journal} {J Synchrotron Rad}\ }\textbf {\bibinfo {volume} {9}},\ \bibinfo
  {pages} {383} (\bibinfo {year} {2002})}\BibitemShut {NoStop}%
\bibitem [{\citenamefont {R{\"o}hlsberger}\ \emph {et~al.}(2010)\citenamefont
  {R{\"o}hlsberger}, \citenamefont {Schlage}, \citenamefont {Sahoo},
  \citenamefont {Couet},\ and\ \citenamefont
  {R{\"u}ffer}}]{rohlsbergerCollectiveLambShift2010}%
  \BibitemOpen
  \bibfield  {author} {\bibinfo {author} {\bibfnamefont {R.}~\bibnamefont
  {R{\"o}hlsberger}}, \bibinfo {author} {\bibfnamefont {K.}~\bibnamefont
  {Schlage}}, \bibinfo {author} {\bibfnamefont {B.}~\bibnamefont {Sahoo}},
  \bibinfo {author} {\bibfnamefont {S.}~\bibnamefont {Couet}},\ and\ \bibinfo
  {author} {\bibfnamefont {R.}~\bibnamefont {R{\"u}ffer}},\ }\bibfield  {title}
  {\bibinfo {title} {Collective {{Lamb Shift}} in {{Single-Photon
  Superradiance}}},\ }\href {https://doi.org/10.1126/science.1187770}
  {\bibfield  {journal} {\bibinfo  {journal} {Science}\ }\textbf {\bibinfo
  {volume} {328}},\ \bibinfo {pages} {1248} (\bibinfo {year} {2010})},\ \Eprint
  {https://arxiv.org/abs/20466883} {20466883} \BibitemShut {NoStop}%
\bibitem [{\citenamefont {R{\"o}hlsberger}\ \emph {et~al.}(2012)\citenamefont
  {R{\"o}hlsberger}, \citenamefont {Wille}, \citenamefont {Schlage},\ and\
  \citenamefont
  {Sahoo}}]{rohlsbergerElectromagneticallyInducedTransparency2012}%
  \BibitemOpen
  \bibfield  {author} {\bibinfo {author} {\bibfnamefont {R.}~\bibnamefont
  {R{\"o}hlsberger}}, \bibinfo {author} {\bibfnamefont {H.-C.}\ \bibnamefont
  {Wille}}, \bibinfo {author} {\bibfnamefont {K.}~\bibnamefont {Schlage}},\
  and\ \bibinfo {author} {\bibfnamefont {B.}~\bibnamefont {Sahoo}},\ }\bibfield
   {title} {\bibinfo {title} {Electromagnetically induced transparency with
  resonant nuclei in a cavity},\ }\href {https://doi.org/10.1038/nature10741}
  {\bibfield  {journal} {\bibinfo  {journal} {Nature}\ }\textbf {\bibinfo
  {volume} {482}},\ \bibinfo {pages} {199} (\bibinfo {year}
  {2012})}\BibitemShut {NoStop}%
\bibitem [{\citenamefont {Heeg}\ \emph {et~al.}(2013)\citenamefont {Heeg},
  \citenamefont {Wille}, \citenamefont {Schlage}, \citenamefont {Guryeva},
  \citenamefont {Schumacher}, \citenamefont {Uschmann}, \citenamefont
  {Schulze}, \citenamefont {Marx}, \citenamefont {K{\"a}mpfer}, \citenamefont
  {Paulus}, \citenamefont {R{\"o}hlsberger},\ and\ \citenamefont
  {Evers}}]{heegVacuumAssistedGenerationControl2013}%
  \BibitemOpen
  \bibfield  {author} {\bibinfo {author} {\bibfnamefont {K.~P.}\ \bibnamefont
  {Heeg}}, \bibinfo {author} {\bibfnamefont {H.-C.}\ \bibnamefont {Wille}},
  \bibinfo {author} {\bibfnamefont {K.}~\bibnamefont {Schlage}}, \bibinfo
  {author} {\bibfnamefont {T.}~\bibnamefont {Guryeva}}, \bibinfo {author}
  {\bibfnamefont {D.}~\bibnamefont {Schumacher}}, \bibinfo {author}
  {\bibfnamefont {I.}~\bibnamefont {Uschmann}}, \bibinfo {author}
  {\bibfnamefont {K.~S.}\ \bibnamefont {Schulze}}, \bibinfo {author}
  {\bibfnamefont {B.}~\bibnamefont {Marx}}, \bibinfo {author} {\bibfnamefont
  {T.}~\bibnamefont {K{\"a}mpfer}}, \bibinfo {author} {\bibfnamefont {G.~G.}\
  \bibnamefont {Paulus}}, \bibinfo {author} {\bibfnamefont {R.}~\bibnamefont
  {R{\"o}hlsberger}},\ and\ \bibinfo {author} {\bibfnamefont {J.}~\bibnamefont
  {Evers}},\ }\bibfield  {title} {\bibinfo {title} {Vacuum-{{Assisted
  Generation}} and {{Control}} of {{Atomic Coherences}} at {{X-Ray
  Energies}}},\ }\href {https://doi.org/10.1103/PhysRevLett.111.073601}
  {\bibfield  {journal} {\bibinfo  {journal} {Phys. Rev. Lett.}\ }\textbf
  {\bibinfo {volume} {111}},\ \bibinfo {pages} {073601} (\bibinfo {year}
  {2013})}\BibitemShut {NoStop}%
\bibitem [{\citenamefont {Heeg}\ \emph
  {et~al.}(2015{\natexlab{a}})\citenamefont {Heeg}, \citenamefont {Ott},
  \citenamefont {Schumacher}, \citenamefont {Wille}, \citenamefont
  {R{\"o}hlsberger}, \citenamefont {Pfeifer},\ and\ \citenamefont
  {Evers}}]{heegInterferometricPhaseDetection2015}%
  \BibitemOpen
  \bibfield  {author} {\bibinfo {author} {\bibfnamefont {K.~P.}\ \bibnamefont
  {Heeg}}, \bibinfo {author} {\bibfnamefont {C.}~\bibnamefont {Ott}}, \bibinfo
  {author} {\bibfnamefont {D.}~\bibnamefont {Schumacher}}, \bibinfo {author}
  {\bibfnamefont {H.-C.}\ \bibnamefont {Wille}}, \bibinfo {author}
  {\bibfnamefont {R.}~\bibnamefont {R{\"o}hlsberger}}, \bibinfo {author}
  {\bibfnamefont {T.}~\bibnamefont {Pfeifer}},\ and\ \bibinfo {author}
  {\bibfnamefont {J.}~\bibnamefont {Evers}},\ }\bibfield  {title} {\bibinfo
  {title} {Interferometric phase detection at {{X-ray}} energies via {{Fano}}
  resonance control},\ }\href {https://doi.org/10.1103/PhysRevLett.114.207401}
  {\bibfield  {journal} {\bibinfo  {journal} {Phys. Rev. Lett.}\ }\textbf
  {\bibinfo {volume} {114}},\ \bibinfo {pages} {207401} (\bibinfo {year}
  {2015}{\natexlab{a}})}\BibitemShut {NoStop}%
\bibitem [{\citenamefont {Heeg}\ \emph
  {et~al.}(2015{\natexlab{b}})\citenamefont {Heeg}, \citenamefont {Haber},
  \citenamefont {Schumacher}, \citenamefont {Bocklage}, \citenamefont {Wille},
  \citenamefont {Schulze}, \citenamefont {Loetzsch}, \citenamefont {Uschmann},
  \citenamefont {Paulus}, \citenamefont {R{\"u}ffer}, \citenamefont
  {R{\"o}hlsberger},\ and\ \citenamefont
  {Evers}}]{heegTunableSubluminalPropagation2015}%
  \BibitemOpen
  \bibfield  {author} {\bibinfo {author} {\bibfnamefont {K.~P.}\ \bibnamefont
  {Heeg}}, \bibinfo {author} {\bibfnamefont {J.}~\bibnamefont {Haber}},
  \bibinfo {author} {\bibfnamefont {D.}~\bibnamefont {Schumacher}}, \bibinfo
  {author} {\bibfnamefont {L.}~\bibnamefont {Bocklage}}, \bibinfo {author}
  {\bibfnamefont {H.-C.}\ \bibnamefont {Wille}}, \bibinfo {author}
  {\bibfnamefont {K.~S.}\ \bibnamefont {Schulze}}, \bibinfo {author}
  {\bibfnamefont {R.}~\bibnamefont {Loetzsch}}, \bibinfo {author}
  {\bibfnamefont {I.}~\bibnamefont {Uschmann}}, \bibinfo {author}
  {\bibfnamefont {G.~G.}\ \bibnamefont {Paulus}}, \bibinfo {author}
  {\bibfnamefont {R.}~\bibnamefont {R{\"u}ffer}}, \bibinfo {author}
  {\bibfnamefont {R.}~\bibnamefont {R{\"o}hlsberger}},\ and\ \bibinfo {author}
  {\bibfnamefont {J.}~\bibnamefont {Evers}},\ }\bibfield  {title} {\bibinfo
  {title} {Tunable {{Subluminal Propagation}} of {{Narrow-band X-Ray
  Pulses}}},\ }\href {https://doi.org/10.1103/PhysRevLett.114.203601}
  {\bibfield  {journal} {\bibinfo  {journal} {Phys. Rev. Lett.}\ }\textbf
  {\bibinfo {volume} {114}},\ \bibinfo {pages} {203601} (\bibinfo {year}
  {2015}{\natexlab{b}})}\BibitemShut {NoStop}%
\bibitem [{\citenamefont {Haber}\ \emph {et~al.}(2016)\citenamefont {Haber},
  \citenamefont {Schulze}, \citenamefont {Schlage}, \citenamefont {Loetzsch},
  \citenamefont {Bocklage}, \citenamefont {Gurieva}, \citenamefont {Bernhardt},
  \citenamefont {Wille}, \citenamefont {R{\"u}ffer}, \citenamefont {Uschmann},
  \citenamefont {Paulus},\ and\ \citenamefont
  {R{\"o}hlsberger}}]{haberCollectiveStrongCoupling2016}%
  \BibitemOpen
  \bibfield  {author} {\bibinfo {author} {\bibfnamefont {J.}~\bibnamefont
  {Haber}}, \bibinfo {author} {\bibfnamefont {K.~S.}\ \bibnamefont {Schulze}},
  \bibinfo {author} {\bibfnamefont {K.}~\bibnamefont {Schlage}}, \bibinfo
  {author} {\bibfnamefont {R.}~\bibnamefont {Loetzsch}}, \bibinfo {author}
  {\bibfnamefont {L.}~\bibnamefont {Bocklage}}, \bibinfo {author}
  {\bibfnamefont {T.}~\bibnamefont {Gurieva}}, \bibinfo {author} {\bibfnamefont
  {H.}~\bibnamefont {Bernhardt}}, \bibinfo {author} {\bibfnamefont {H.-C.}\
  \bibnamefont {Wille}}, \bibinfo {author} {\bibfnamefont {R.}~\bibnamefont
  {R{\"u}ffer}}, \bibinfo {author} {\bibfnamefont {I.}~\bibnamefont
  {Uschmann}}, \bibinfo {author} {\bibfnamefont {G.~G.}\ \bibnamefont
  {Paulus}},\ and\ \bibinfo {author} {\bibfnamefont {R.}~\bibnamefont
  {R{\"o}hlsberger}},\ }\bibfield  {title} {\bibinfo {title} {Collective strong
  coupling of {{X-rays}} and nuclei in a nuclear optical lattice},\ }\href
  {https://doi.org/10.1038/nphoton.2016.77} {\bibfield  {journal} {\bibinfo
  {journal} {Nature Photon}\ }\textbf {\bibinfo {volume} {10}},\ \bibinfo
  {pages} {445} (\bibinfo {year} {2016})}\BibitemShut {NoStop}%
\bibitem [{\citenamefont {Haber}\ \emph {et~al.}(2017)\citenamefont {Haber},
  \citenamefont {Kong}, \citenamefont {Strohm}, \citenamefont {Willing},
  \citenamefont {Gollwitzer}, \citenamefont {Bocklage}, \citenamefont
  {R{\"u}ffer}, \citenamefont {P{\'a}lffy},\ and\ \citenamefont
  {R{\"o}hlsberger}}]{haberRabiOscillationsXray2017}%
  \BibitemOpen
  \bibfield  {author} {\bibinfo {author} {\bibfnamefont {J.}~\bibnamefont
  {Haber}}, \bibinfo {author} {\bibfnamefont {X.}~\bibnamefont {Kong}},
  \bibinfo {author} {\bibfnamefont {C.}~\bibnamefont {Strohm}}, \bibinfo
  {author} {\bibfnamefont {S.}~\bibnamefont {Willing}}, \bibinfo {author}
  {\bibfnamefont {J.}~\bibnamefont {Gollwitzer}}, \bibinfo {author}
  {\bibfnamefont {L.}~\bibnamefont {Bocklage}}, \bibinfo {author}
  {\bibfnamefont {R.}~\bibnamefont {R{\"u}ffer}}, \bibinfo {author}
  {\bibfnamefont {A.}~\bibnamefont {P{\'a}lffy}},\ and\ \bibinfo {author}
  {\bibfnamefont {R.}~\bibnamefont {R{\"o}hlsberger}},\ }\bibfield  {title}
  {\bibinfo {title} {Rabi oscillations of {{X-ray}} radiation between two
  nuclear ensembles},\ }\href {https://doi.org/10.1038/s41566-017-0013-3}
  {\bibfield  {journal} {\bibinfo  {journal} {Nat. Photon.}\ }\textbf {\bibinfo
  {volume} {11}},\ \bibinfo {pages} {720} (\bibinfo {year} {2017})}\BibitemShut
  {NoStop}%
\bibitem [{\citenamefont {Lentrodt}\ \emph {et~al.}(2023)\citenamefont
  {Lentrodt}, \citenamefont {Diekmann}, \citenamefont {Keitel}, \citenamefont
  {Rotter},\ and\ \citenamefont
  {Evers}}]{lentrodtCertifyingMultimodeLightMatter2023}%
  \BibitemOpen
  \bibfield  {author} {\bibinfo {author} {\bibfnamefont {D.}~\bibnamefont
  {Lentrodt}}, \bibinfo {author} {\bibfnamefont {O.}~\bibnamefont {Diekmann}},
  \bibinfo {author} {\bibfnamefont {C.~H.}\ \bibnamefont {Keitel}}, \bibinfo
  {author} {\bibfnamefont {S.}~\bibnamefont {Rotter}},\ and\ \bibinfo {author}
  {\bibfnamefont {J.}~\bibnamefont {Evers}},\ }\bibfield  {title} {\bibinfo
  {title} {Certifying {{Multimode Light-Matter Interaction}} in {{Lossy
  Resonators}}},\ }\href {https://doi.org/10.1103/PhysRevLett.130.263602}
  {\bibfield  {journal} {\bibinfo  {journal} {Phys. Rev. Lett.}\ }\textbf
  {\bibinfo {volume} {130}},\ \bibinfo {pages} {263602} (\bibinfo {year}
  {2023})}\BibitemShut {NoStop}%
\bibitem [{\citenamefont {Chen}\ \emph {et~al.}(2022)\citenamefont {Chen},
  \citenamefont {Lin}, \citenamefont {Wang}, \citenamefont {P{\'a}lffy},\ and\
  \citenamefont {Liao}}]{chenTransientNuclearInversion2022}%
  \BibitemOpen
  \bibfield  {author} {\bibinfo {author} {\bibfnamefont {Y.-H.}\ \bibnamefont
  {Chen}}, \bibinfo {author} {\bibfnamefont {P.-H.}\ \bibnamefont {Lin}},
  \bibinfo {author} {\bibfnamefont {G.-Y.}\ \bibnamefont {Wang}}, \bibinfo
  {author} {\bibfnamefont {A.}~\bibnamefont {P{\'a}lffy}},\ and\ \bibinfo
  {author} {\bibfnamefont {W.-T.}\ \bibnamefont {Liao}},\ }\bibfield  {title}
  {\bibinfo {title} {Transient nuclear inversion by {{X-ray}} free electron
  laser in a tapered x-{{Ray}} waveguide},\ }\href
  {https://doi.org/10.1103/PhysRevResearch.4.L032007} {\bibfield  {journal}
  {\bibinfo  {journal} {Phys. Rev. Research}\ }\textbf {\bibinfo {volume}
  {4}},\ \bibinfo {pages} {L032007} (\bibinfo {year} {2022})}\BibitemShut
  {NoStop}%
\bibitem [{\citenamefont {Lee}\ \emph {et~al.}(2023)\citenamefont {Lee},
  \citenamefont {Ahrens},\ and\ \citenamefont
  {Liao}}]{leeGravitationallySensitiveStructured2023}%
  \BibitemOpen
  \bibfield  {author} {\bibinfo {author} {\bibfnamefont {S.-Y.}\ \bibnamefont
  {Lee}}, \bibinfo {author} {\bibfnamefont {S.}~\bibnamefont {Ahrens}},\ and\
  \bibinfo {author} {\bibfnamefont {W.-T.}\ \bibnamefont {Liao}},\ }\href
  {https://doi.org/10.48550/arXiv.2305.00613} {\bibinfo {title}
  {Gravitationally sensitive structured x-ray optics using nuclear resonances}}
  (\bibinfo {year} {2023}),\ \Eprint {https://arxiv.org/abs/2305.00613}
  {arxiv:2305.00613 [gr-qc, physics:physics]} \BibitemShut {NoStop}%
\bibitem [{\citenamefont {Gruner}\ and\ \citenamefont
  {Welsch}(1996)}]{grunerGreenfunctionApproachRadiationfield1996}%
  \BibitemOpen
  \bibfield  {author} {\bibinfo {author} {\bibfnamefont {T.}~\bibnamefont
  {Gruner}}\ and\ \bibinfo {author} {\bibfnamefont {D.-G.}\ \bibnamefont
  {Welsch}},\ }\bibfield  {title} {\bibinfo {title} {Green-function approach to
  the radiation-field quantization for homogeneous and inhomogeneous
  {{Kramers-Kronig}} dielectrics},\ }\href
  {https://doi.org/10.1103/PhysRevA.53.1818} {\bibfield  {journal} {\bibinfo
  {journal} {Phys. Rev. A}\ }\textbf {\bibinfo {volume} {53}},\ \bibinfo
  {pages} {1818} (\bibinfo {year} {1996})}\BibitemShut {NoStop}%
\bibitem [{\citenamefont {Dung}\ \emph {et~al.}(1998)\citenamefont {Dung},
  \citenamefont {Kn{\"o}ll},\ and\ \citenamefont
  {Welsch}}]{dungThreedimensionalQuantizationElectromagnetic1998}%
  \BibitemOpen
  \bibfield  {author} {\bibinfo {author} {\bibfnamefont {H.~T.}\ \bibnamefont
  {Dung}}, \bibinfo {author} {\bibfnamefont {L.}~\bibnamefont {Kn{\"o}ll}},\
  and\ \bibinfo {author} {\bibfnamefont {D.-G.}\ \bibnamefont {Welsch}},\
  }\bibfield  {title} {\bibinfo {title} {Three-dimensional quantization of the
  electromagnetic field in dispersive and absorbing inhomogeneous
  dielectrics},\ }\href {https://doi.org/10.1103/PhysRevA.57.3931} {\bibfield
  {journal} {\bibinfo  {journal} {Phys. Rev. A}\ }\textbf {\bibinfo {volume}
  {57}},\ \bibinfo {pages} {3931} (\bibinfo {year} {1998})}\BibitemShut
  {NoStop}%
\bibitem [{\citenamefont {Dung}\ \emph {et~al.}(2002)\citenamefont {Dung},
  \citenamefont {Kn{\"o}ll},\ and\ \citenamefont
  {Welsch}}]{dungResonantDipoledipoleInteraction2002}%
  \BibitemOpen
  \bibfield  {author} {\bibinfo {author} {\bibfnamefont {H.~T.}\ \bibnamefont
  {Dung}}, \bibinfo {author} {\bibfnamefont {L.}~\bibnamefont {Kn{\"o}ll}},\
  and\ \bibinfo {author} {\bibfnamefont {D.-G.}\ \bibnamefont {Welsch}},\
  }\bibfield  {title} {\bibinfo {title} {Resonant dipole-dipole interaction in
  the presence of dispersing and absorbing surroundings},\ }\href
  {https://doi.org/10.1103/PhysRevA.66.063810} {\bibfield  {journal} {\bibinfo
  {journal} {Phys. Rev. A}\ }\textbf {\bibinfo {volume} {66}},\ \bibinfo
  {pages} {063810} (\bibinfo {year} {2002})}\BibitemShut {NoStop}%
\bibitem [{\citenamefont {Trost}\ \emph {et~al.}(2023)\citenamefont {Trost},
  \citenamefont {Ayyer}, \citenamefont {Prasciolu}, \citenamefont
  {Fleckenstein}, \citenamefont {Barthelmess}, \citenamefont {Yefanov},
  \citenamefont {Dresselhaus}, \citenamefont {Li}, \citenamefont {Bajt},
  \citenamefont {Carnis}, \citenamefont {Wollweber}, \citenamefont {Mall},
  \citenamefont {Shen}, \citenamefont {Zhuang}, \citenamefont {Richter},
  \citenamefont {Karl}, \citenamefont {Cardoch}, \citenamefont {Patra},
  \citenamefont {M{\"o}ller}, \citenamefont {Zozulya}, \citenamefont {Shayduk},
  \citenamefont {Lu}, \citenamefont {Brau{\ss}e}, \citenamefont {Friedrich},
  \citenamefont {Boesenberg}, \citenamefont {Petrov}, \citenamefont {Tomin},
  \citenamefont {Guetg}, \citenamefont {Madsen}, \citenamefont {Timneanu},
  \citenamefont {Caleman}, \citenamefont {R{\"o}hlsberger}, \citenamefont {{von
  Zanthier}},\ and\ \citenamefont {Chapman}}]{trostImagingCorrelationXRay2023}%
  \BibitemOpen
  \bibfield  {author} {\bibinfo {author} {\bibfnamefont {F.}~\bibnamefont
  {Trost}}, \bibinfo {author} {\bibfnamefont {K.}~\bibnamefont {Ayyer}},
  \bibinfo {author} {\bibfnamefont {M.}~\bibnamefont {Prasciolu}}, \bibinfo
  {author} {\bibfnamefont {H.}~\bibnamefont {Fleckenstein}}, \bibinfo {author}
  {\bibfnamefont {M.}~\bibnamefont {Barthelmess}}, \bibinfo {author}
  {\bibfnamefont {O.}~\bibnamefont {Yefanov}}, \bibinfo {author} {\bibfnamefont
  {J.~L.}\ \bibnamefont {Dresselhaus}}, \bibinfo {author} {\bibfnamefont
  {C.}~\bibnamefont {Li}}, \bibinfo {author} {\bibfnamefont {S.}~\bibnamefont
  {Bajt}}, \bibinfo {author} {\bibfnamefont {J.}~\bibnamefont {Carnis}},
  \bibinfo {author} {\bibfnamefont {T.}~\bibnamefont {Wollweber}}, \bibinfo
  {author} {\bibfnamefont {A.}~\bibnamefont {Mall}}, \bibinfo {author}
  {\bibfnamefont {Z.}~\bibnamefont {Shen}}, \bibinfo {author} {\bibfnamefont
  {Y.}~\bibnamefont {Zhuang}}, \bibinfo {author} {\bibfnamefont
  {S.}~\bibnamefont {Richter}}, \bibinfo {author} {\bibfnamefont
  {S.}~\bibnamefont {Karl}}, \bibinfo {author} {\bibfnamefont {S.}~\bibnamefont
  {Cardoch}}, \bibinfo {author} {\bibfnamefont {K.~K.}\ \bibnamefont {Patra}},
  \bibinfo {author} {\bibfnamefont {J.}~\bibnamefont {M{\"o}ller}}, \bibinfo
  {author} {\bibfnamefont {A.}~\bibnamefont {Zozulya}}, \bibinfo {author}
  {\bibfnamefont {R.}~\bibnamefont {Shayduk}}, \bibinfo {author} {\bibfnamefont
  {W.}~\bibnamefont {Lu}}, \bibinfo {author} {\bibfnamefont {F.}~\bibnamefont
  {Brau{\ss}e}}, \bibinfo {author} {\bibfnamefont {B.}~\bibnamefont
  {Friedrich}}, \bibinfo {author} {\bibfnamefont {U.}~\bibnamefont
  {Boesenberg}}, \bibinfo {author} {\bibfnamefont {I.}~\bibnamefont {Petrov}},
  \bibinfo {author} {\bibfnamefont {S.}~\bibnamefont {Tomin}}, \bibinfo
  {author} {\bibfnamefont {M.}~\bibnamefont {Guetg}}, \bibinfo {author}
  {\bibfnamefont {A.}~\bibnamefont {Madsen}}, \bibinfo {author} {\bibfnamefont
  {N.}~\bibnamefont {Timneanu}}, \bibinfo {author} {\bibfnamefont
  {C.}~\bibnamefont {Caleman}}, \bibinfo {author} {\bibfnamefont
  {R.}~\bibnamefont {R{\"o}hlsberger}}, \bibinfo {author} {\bibfnamefont
  {J.}~\bibnamefont {{von Zanthier}}},\ and\ \bibinfo {author} {\bibfnamefont
  {H.~N.}\ \bibnamefont {Chapman}},\ }\bibfield  {title} {\bibinfo {title}
  {Imaging via {{Correlation}} of {{X-Ray Fluorescence Photons}}},\ }\href
  {https://doi.org/10.1103/PhysRevLett.130.173201} {\bibfield  {journal}
  {\bibinfo  {journal} {Phys. Rev. Lett.}\ }\textbf {\bibinfo {volume} {130}},\
  \bibinfo {pages} {173201} (\bibinfo {year} {2023})}\BibitemShut {NoStop}%
\bibitem [{\citenamefont {Mili{\'c}evi{\'c}}\ \emph {et~al.}(2015)\citenamefont
  {Mili{\'c}evi{\'c}}, \citenamefont {Ozawa}, \citenamefont {Andreakou},
  \citenamefont {Carusotto}, \citenamefont {Jacqmin}, \citenamefont {Galopin},
  \citenamefont {Lema{\^i}tre}, \citenamefont {Le~Gratiet}, \citenamefont
  {Sagnes}, \citenamefont {Bloch},\ and\ \citenamefont
  {Amo}}]{milicevicEdgeStatesPolariton2015}%
  \BibitemOpen
  \bibfield  {author} {\bibinfo {author} {\bibfnamefont {M.}~\bibnamefont
  {Mili{\'c}evi{\'c}}}, \bibinfo {author} {\bibfnamefont {T.}~\bibnamefont
  {Ozawa}}, \bibinfo {author} {\bibfnamefont {P.}~\bibnamefont {Andreakou}},
  \bibinfo {author} {\bibfnamefont {I.}~\bibnamefont {Carusotto}}, \bibinfo
  {author} {\bibfnamefont {T.}~\bibnamefont {Jacqmin}}, \bibinfo {author}
  {\bibfnamefont {E.}~\bibnamefont {Galopin}}, \bibinfo {author} {\bibfnamefont
  {A.}~\bibnamefont {Lema{\^i}tre}}, \bibinfo {author} {\bibfnamefont
  {L.}~\bibnamefont {Le~Gratiet}}, \bibinfo {author} {\bibfnamefont
  {I.}~\bibnamefont {Sagnes}}, \bibinfo {author} {\bibfnamefont
  {J.}~\bibnamefont {Bloch}},\ and\ \bibinfo {author} {\bibfnamefont
  {A.}~\bibnamefont {Amo}},\ }\bibfield  {title} {\bibinfo {title} {Edge states
  in polariton honeycomb lattices},\ }\href
  {https://doi.org/10.1088/2053-1583/2/3/034012} {\bibfield  {journal}
  {\bibinfo  {journal} {2D Materials}\ }\textbf {\bibinfo {volume} {2}},\
  \bibinfo {pages} {034012} (\bibinfo {year} {2015})}\BibitemShut {NoStop}%
\bibitem [{\citenamefont {Sala}\ \emph {et~al.}(2015)\citenamefont {Sala},
  \citenamefont {Solnyshkov}, \citenamefont {Carusotto}, \citenamefont
  {Jacqmin}, \citenamefont {Lema{\^i}tre}, \citenamefont {Ter{\c c}as},
  \citenamefont {Nalitov}, \citenamefont {Abbarchi}, \citenamefont {Galopin},
  \citenamefont {Sagnes}, \citenamefont {Bloch}, \citenamefont {Malpuech},\
  and\ \citenamefont {Amo}}]{salaSpinOrbitCouplingPhotons2015}%
  \BibitemOpen
  \bibfield  {author} {\bibinfo {author} {\bibfnamefont {V.~G.}\ \bibnamefont
  {Sala}}, \bibinfo {author} {\bibfnamefont {D.~D.}\ \bibnamefont
  {Solnyshkov}}, \bibinfo {author} {\bibfnamefont {I.}~\bibnamefont
  {Carusotto}}, \bibinfo {author} {\bibfnamefont {T.}~\bibnamefont {Jacqmin}},
  \bibinfo {author} {\bibfnamefont {A.}~\bibnamefont {Lema{\^i}tre}}, \bibinfo
  {author} {\bibfnamefont {H.}~\bibnamefont {Ter{\c c}as}}, \bibinfo {author}
  {\bibfnamefont {A.}~\bibnamefont {Nalitov}}, \bibinfo {author} {\bibfnamefont
  {M.}~\bibnamefont {Abbarchi}}, \bibinfo {author} {\bibfnamefont
  {E.}~\bibnamefont {Galopin}}, \bibinfo {author} {\bibfnamefont
  {I.}~\bibnamefont {Sagnes}}, \bibinfo {author} {\bibfnamefont
  {J.}~\bibnamefont {Bloch}}, \bibinfo {author} {\bibfnamefont
  {G.}~\bibnamefont {Malpuech}},\ and\ \bibinfo {author} {\bibfnamefont
  {A.}~\bibnamefont {Amo}},\ }\bibfield  {title} {\bibinfo {title}
  {Spin-{{Orbit Coupling}} for {{Photons}} and {{Polaritons}} in
  {{Microstructures}}},\ }\href {https://doi.org/10.1103/PhysRevX.5.011034}
  {\bibfield  {journal} {\bibinfo  {journal} {Phys. Rev. X}\ }\textbf {\bibinfo
  {volume} {5}},\ \bibinfo {pages} {011034} (\bibinfo {year}
  {2015})}\BibitemShut {NoStop}%
\bibitem [{\citenamefont {{St-Jean}}\ \emph {et~al.}(2017)\citenamefont
  {{St-Jean}}, \citenamefont {Goblot}, \citenamefont {Galopin}, \citenamefont
  {Lema{\^i}tre}, \citenamefont {Ozawa}, \citenamefont {Gratiet}, \citenamefont
  {Sagnes}, \citenamefont {Bloch},\ and\ \citenamefont
  {Amo}}]{st-jeanLasingTopologicalEdge2017}%
  \BibitemOpen
  \bibfield  {author} {\bibinfo {author} {\bibfnamefont {P.}~\bibnamefont
  {{St-Jean}}}, \bibinfo {author} {\bibfnamefont {V.}~\bibnamefont {Goblot}},
  \bibinfo {author} {\bibfnamefont {E.}~\bibnamefont {Galopin}}, \bibinfo
  {author} {\bibfnamefont {A.}~\bibnamefont {Lema{\^i}tre}}, \bibinfo {author}
  {\bibfnamefont {T.}~\bibnamefont {Ozawa}}, \bibinfo {author} {\bibfnamefont
  {L.~L.}\ \bibnamefont {Gratiet}}, \bibinfo {author} {\bibfnamefont
  {I.}~\bibnamefont {Sagnes}}, \bibinfo {author} {\bibfnamefont
  {J.}~\bibnamefont {Bloch}},\ and\ \bibinfo {author} {\bibfnamefont
  {A.}~\bibnamefont {Amo}},\ }\bibfield  {title} {\bibinfo {title} {Lasing in
  topological edge states of a one-dimensional lattice},\ }\href
  {https://doi.org/10.1038/s41566-017-0006-2} {\bibfield  {journal} {\bibinfo
  {journal} {Nature Photonics}\ }\textbf {\bibinfo {volume} {11}},\ \bibinfo
  {pages} {651} (\bibinfo {year} {2017})}\BibitemShut {NoStop}%
\bibitem [{\citenamefont {{Asenjo-Garcia}}\ \emph
  {et~al.}(2017{\natexlab{a}})\citenamefont {{Asenjo-Garcia}}, \citenamefont
  {{Moreno-Cardoner}}, \citenamefont {Albrecht}, \citenamefont {Kimble},\ and\
  \citenamefont {Chang}}]{asenjo-garciaExponentialImprovementPhoton2017}%
  \BibitemOpen
  \bibfield  {author} {\bibinfo {author} {\bibfnamefont {A.}~\bibnamefont
  {{Asenjo-Garcia}}}, \bibinfo {author} {\bibfnamefont {M.}~\bibnamefont
  {{Moreno-Cardoner}}}, \bibinfo {author} {\bibfnamefont {A.}~\bibnamefont
  {Albrecht}}, \bibinfo {author} {\bibfnamefont {H.~J.}\ \bibnamefont
  {Kimble}},\ and\ \bibinfo {author} {\bibfnamefont {D.~E.}\ \bibnamefont
  {Chang}},\ }\bibfield  {title} {\bibinfo {title} {Exponential improvement in
  photon storage fidelities using subradiance and ``{{Selective}} radiance'' in
  atomic arrays},\ }\href {https://doi.org/10.1103/PhysRevX.7.031024}
  {\bibfield  {journal} {\bibinfo  {journal} {Phys. Rev. X}\ }\textbf {\bibinfo
  {volume} {7}},\ \bibinfo {pages} {031024} (\bibinfo {year}
  {2017}{\natexlab{a}})}\BibitemShut {NoStop}%
\bibitem [{\citenamefont {Bukreeva}\ \emph {et~al.}(2006)\citenamefont
  {Bukreeva}, \citenamefont {Popov}, \citenamefont {Pelliccia}, \citenamefont
  {Cedola}, \citenamefont {Dabagov},\ and\ \citenamefont
  {Lagomarsino}}]{bukreevaWaveFieldFormationHollow2006}%
  \BibitemOpen
  \bibfield  {author} {\bibinfo {author} {\bibfnamefont {I.}~\bibnamefont
  {Bukreeva}}, \bibinfo {author} {\bibfnamefont {A.}~\bibnamefont {Popov}},
  \bibinfo {author} {\bibfnamefont {D.}~\bibnamefont {Pelliccia}}, \bibinfo
  {author} {\bibfnamefont {A.}~\bibnamefont {Cedola}}, \bibinfo {author}
  {\bibfnamefont {S.~B.}\ \bibnamefont {Dabagov}},\ and\ \bibinfo {author}
  {\bibfnamefont {S.}~\bibnamefont {Lagomarsino}},\ }\bibfield  {title}
  {\bibinfo {title} {Wave-{{Field Formation}} in a {{Hollow X-Ray
  Waveguide}}},\ }\href {https://doi.org/10.1103/PhysRevLett.97.184801}
  {\bibfield  {journal} {\bibinfo  {journal} {Phys. Rev. Lett.}\ }\textbf
  {\bibinfo {volume} {97}},\ \bibinfo {pages} {184801} (\bibinfo {year}
  {2006})}\BibitemShut {NoStop}%
\bibitem [{\citenamefont {Osterhoff}\ and\ \citenamefont
  {Salditt}(2011)}]{osterhoffCoherenceFilteringXray2011}%
  \BibitemOpen
  \bibfield  {author} {\bibinfo {author} {\bibfnamefont {M.}~\bibnamefont
  {Osterhoff}}\ and\ \bibinfo {author} {\bibfnamefont {T.}~\bibnamefont
  {Salditt}},\ }\bibfield  {title} {\bibinfo {title} {Coherence filtering of
  x-ray waveguides: Analytical and numerical approach},\ }\href
  {https://doi.org/10.1088/1367-2630/13/10/103026} {\bibfield  {journal}
  {\bibinfo  {journal} {New J. Phys.}\ }\textbf {\bibinfo {volume} {13}},\
  \bibinfo {pages} {103026} (\bibinfo {year} {2011})}\BibitemShut {NoStop}%
\bibitem [{\citenamefont {Lagomarsino}\ \emph {et~al.}(1996)\citenamefont
  {Lagomarsino}, \citenamefont {Jark}, \citenamefont {Di~Fonzo}, \citenamefont
  {Cedola}, \citenamefont {Mueller}, \citenamefont {Engstr{\"o}m},\ and\
  \citenamefont {Riekel}}]{lagomarsinoSubmicrometerRayBeam1996}%
  \BibitemOpen
  \bibfield  {author} {\bibinfo {author} {\bibfnamefont {S.}~\bibnamefont
  {Lagomarsino}}, \bibinfo {author} {\bibfnamefont {W.}~\bibnamefont {Jark}},
  \bibinfo {author} {\bibfnamefont {S.}~\bibnamefont {Di~Fonzo}}, \bibinfo
  {author} {\bibfnamefont {A.}~\bibnamefont {Cedola}}, \bibinfo {author}
  {\bibfnamefont {B.}~\bibnamefont {Mueller}}, \bibinfo {author} {\bibfnamefont
  {P.}~\bibnamefont {Engstr{\"o}m}},\ and\ \bibinfo {author} {\bibfnamefont
  {C.}~\bibnamefont {Riekel}},\ }\bibfield  {title} {\bibinfo {title}
  {Submicrometer x-ray beam production by a thin film waveguide},\ }\href
  {https://doi.org/10.1063/1.361761} {\bibfield  {journal} {\bibinfo  {journal}
  {Journal of Applied Physics}\ }\textbf {\bibinfo {volume} {79}},\ \bibinfo
  {pages} {4471} (\bibinfo {year} {1996})}\BibitemShut {NoStop}%
\bibitem [{\citenamefont {Fuhse}\ and\ \citenamefont
  {Salditt}(2005)}]{fuhseFinitedifferenceFieldCalculations2005}%
  \BibitemOpen
  \bibfield  {author} {\bibinfo {author} {\bibfnamefont {C.}~\bibnamefont
  {Fuhse}}\ and\ \bibinfo {author} {\bibfnamefont {T.}~\bibnamefont
  {Salditt}},\ }\bibfield  {title} {\bibinfo {title} {Finite-difference field
  calculations for one-dimensionally confined {{X-ray}} waveguides},\ }\href
  {https://doi.org/10.1016/j.physb.2004.11.019} {\bibfield  {journal} {\bibinfo
   {journal} {Physica B: Condensed Matter}\ }\textbf {\bibinfo {volume}
  {357}},\ \bibinfo {pages} {57} (\bibinfo {year} {2005})}\BibitemShut
  {NoStop}%
\bibitem [{\citenamefont {Fuhse}\ and\ \citenamefont
  {Salditt}(2006)}]{Fuhse_OC_2006}%
  \BibitemOpen
  \bibfield  {author} {\bibinfo {author} {\bibfnamefont {C.}~\bibnamefont
  {Fuhse}}\ and\ \bibinfo {author} {\bibfnamefont {T.}~\bibnamefont
  {Salditt}},\ }\bibfield  {title} {\bibinfo {title} {Propagation of x-rays in
  ultra-narrow slits},\ }\href {https://doi.org/10.1016/j.optcom.2006.03.011}
  {\bibfield  {journal} {\bibinfo  {journal} {Optics Communications}\ }\textbf
  {\bibinfo {volume} {265}},\ \bibinfo {pages} {140} (\bibinfo {year}
  {2006})}\BibitemShut {NoStop}%
\bibitem [{\citenamefont {Pelliccia}\ \emph {et~al.}(2007)\citenamefont
  {Pelliccia}, \citenamefont {Bukreeva}, \citenamefont {Ilie}, \citenamefont
  {Jark}, \citenamefont {Cedola}, \citenamefont {Scarinci},\ and\ \citenamefont
  {Lagomarsino}}]{pellicciaComputerSimulationsExperimental2007}%
  \BibitemOpen
  \bibfield  {author} {\bibinfo {author} {\bibfnamefont {D.}~\bibnamefont
  {Pelliccia}}, \bibinfo {author} {\bibfnamefont {I.}~\bibnamefont {Bukreeva}},
  \bibinfo {author} {\bibfnamefont {M.}~\bibnamefont {Ilie}}, \bibinfo {author}
  {\bibfnamefont {W.}~\bibnamefont {Jark}}, \bibinfo {author} {\bibfnamefont
  {A.}~\bibnamefont {Cedola}}, \bibinfo {author} {\bibfnamefont
  {F.}~\bibnamefont {Scarinci}},\ and\ \bibinfo {author} {\bibfnamefont
  {S.}~\bibnamefont {Lagomarsino}},\ }\bibfield  {title} {\bibinfo {title}
  {Computer simulations and experimental results on air-gap {{X-ray}}
  waveguides},\ }\href {https://doi.org/10.1016/j.sab.2007.02.018} {\bibfield
  {journal} {\bibinfo  {journal} {Spectrochimica Acta Part B: Atomic
  Spectroscopy}\ }\textbf {\bibinfo {volume} {62}},\ \bibinfo {pages} {615}
  (\bibinfo {year} {2007})}\BibitemShut {NoStop}%
\bibitem [{\citenamefont {Lohse}\ and\ \citenamefont
  {Andreji{\'c}}(2023)}]{Lohse__2023}%
  \BibitemOpen
  \bibfield  {author} {\bibinfo {author} {\bibfnamefont {L.~M.}\ \bibnamefont
  {Lohse}}\ and\ \bibinfo {author} {\bibfnamefont {P.}~\bibnamefont
  {Andreji{\'c}}},\ }\href {https://doi.org/10.1364/opticaopen.24028686.v1}
  {\bibinfo {title} {Nano-optical theory of planar x-ray waveguides}},\
  \bibinfo {howpublished} {accepted for publication} (\bibinfo {year}
  {2023})\BibitemShut {NoStop}%
\bibitem [{\citenamefont {Adams}\ \emph {et~al.}(2019)\citenamefont {Adams},
  \citenamefont {Aeppli}, \citenamefont {Allison}, \citenamefont {Baron},
  \citenamefont {Bucksbaum}, \citenamefont {Chumakov}, \citenamefont {Corder},
  \citenamefont {Cramer}, \citenamefont {DeBeer}, \citenamefont {Ding},
  \citenamefont {Evers}, \citenamefont {Frisch}, \citenamefont {Fuchs},
  \citenamefont {Gr{\"u}bel}, \citenamefont {Hastings}, \citenamefont {Heyl},
  \citenamefont {Holberg}, \citenamefont {Huang}, \citenamefont {Ishikawa},
  \citenamefont {Kaldun}, \citenamefont {Kim}, \citenamefont {Kolodziej},
  \citenamefont {Krzywinski}, \citenamefont {Li}, \citenamefont {Liao},
  \citenamefont {Lindberg}, \citenamefont {Madsen}, \citenamefont {Maxwell},
  \citenamefont {Monaco}, \citenamefont {Nelson}, \citenamefont {Palffy},
  \citenamefont {Porat}, \citenamefont {Qin}, \citenamefont {Raubenheimer},
  \citenamefont {Reis}, \citenamefont {R{\"o}hlsberger}, \citenamefont
  {Santra}, \citenamefont {Schoenlein}, \citenamefont {Sch{\"u}nemann},
  \citenamefont {Shpyrko}, \citenamefont {Shvyd'ko}, \citenamefont {Shwartz},
  \citenamefont {Singer}, \citenamefont {Sinha}, \citenamefont {Sutton},
  \citenamefont {Tamasaku}, \citenamefont {Wille}, \citenamefont {Yabashi},
  \citenamefont {Ye},\ and\ \citenamefont
  {Zhu}}]{adamsScientificOpportunitiesXray2019}%
  \BibitemOpen
  \bibfield  {author} {\bibinfo {author} {\bibfnamefont {B.}~\bibnamefont
  {Adams}}, \bibinfo {author} {\bibfnamefont {G.}~\bibnamefont {Aeppli}},
  \bibinfo {author} {\bibfnamefont {T.}~\bibnamefont {Allison}}, \bibinfo
  {author} {\bibfnamefont {A.~Q.~R.}\ \bibnamefont {Baron}}, \bibinfo {author}
  {\bibfnamefont {P.}~\bibnamefont {Bucksbaum}}, \bibinfo {author}
  {\bibfnamefont {A.~I.}\ \bibnamefont {Chumakov}}, \bibinfo {author}
  {\bibfnamefont {C.}~\bibnamefont {Corder}}, \bibinfo {author} {\bibfnamefont
  {S.~P.}\ \bibnamefont {Cramer}}, \bibinfo {author} {\bibfnamefont
  {S.}~\bibnamefont {DeBeer}}, \bibinfo {author} {\bibfnamefont
  {Y.}~\bibnamefont {Ding}}, \bibinfo {author} {\bibfnamefont {J.}~\bibnamefont
  {Evers}}, \bibinfo {author} {\bibfnamefont {J.}~\bibnamefont {Frisch}},
  \bibinfo {author} {\bibfnamefont {M.}~\bibnamefont {Fuchs}}, \bibinfo
  {author} {\bibfnamefont {G.}~\bibnamefont {Gr{\"u}bel}}, \bibinfo {author}
  {\bibfnamefont {J.~B.}\ \bibnamefont {Hastings}}, \bibinfo {author}
  {\bibfnamefont {C.~M.}\ \bibnamefont {Heyl}}, \bibinfo {author}
  {\bibfnamefont {L.}~\bibnamefont {Holberg}}, \bibinfo {author} {\bibfnamefont
  {Z.}~\bibnamefont {Huang}}, \bibinfo {author} {\bibfnamefont
  {T.}~\bibnamefont {Ishikawa}}, \bibinfo {author} {\bibfnamefont
  {A.}~\bibnamefont {Kaldun}}, \bibinfo {author} {\bibfnamefont {K.-J.}\
  \bibnamefont {Kim}}, \bibinfo {author} {\bibfnamefont {T.}~\bibnamefont
  {Kolodziej}}, \bibinfo {author} {\bibfnamefont {J.}~\bibnamefont
  {Krzywinski}}, \bibinfo {author} {\bibfnamefont {Z.}~\bibnamefont {Li}},
  \bibinfo {author} {\bibfnamefont {W.-T.}\ \bibnamefont {Liao}}, \bibinfo
  {author} {\bibfnamefont {R.}~\bibnamefont {Lindberg}}, \bibinfo {author}
  {\bibfnamefont {A.}~\bibnamefont {Madsen}}, \bibinfo {author} {\bibfnamefont
  {T.}~\bibnamefont {Maxwell}}, \bibinfo {author} {\bibfnamefont
  {G.}~\bibnamefont {Monaco}}, \bibinfo {author} {\bibfnamefont
  {K.}~\bibnamefont {Nelson}}, \bibinfo {author} {\bibfnamefont
  {A.}~\bibnamefont {Palffy}}, \bibinfo {author} {\bibfnamefont
  {G.}~\bibnamefont {Porat}}, \bibinfo {author} {\bibfnamefont
  {W.}~\bibnamefont {Qin}}, \bibinfo {author} {\bibfnamefont {T.}~\bibnamefont
  {Raubenheimer}}, \bibinfo {author} {\bibfnamefont {D.~A.}\ \bibnamefont
  {Reis}}, \bibinfo {author} {\bibfnamefont {R.}~\bibnamefont
  {R{\"o}hlsberger}}, \bibinfo {author} {\bibfnamefont {R.}~\bibnamefont
  {Santra}}, \bibinfo {author} {\bibfnamefont {R.}~\bibnamefont {Schoenlein}},
  \bibinfo {author} {\bibfnamefont {V.}~\bibnamefont {Sch{\"u}nemann}},
  \bibinfo {author} {\bibfnamefont {O.}~\bibnamefont {Shpyrko}}, \bibinfo
  {author} {\bibfnamefont {Y.}~\bibnamefont {Shvyd'ko}}, \bibinfo {author}
  {\bibfnamefont {S.}~\bibnamefont {Shwartz}}, \bibinfo {author} {\bibfnamefont
  {A.}~\bibnamefont {Singer}}, \bibinfo {author} {\bibfnamefont {S.~K.}\
  \bibnamefont {Sinha}}, \bibinfo {author} {\bibfnamefont {M.}~\bibnamefont
  {Sutton}}, \bibinfo {author} {\bibfnamefont {K.}~\bibnamefont {Tamasaku}},
  \bibinfo {author} {\bibfnamefont {H.-C.}\ \bibnamefont {Wille}}, \bibinfo
  {author} {\bibfnamefont {M.}~\bibnamefont {Yabashi}}, \bibinfo {author}
  {\bibfnamefont {J.}~\bibnamefont {Ye}},\ and\ \bibinfo {author}
  {\bibfnamefont {D.}~\bibnamefont {Zhu}},\ }\href
  {https://doi.org/10.48550/arXiv.1903.09317} {\bibinfo {title} {Scientific
  {{Opportunities}} with an {{X-ray Free-Electron Laser Oscillator}}}}
  (\bibinfo {year} {2019}),\ \Eprint {https://arxiv.org/abs/1903.09317}
  {arxiv:1903.09317 [physics]} \BibitemShut {NoStop}%
\bibitem [{\citenamefont {{Asenjo-Garcia}}\ \emph
  {et~al.}(2017{\natexlab{b}})\citenamefont {{Asenjo-Garcia}}, \citenamefont
  {Hood}, \citenamefont {Chang},\ and\ \citenamefont
  {Kimble}}]{asenjo-garciaAtomlightInteractionsQuasionedimensional2017}%
  \BibitemOpen
  \bibfield  {author} {\bibinfo {author} {\bibfnamefont {A.}~\bibnamefont
  {{Asenjo-Garcia}}}, \bibinfo {author} {\bibfnamefont {J.~D.}\ \bibnamefont
  {Hood}}, \bibinfo {author} {\bibfnamefont {D.~E.}\ \bibnamefont {Chang}},\
  and\ \bibinfo {author} {\bibfnamefont {H.~J.}\ \bibnamefont {Kimble}},\
  }\bibfield  {title} {\bibinfo {title} {Atom-light interactions in
  quasi-one-dimensional nanostructures: {{A Green}}'s-function perspective},\
  }\href {https://doi.org/10.1103/PhysRevA.95.033818} {\bibfield  {journal}
  {\bibinfo  {journal} {Phys. Rev. A}\ }\textbf {\bibinfo {volume} {95}},\
  \bibinfo {pages} {033818} (\bibinfo {year} {2017}{\natexlab{b}})}\BibitemShut
  {NoStop}%
\bibitem [{\citenamefont {Chang}\ \emph {et~al.}(2012)\citenamefont {Chang},
  \citenamefont {Jiang}, \citenamefont {Gorshkov},\ and\ \citenamefont
  {Kimble}}]{changCavityQEDAtomic2012}%
  \BibitemOpen
  \bibfield  {author} {\bibinfo {author} {\bibfnamefont {D.~E.}\ \bibnamefont
  {Chang}}, \bibinfo {author} {\bibfnamefont {L.}~\bibnamefont {Jiang}},
  \bibinfo {author} {\bibfnamefont {A.~V.}\ \bibnamefont {Gorshkov}},\ and\
  \bibinfo {author} {\bibfnamefont {H.~J.}\ \bibnamefont {Kimble}},\ }\bibfield
   {title} {\bibinfo {title} {Cavity {{QED}} with atomic mirrors},\ }\href
  {https://doi.org/10.1088/1367-2630/14/6/063003} {\bibfield  {journal}
  {\bibinfo  {journal} {New J. Phys.}\ }\textbf {\bibinfo {volume} {14}},\
  \bibinfo {pages} {063003} (\bibinfo {year} {2012})}\BibitemShut {NoStop}%
\bibitem [{\citenamefont {Svidzinsky}\ and\ \citenamefont
  {Chang}(2008)}]{svidzinskyCooperativeSpontaneousEmission2008}%
  \BibitemOpen
  \bibfield  {author} {\bibinfo {author} {\bibfnamefont {A.}~\bibnamefont
  {Svidzinsky}}\ and\ \bibinfo {author} {\bibfnamefont {J.-T.}\ \bibnamefont
  {Chang}},\ }\bibfield  {title} {\bibinfo {title} {Cooperative spontaneous
  emission as a many-body eigenvalue problem},\ }\href
  {https://doi.org/10.1103/PhysRevA.77.043833} {\bibfield  {journal} {\bibinfo
  {journal} {Phys. Rev. A}\ }\textbf {\bibinfo {volume} {77}},\ \bibinfo
  {pages} {043833} (\bibinfo {year} {2008})}\BibitemShut {NoStop}%
\bibitem [{\citenamefont {Svidzinsky}\ \emph {et~al.}(2010)\citenamefont
  {Svidzinsky}, \citenamefont {Chang},\ and\ \citenamefont
  {Scully}}]{svidzinskyCooperativeSpontaneousEmission2010}%
  \BibitemOpen
  \bibfield  {author} {\bibinfo {author} {\bibfnamefont {A.~A.}\ \bibnamefont
  {Svidzinsky}}, \bibinfo {author} {\bibfnamefont {J.-T.}\ \bibnamefont
  {Chang}},\ and\ \bibinfo {author} {\bibfnamefont {M.~O.}\ \bibnamefont
  {Scully}},\ }\bibfield  {title} {\bibinfo {title} {Cooperative {{Spontaneous
  Emission}} of {{N Atoms}}: {{Many-Body Eigenstates}}, the {{Effect}} of
  {{Virtual Lamb Shift Processes}}, and {{Analogy}} with {{Radiation}} of {{N
  Classical Oscillators}}},\ }\href
  {https://doi.org/10.1103/PhysRevA.81.053821} {\bibfield  {journal} {\bibinfo
  {journal} {Phys. Rev. A}\ }\textbf {\bibinfo {volume} {81}},\ \bibinfo
  {pages} {053821} (\bibinfo {year} {2010})}\BibitemShut {NoStop}%
\bibitem [{\citenamefont {Ma}\ and\ \citenamefont
  {Yelin}(2023)}]{maCollectiveLambShift2023}%
  \BibitemOpen
  \bibfield  {author} {\bibinfo {author} {\bibfnamefont {H.}~\bibnamefont
  {Ma}}\ and\ \bibinfo {author} {\bibfnamefont {S.~F.}\ \bibnamefont {Yelin}},\
  }\href@noop {} {\bibinfo {title} {Collective {{Lamb Shift}} and {{Spontaneous
  Emission}} of {{A Dense Atomic Gas}}}} (\bibinfo {year} {2023}),\ \Eprint
  {https://arxiv.org/abs/2305.01865} {arxiv:2305.01865 [quant-ph]} \BibitemShut
  {NoStop}%
\bibitem [{\citenamefont {Javanainen}\ \emph {et~al.}(2017)\citenamefont
  {Javanainen}, \citenamefont {Ruostekoski}, \citenamefont {Li},\ and\
  \citenamefont {Yoo}}]{javanainenExactElectrodynamicsStandard2017}%
  \BibitemOpen
  \bibfield  {author} {\bibinfo {author} {\bibfnamefont {J.}~\bibnamefont
  {Javanainen}}, \bibinfo {author} {\bibfnamefont {J.}~\bibnamefont
  {Ruostekoski}}, \bibinfo {author} {\bibfnamefont {Y.}~\bibnamefont {Li}},\
  and\ \bibinfo {author} {\bibfnamefont {S.-M.}\ \bibnamefont {Yoo}},\
  }\bibfield  {title} {\bibinfo {title} {Exact electrodynamics versus standard
  optics for a slab of cold dense gas},\ }\href
  {https://doi.org/10.1103/PhysRevA.96.033835} {\bibfield  {journal} {\bibinfo
  {journal} {Phys. Rev. A}\ }\textbf {\bibinfo {volume} {96}},\ \bibinfo
  {pages} {033835} (\bibinfo {year} {2017})}\BibitemShut {NoStop}%
\bibitem [{\citenamefont {Javanainen}\ \emph {et~al.}(2014)\citenamefont
  {Javanainen}, \citenamefont {Ruostekoski}, \citenamefont {Li},\ and\
  \citenamefont {Yoo}}]{javanainenShiftsResonanceLine2014}%
  \BibitemOpen
  \bibfield  {author} {\bibinfo {author} {\bibfnamefont {J.}~\bibnamefont
  {Javanainen}}, \bibinfo {author} {\bibfnamefont {J.}~\bibnamefont
  {Ruostekoski}}, \bibinfo {author} {\bibfnamefont {Y.}~\bibnamefont {Li}},\
  and\ \bibinfo {author} {\bibfnamefont {S.-M.}\ \bibnamefont {Yoo}},\
  }\bibfield  {title} {\bibinfo {title} {Shifts of a {{Resonance Line}} in a
  {{Dense Atomic Sample}}},\ }\href
  {https://doi.org/10.1103/PhysRevLett.112.113603} {\bibfield  {journal}
  {\bibinfo  {journal} {Phys. Rev. Lett.}\ }\textbf {\bibinfo {volume} {112}},\
  \bibinfo {pages} {113603} (\bibinfo {year} {2014})}\BibitemShut {NoStop}%
\bibitem [{\citenamefont {Corman}\ \emph {et~al.}(2017)\citenamefont {Corman},
  \citenamefont {Ville}, \citenamefont {{Saint-Jalm}}, \citenamefont
  {Aidelsburger}, \citenamefont {Bienaim{\'e}}, \citenamefont {Nascimb{\`e}ne},
  \citenamefont {Dalibard},\ and\ \citenamefont
  {Beugnon}}]{cormanTransmissionNearresonantLight2017}%
  \BibitemOpen
  \bibfield  {author} {\bibinfo {author} {\bibfnamefont {L.}~\bibnamefont
  {Corman}}, \bibinfo {author} {\bibfnamefont {J.~L.}\ \bibnamefont {Ville}},
  \bibinfo {author} {\bibfnamefont {R.}~\bibnamefont {{Saint-Jalm}}}, \bibinfo
  {author} {\bibfnamefont {M.}~\bibnamefont {Aidelsburger}}, \bibinfo {author}
  {\bibfnamefont {T.}~\bibnamefont {Bienaim{\'e}}}, \bibinfo {author}
  {\bibfnamefont {S.}~\bibnamefont {Nascimb{\`e}ne}}, \bibinfo {author}
  {\bibfnamefont {J.}~\bibnamefont {Dalibard}},\ and\ \bibinfo {author}
  {\bibfnamefont {J.}~\bibnamefont {Beugnon}},\ }\bibfield  {title} {\bibinfo
  {title} {Transmission of near-resonant light through a dense slab of cold
  atoms},\ }\href {https://doi.org/10.1103/PhysRevA.96.053629} {\bibfield
  {journal} {\bibinfo  {journal} {Phys. Rev. A}\ }\textbf {\bibinfo {volume}
  {96}},\ \bibinfo {pages} {053629} (\bibinfo {year} {2017})}\BibitemShut
  {NoStop}%
\bibitem [{\citenamefont {Jenkins}\ \emph {et~al.}(2016)\citenamefont
  {Jenkins}, \citenamefont {Ruostekoski}, \citenamefont {Javanainen},
  \citenamefont {Bourgain}, \citenamefont {Jennewein}, \citenamefont
  {Sortais},\ and\ \citenamefont
  {Browaeys}}]{jenkinsOpticalResonanceShifts2016}%
  \BibitemOpen
  \bibfield  {author} {\bibinfo {author} {\bibfnamefont {S.~D.}\ \bibnamefont
  {Jenkins}}, \bibinfo {author} {\bibfnamefont {J.}~\bibnamefont
  {Ruostekoski}}, \bibinfo {author} {\bibfnamefont {J.}~\bibnamefont
  {Javanainen}}, \bibinfo {author} {\bibfnamefont {R.}~\bibnamefont
  {Bourgain}}, \bibinfo {author} {\bibfnamefont {S.}~\bibnamefont {Jennewein}},
  \bibinfo {author} {\bibfnamefont {Y.~R.~P.}\ \bibnamefont {Sortais}},\ and\
  \bibinfo {author} {\bibfnamefont {A.}~\bibnamefont {Browaeys}},\ }\bibfield
  {title} {\bibinfo {title} {Optical {{Resonance Shifts}} in the
  {{Fluorescence}} of {{Thermal}} and {{Cold Atomic Gases}}},\ }\href
  {https://doi.org/10.1103/PhysRevLett.116.183601} {\bibfield  {journal}
  {\bibinfo  {journal} {Phys. Rev. Lett.}\ }\textbf {\bibinfo {volume} {116}},\
  \bibinfo {pages} {183601} (\bibinfo {year} {2016})}\BibitemShut {NoStop}%
\bibitem [{\citenamefont {Schneider}\ \emph {et~al.}(2016)\citenamefont
  {Schneider}, \citenamefont {Sproll}, \citenamefont {Stawiarski},
  \citenamefont {Schmitteckert},\ and\ \citenamefont
  {Busch}}]{schneider_PRA_2016}%
  \BibitemOpen
  \bibfield  {author} {\bibinfo {author} {\bibfnamefont {M.~P.}\ \bibnamefont
  {Schneider}}, \bibinfo {author} {\bibfnamefont {T.}~\bibnamefont {Sproll}},
  \bibinfo {author} {\bibfnamefont {C.}~\bibnamefont {Stawiarski}}, \bibinfo
  {author} {\bibfnamefont {P.}~\bibnamefont {Schmitteckert}},\ and\ \bibinfo
  {author} {\bibfnamefont {K.}~\bibnamefont {Busch}},\ }\bibfield  {title}
  {\bibinfo {title} {Green's-function formalism for waveguide qed
  applications},\ }\href {https://doi.org/10.1103/PhysRevA.93.013828}
  {\bibfield  {journal} {\bibinfo  {journal} {Phys. Rev. A}\ }\textbf {\bibinfo
  {volume} {93}},\ \bibinfo {pages} {013828} (\bibinfo {year}
  {2016})}\BibitemShut {NoStop}%
\bibitem [{\citenamefont {Lentrodt}\ \emph {et~al.}(2020)\citenamefont
  {Lentrodt}, \citenamefont {Heeg}, \citenamefont {Keitel},\ and\ \citenamefont
  {Evers}}]{lentrodtInitioQuantumModels2020}%
  \BibitemOpen
  \bibfield  {author} {\bibinfo {author} {\bibfnamefont {D.}~\bibnamefont
  {Lentrodt}}, \bibinfo {author} {\bibfnamefont {K.~P.}\ \bibnamefont {Heeg}},
  \bibinfo {author} {\bibfnamefont {C.~H.}\ \bibnamefont {Keitel}},\ and\
  \bibinfo {author} {\bibfnamefont {J.}~\bibnamefont {Evers}},\ }\bibfield
  {title} {\bibinfo {title} {Ab initio quantum models for thin-film x-{{Ray}}
  cavity {{QED}}},\ }\href {https://doi.org/10.1103/PhysRevResearch.2.023396}
  {\bibfield  {journal} {\bibinfo  {journal} {Phys. Rev. Research}\ }\textbf
  {\bibinfo {volume} {2}},\ \bibinfo {pages} {023396} (\bibinfo {year}
  {2020})}\BibitemShut {NoStop}%
\bibitem [{\citenamefont {Kong}\ \emph {et~al.}(2020)\citenamefont {Kong},
  \citenamefont {Chang},\ and\ \citenamefont
  {P{\'a}lffy}}]{kongGreenSfunctionFormalism2020}%
  \BibitemOpen
  \bibfield  {author} {\bibinfo {author} {\bibfnamefont {X.}~\bibnamefont
  {Kong}}, \bibinfo {author} {\bibfnamefont {D.~E.}\ \bibnamefont {Chang}},\
  and\ \bibinfo {author} {\bibfnamefont {A.}~\bibnamefont {P{\'a}lffy}},\
  }\bibfield  {title} {\bibinfo {title} {Green's-function formalism for
  resonant interaction of x rays with nuclei in structured media},\ }\href
  {https://doi.org/10.1103/PhysRevA.102.033710} {\bibfield  {journal} {\bibinfo
   {journal} {Phys. Rev. A}\ }\textbf {\bibinfo {volume} {102}},\ \bibinfo
  {pages} {033710} (\bibinfo {year} {2020})}\BibitemShut {NoStop}%
\bibitem [{\citenamefont {Kagan}\ \emph {et~al.}(1979)\citenamefont {Kagan},
  \citenamefont {Afanas'ev},\ and\ \citenamefont
  {Kohn}}]{kaganExcitationIsomericNuclear1979}%
  \BibitemOpen
  \bibfield  {author} {\bibinfo {author} {\bibfnamefont {{\relax
  Yu}.}~\bibnamefont {Kagan}}, \bibinfo {author} {\bibfnamefont {A.~M.}\
  \bibnamefont {Afanas'ev}},\ and\ \bibinfo {author} {\bibfnamefont {V.~G.}\
  \bibnamefont {Kohn}},\ }\bibfield  {title} {\bibinfo {title} {On excitation
  of isomeric nuclear states in a crystal by synchrotron radiation},\ }\href
  {https://doi.org/10.1088/0022-3719/12/3/027} {\bibfield  {journal} {\bibinfo
  {journal} {J. Phys. C: Solid State Phys.}\ }\textbf {\bibinfo {volume}
  {12}},\ \bibinfo {pages} {615} (\bibinfo {year} {1979})}\BibitemShut
  {NoStop}%
\bibitem [{\citenamefont {Shvyd'ko}(1999)}]{shvydkoNuclearResonantForward1999}%
  \BibitemOpen
  \bibfield  {author} {\bibinfo {author} {\bibfnamefont {Y.~V.}\ \bibnamefont
  {Shvyd'ko}},\ }\bibfield  {title} {\bibinfo {title} {Nuclear resonant forward
  scattering of x rays: {{Time}} and space picture},\ }\href
  {https://doi.org/10.1103/PhysRevB.59.9132} {\bibfield  {journal} {\bibinfo
  {journal} {Phys. Rev. B}\ }\textbf {\bibinfo {volume} {59}},\ \bibinfo
  {pages} {9132} (\bibinfo {year} {1999})}\BibitemShut {NoStop}%
\bibitem [{\citenamefont {Hannon}\ and\ \citenamefont
  {Trammell}(1968)}]{hannonMossbauerDiffractionQuantum1968}%
  \BibitemOpen
  \bibfield  {author} {\bibinfo {author} {\bibfnamefont {J.~P.}\ \bibnamefont
  {Hannon}}\ and\ \bibinfo {author} {\bibfnamefont {G.~T.}\ \bibnamefont
  {Trammell}},\ }\bibfield  {title} {\bibinfo {title} {M\"ossbauer
  {{Diffraction}}. {{I}}. {{Quantum Theory}} of {{Gamma-Ray}} and {{X-Ray
  Optics}}},\ }\href {https://doi.org/10.1103/PhysRev.169.315} {\bibfield
  {journal} {\bibinfo  {journal} {Phys. Rev.}\ }\textbf {\bibinfo {volume}
  {169}},\ \bibinfo {pages} {315} (\bibinfo {year} {1968})}\BibitemShut
  {NoStop}%
\bibitem [{\citenamefont {Hannon}\ and\ \citenamefont
  {Trammell}(1969)}]{hannonMossbauerDiffractionII1969}%
  \BibitemOpen
  \bibfield  {author} {\bibinfo {author} {\bibfnamefont {J.~P.}\ \bibnamefont
  {Hannon}}\ and\ \bibinfo {author} {\bibfnamefont {G.~T.}\ \bibnamefont
  {Trammell}},\ }\bibfield  {title} {\bibinfo {title} {M\"ossbauer
  {{Diffraction}}. {{II}}. {{Dynamical Theory}} of {{M\"ossbauer Optics}}},\
  }\href {https://doi.org/10.1103/PhysRev.186.306} {\bibfield  {journal}
  {\bibinfo  {journal} {Phys. Rev.}\ }\textbf {\bibinfo {volume} {186}},\
  \bibinfo {pages} {306} (\bibinfo {year} {1969})}\BibitemShut {NoStop}%
\bibitem [{\citenamefont {Hannon}\ \emph {et~al.}(1974)\citenamefont {Hannon},
  \citenamefont {Carron},\ and\ \citenamefont
  {Trammell}}]{hannonMossbauerDiffractionIII1974}%
  \BibitemOpen
  \bibfield  {author} {\bibinfo {author} {\bibfnamefont {J.~P.}\ \bibnamefont
  {Hannon}}, \bibinfo {author} {\bibfnamefont {N.~J.}\ \bibnamefont {Carron}},\
  and\ \bibinfo {author} {\bibfnamefont {G.~T.}\ \bibnamefont {Trammell}},\
  }\bibfield  {title} {\bibinfo {title} {M\"ossbauer diffraction. {{III}}.
  {{Emission}} of {{M\"ossbauer}} gamma rays from crystals. {{A}}. {{General}}
  theory},\ }\href {https://doi.org/10.1103/PhysRevB.9.2791} {\bibfield
  {journal} {\bibinfo  {journal} {Phys. Rev. B}\ }\textbf {\bibinfo {volume}
  {9}},\ \bibinfo {pages} {2791} (\bibinfo {year} {1974})}\BibitemShut
  {NoStop}%
\bibitem [{\citenamefont {Buhmann}(2012)}]{buhmannDispersionForces2012}%
  \BibitemOpen
  \bibfield  {author} {\bibinfo {author} {\bibfnamefont {S.~Y.}\ \bibnamefont
  {Buhmann}},\ }\href {https://doi.org/10.1007/978-3-642-32484-0} {\emph
  {\bibinfo {title} {Dispersion {{Forces I}}}}},\ \bibinfo {series} {Springer
  {{Tracts}} in {{Modern Physics}}}, Vol.\ \bibinfo {volume} {247}\ (\bibinfo
  {publisher} {{Springer, Berlin, Heidelberg}},\ \bibinfo {year}
  {2012})\BibitemShut {NoStop}%
\bibitem [{\citenamefont {Andreji{\'c}}\ and\ \citenamefont
  {P{\'a}lffy}(2021)}]{andrejicSuperradianceAnomalousHyperfine2021}%
  \BibitemOpen
  \bibfield  {author} {\bibinfo {author} {\bibfnamefont {P.}~\bibnamefont
  {Andreji{\'c}}}\ and\ \bibinfo {author} {\bibfnamefont {A.}~\bibnamefont
  {P{\'a}lffy}},\ }\bibfield  {title} {\bibinfo {title} {Superradiance and
  anomalous hyperfine splitting in inhomogeneous ensembles},\ }\href
  {https://doi.org/10.1103/PhysRevA.104.033702} {\bibfield  {journal} {\bibinfo
   {journal} {Phys. Rev. A}\ }\textbf {\bibinfo {volume} {104}},\ \bibinfo
  {pages} {033702} (\bibinfo {year} {2021})}\BibitemShut {NoStop}%
\bibitem [{\citenamefont
  {Andreji{\'c}}(2023)}]{andrejicControlHighFrequency2023}%
  \BibitemOpen
  \bibfield  {author} {\bibinfo {author} {\bibfnamefont {P.}~\bibnamefont
  {Andreji{\'c}}},\ }\emph {\bibinfo {title} {Control of High Frequency
  Electromagnetic Radiation}},\ \href@noop {} {Ph.D. thesis},\ \bibinfo
  {school} {Heidelberg University} (\bibinfo {year} {2023})\BibitemShut
  {NoStop}%
\bibitem [{\citenamefont {Toma{\v
  s}}(1995)}]{tomasGreenFunctionMultilayers1995}%
  \BibitemOpen
  \bibfield  {author} {\bibinfo {author} {\bibfnamefont {M.~S.}\ \bibnamefont
  {Toma{\v s}}},\ }\bibfield  {title} {\bibinfo {title} {Green function for
  multilayers: {{Light}} scattering in planar cavities},\ }\href
  {https://doi.org/10.1103/PhysRevA.51.2545} {\bibfield  {journal} {\bibinfo
  {journal} {Phys. Rev. A}\ }\textbf {\bibinfo {volume} {51}},\ \bibinfo
  {pages} {2545} (\bibinfo {year} {1995})}\BibitemShut {NoStop}%
\bibitem [{\citenamefont
  {Johansson}(2011)}]{johanssonElectromagneticGreenFunction2011}%
  \BibitemOpen
  \bibfield  {author} {\bibinfo {author} {\bibfnamefont {P.}~\bibnamefont
  {Johansson}},\ }\bibfield  {title} {\bibinfo {title} {Electromagnetic
  {{Green}}'s function for layered systems: {{Applications}} to nanohole
  interactions in thin metal films},\ }\href
  {https://doi.org/10.1103/PhysRevB.83.195408} {\bibfield  {journal} {\bibinfo
  {journal} {Phys. Rev. B}\ }\textbf {\bibinfo {volume} {83}},\ \bibinfo
  {pages} {195408} (\bibinfo {year} {2011})}\BibitemShut {NoStop}%
\bibitem [{\citenamefont {Hanson}\ and\ \citenamefont
  {Yakovlev}(2002)}]{hansonOperatorTheoryElectromagnetics2002}%
  \BibitemOpen
  \bibfield  {author} {\bibinfo {author} {\bibfnamefont {G.~W.}\ \bibnamefont
  {Hanson}}\ and\ \bibinfo {author} {\bibfnamefont {A.~B.}\ \bibnamefont
  {Yakovlev}},\ }\href@noop {} {\emph {\bibinfo {title} {Operator {{Theory}}
  for {{Electromagnetics}}: {{An Introduction}}}}}\ (\bibinfo  {publisher}
  {{Springer New York}},\ \bibinfo {address} {{New York, NY}},\ \bibinfo {year}
  {2002})\BibitemShut {NoStop}%
\bibitem [{\citenamefont {Pelliccia}\ \emph {et~al.}(2006)\citenamefont
  {Pelliccia}, \citenamefont {Bukreeva}, \citenamefont {Cedola},\ and\
  \citenamefont {Lagomarsino}}]{pellicciaDispersionPropertiesXray2006}%
  \BibitemOpen
  \bibfield  {author} {\bibinfo {author} {\bibfnamefont {D.}~\bibnamefont
  {Pelliccia}}, \bibinfo {author} {\bibfnamefont {I.}~\bibnamefont {Bukreeva}},
  \bibinfo {author} {\bibfnamefont {A.}~\bibnamefont {Cedola}},\ and\ \bibinfo
  {author} {\bibfnamefont {S.}~\bibnamefont {Lagomarsino}},\ }\bibfield
  {title} {\bibinfo {title} {Dispersion properties of x-ray waveguides},\
  }\href {https://doi.org/10.1364/AO.45.002821} {\bibfield  {journal} {\bibinfo
   {journal} {Appl. Opt.}\ }\textbf {\bibinfo {volume} {45}},\ \bibinfo {pages}
  {2821} (\bibinfo {year} {2006})}\BibitemShut {NoStop}%
\bibitem [{\citenamefont {Leung}\ and\ \citenamefont
  {Pang}(1996)}]{leungCompletenessTimeindependentPerturbation1996}%
  \BibitemOpen
  \bibfield  {author} {\bibinfo {author} {\bibfnamefont {P.~T.}\ \bibnamefont
  {Leung}}\ and\ \bibinfo {author} {\bibfnamefont {K.~M.}\ \bibnamefont
  {Pang}},\ }\bibfield  {title} {\bibinfo {title} {Completeness and
  time-independent perturbation of morphology-dependent resonances in
  dielectric spheres},\ }\href {https://doi.org/10.1364/JOSAB.13.000805}
  {\bibfield  {journal} {\bibinfo  {journal} {J. Opt. Soc. Am. B}\ }\textbf
  {\bibinfo {volume} {13}},\ \bibinfo {pages} {805} (\bibinfo {year}
  {1996})}\BibitemShut {NoStop}%
\bibitem [{\citenamefont {Kristensen}\ \emph {et~al.}(2012)\citenamefont
  {Kristensen}, \citenamefont {Van~Vlack},\ and\ \citenamefont
  {Hughes}}]{kristensenGeneralizedEffectiveMode2012}%
  \BibitemOpen
  \bibfield  {author} {\bibinfo {author} {\bibfnamefont {P.~T.}\ \bibnamefont
  {Kristensen}}, \bibinfo {author} {\bibfnamefont {C.}~\bibnamefont
  {Van~Vlack}},\ and\ \bibinfo {author} {\bibfnamefont {S.}~\bibnamefont
  {Hughes}},\ }\bibfield  {title} {\bibinfo {title} {Generalized effective mode
  volume for leaky optical cavities},\ }\href
  {https://doi.org/10.1364/OL.37.001649} {\bibfield  {journal} {\bibinfo
  {journal} {Opt. Lett.}\ }\textbf {\bibinfo {volume} {37}},\ \bibinfo {pages}
  {1649} (\bibinfo {year} {2012})}\BibitemShut {NoStop}%
\bibitem [{\citenamefont
  {Weisstein}({\natexlab{a}})}]{weissteinShermanMorrisonFormula}%
  \BibitemOpen
  \bibfield  {author} {\bibinfo {author} {\bibfnamefont {E.~W.}\ \bibnamefont
  {Weisstein}},\ }\href@noop {} {\bibinfo {title} {Sherman-{{Morrison
  Formula}}}},\ \bibinfo {howpublished} {https://mathworld.wolfram.com/}
  ({\natexlab{a}})\BibitemShut {NoStop}%
\bibitem [{\citenamefont {Heeg}\ and\ \citenamefont
  {Evers}(2013)}]{heegXrayQuantumOptics2013}%
  \BibitemOpen
  \bibfield  {author} {\bibinfo {author} {\bibfnamefont {K.~P.}\ \bibnamefont
  {Heeg}}\ and\ \bibinfo {author} {\bibfnamefont {J.}~\bibnamefont {Evers}},\
  }\bibfield  {title} {\bibinfo {title} {X-ray quantum optics with
  {{M\"ossbauer}} nuclei embedded in thin-film cavities},\ }\href
  {https://doi.org/10.1103/PhysRevA.88.043828} {\bibfield  {journal} {\bibinfo
  {journal} {Phys. Rev. A}\ }\textbf {\bibinfo {volume} {88}},\ \bibinfo
  {pages} {043828} (\bibinfo {year} {2013})}\BibitemShut {NoStop}%
\bibitem [{\citenamefont
  {Weisstein}({\natexlab{b}})}]{weissteinSokhotskyFormula}%
  \BibitemOpen
  \bibfield  {author} {\bibinfo {author} {\bibfnamefont {E.~W.}\ \bibnamefont
  {Weisstein}},\ }\href@noop {} {\bibinfo {title} {Sokhotsky's {{Formula}}}},\
  \bibinfo {howpublished}
  {https://mathworld.wolfram.com/SokhotskysFormula.html}
  ({\natexlab{b}})\BibitemShut {NoStop}%
\end{thebibliography}%
\cleardoublepage
\appendix
\onecolumngrid
\section{Equation of motion for macroscopic quantized field}\label{app:kramers-kronig}
In this appendix, we demonstrate that the quantized field obeys the operator form of the macroscopic Maxwell's equations,
\begin{equation}
	\hat{B}(\vec{r},\omega) = \hat{B}_{in}(\vec{r},\omega) - 
	\mu_0  \int\dd[3]{r'}\rho(r') \dyarr{G}_{mm}(\vec{r},\vec{r}',\omega) \cdot \hat{m}(\vec{r}',\omega).
\end{equation}
We begin with the Heisenberg equation of motion for a given noise-frequency mode. Since the noise-frequency is a formal parameter, we must explicitly keep track of both the noise frequency $\nu$ as well as the time dependence $t$,
\begin{equation}
	\hat{B}_+(\vec{r},\nu) \to \hat{B}_+(\vec{r},\nu,t).
\end{equation}
To evaluate the equation of motion, we require the equal time commutator of magnetic field components. This is given by
\begin{equation}
	[\hat{B}_+(\vec{r},\nu,t), \hat{B}_-(\vec{r}',\nu',t)]=-\frac{\hbar \mu_0}{\pi} \im\left\{\dyarr{G}_{mm}(\vec{r},\vec{r}',\nu)\right\}\delta(\nu-\nu').
\end{equation}
We note that this holds even in the interaction picture.
The equation of motion in interaction picture reads
\begin{align}
	\partial_t \hat{B}_+(\vec{r},\nu,t) =& -\frac{i}{\hbar}[\hat{B}_+(\vec{r},\nu,t), H_F - H_{F,T} + H_I]
	\nonumber
	\\
	=& i(\omega_0 - \nu)\hat{B}(\vec{r},\nu,t) -\frac{i\mu_0}{\pi}e^{i\omega_0 t}\int \dd[3]{r'} \rho(\vec{r}')\im\left\{\dyarr{G}_{mm}(\vec{r},\vec{r}',\nu)\right\}\cdot \hat{m}(\vec{r}',t).
\end{align}
The formal solution is given by
\begin{equation}\label{eq:eom-field-solution-1}
	\hat{B}_+(\vec{r},\nu,t) = 
	e^{i(\omega_0 - \nu)t}\hat{B}(\vec{r},\nu,-\infty)
	-\frac{i\mu_0}{\pi}e^{i\omega_0 t}\int_{-\infty}^{t}\dd{t'} e^{-i\nu(t-t')}\int \dd[3]{r'} \rho(\vec{r}')\im\left\{\dyarr{G}_{mm}(\vec{r},\vec{r}',\nu)\right\}\cdot \hat{m}(\vec{r}',t').
\end{equation}
Recalling that the full field in interaction picture is given by
\begin{equation}
	\hat{B}(\vec{r},t) = \int_0^{\infty}\dd{\nu}e^{-i\omega_0 t}\hat{B}_+(\vec{r},\nu,t) + \mathrm{h.c.},
\end{equation}
we add \eqref{eq:eom-field-solution-1} and its Hermitian conjugate together, and integrate over the noise frequencies to obtain the following solution for the total field,
\begin{equation}\label{eq:eom-field-solution-2}
	\hat{B}(\vec{r},t) = 
	\hat{B}_{in}(\vec{r},t)
	+\frac{\mu_0}{\pi}\int_{0}^{\infty}\dd{\nu}\int_{-\infty}^{t}\dd{t'}\left(ie^{i\nu(t-t')}-ie^{-i\nu(t-t')}\right)
	\int \dd[3]{r'}
	\im\left\{\dyarr{G}_{mm}(\vec{r},\vec{r}',\nu)\right\}\cdot \hat{m}(\vec{r}',t').
\end{equation}
Here, we have defined the input field
\begin{equation}
	\hat{B}_{in}(\vec{r},t) = \int_0^{\infty}e^{-i\nu t}\hat{B}(\vec{r},\nu,-\infty) + \mathrm{h.c.}
\end{equation}
It is the homogeneous solution for the free field equations of motion in the absence of the resonant nuclei.
To simplify \eqref{eq:eom-field-solution-2}, we note that the Green's function obeys the Schwarz reflection principle~\cite{buhmannDispersionForces2012},
\begin{equation}
	\dyarr{G}_{mm}(\vec{r},\vec{r}',\omega)^* = \dyarr{G}_{mm}(\vec{r},\vec{r}',-\omega^*),
\end{equation}
and thus
\begin{equation}
	\im\left\{\dyarr{G}_{mm}(\vec{r},\vec{r}',\nu)\right\} 
	=-\im\left\{\dyarr{G}_{mm}(\vec{r},\vec{r}',-\nu)\right\}.
\end{equation}
This then gives
\begin{equation}
	\int_{0}^{\infty}\dd{\nu}\int_{-\infty}^{t}\dd{t'}\left(ie^{i\nu(t-t')}-ie^{-i\nu(t-t')}\right)
	\im\left\{\dyarr{G}_{mm}(\vec{r},\vec{r}',\nu)\right\}
	=
	i\int_{-\infty}^{\infty}\dd{\nu}\int_{-\infty}^{t}\dd{t'}e^{i\nu(t-t')}
	\im\left\{\dyarr{G}_{mm}(\vec{r},\vec{r}',\nu)\right\}.
\end{equation}
We can then rewrite the scattered part of \eqref{eq:eom-field-solution-2} as
\begin{equation}\label{eq:eom-field-solution-3}
	\hat{B}_{sc}(\vec{r},t) = \frac{i\mu_0}{\pi}\int_{-\infty}^{\infty}\dd{\nu}\int_{-\infty}^{\infty}\dd{t'}\Theta(t-t')e^{i\nu(t-t')}
	\int \dd[3]{r'}\rho(\vec{r}')\im\left\{\dyarr{G}_{mm}(\vec{r},\vec{r}',\nu)\right\}\cdot \hat{m}(\vec{r}',t'),
\end{equation}
where $\Theta(t)$ is the Heaviside step distribution.
Taking a temporal Fourier transform of \eqref{eq:eom-field-solution-3}, we obtain
\begin{align}
	\hat{B}_{sc}(\vec{r},\omega) =& \int \dd{t}e^{i\omega t}  \hat{B}_{sc}(\vec{r},t)
	\nonumber
	\\
	=& \frac{\mu_0}{\pi}\int_{-\infty}^{\infty}\dd{\nu}\frac{1}{\nu-\omega + i0^+}\int \dd[3]{r'}\rho(\vec{r}')\im\left\{\dyarr{G}_{mm}(\vec{r},\vec{r}',\nu)\right\}\cdot \hat{m}(\vec{r}',\omega).
\end{align}
We can then use the Sokhotski-Plemelj formula~\cite{weissteinSokhotskyFormula}
\begin{equation}
	\frac{1}{\omega+i0^+} = -i\pi \delta(\omega) +\mathcal{P}\frac{1}{\omega},
\end{equation}
where $\mathcal{P}$ denotes the Cauchy principle value, and the Kramers-Kronig relations
\begin{equation}
	\pi \re\left\{\dyarr{G}_{mm}(\vec{r},\vec{r}',\omega)\right\} =\mathcal{P}\int_{-\infty}^\infty \dd{\nu}\frac{1}{\omega - \nu}
	\im\left\{\dyarr{G}_{mm}(\vec{r},\vec{r}',\nu)\right\},
\end{equation}
to obtain
\begin{equation}
	\hat{B}_{sc}(\vec{r},\omega) = -\mu_0  \int\dd[3]{r'}\rho(r') \dyarr{G}_{mm}(\vec{r},\vec{r}',\omega) \cdot \hat{m}(\vec{r}',\omega),
\end{equation}
as required.

\section{Optical Bloch equations for nuclear transitions}\label{app:optical-bloch}
In this appendix, we will evaluate the equation of motion for the nuclear transition operators, and derive the linear susceptibility.
We begin with the master equation for an arbitrary operator $\hat{O}$,
\begin{equation}
	\partial_t \hat{O} = \frac{i}{\hbar}[\hat{H}_N + \hat{H}_F - H_T + \hat{H}_I(t), \hat{O}] + L^H[\hat{O}],
\end{equation}
where $L^H$ is the Heisenberg form of the Lindblad terms.
For the transition operators, we find the following commutators with the interaction Hamiltonian,
\begin{align}
	[\hat{H}_I(t), \trfi{\mu \nu}(\vec{r}, t)] &= m_0\sum_j \left(
		\trfi{\mu j}(\vec{r},t)\vec{d}_{\nu j}e^{i\omega_0 t}-\trfi{j \nu}(\vec{r},t)\vec{d}_{\mu j}^* e^{-i\omega_0 t}
	\right)
	\cdot 
	\hat{B}(\vec{r},t)
	\\
	[\hat{H}_I(t), \trfi{j k}(\vec{r}, t)] &= -m_0 \sum_\mu \left(
		\trfi{\mu k}(\vec{r},t)\vec{d}_{\mu j}e^{i\omega_0 t} - \trfi{j \mu}(\vec{r},t) \vec{d}_{\mu k}^* e^{-i\omega_0 t}
	\right)
	\cdot 
	\hat{B}(\vec{r},t),
	\\
	[\hat{H}_I(t), \trfi{\mu j}(\vec{r}, t)] &= m_0
	\left(
		\sum_{\nu}\trfi{\mu\nu}(\vec{r},t)\vec{d}_{\nu j}^*e^{-i \omega_0 t} -\sum_k \trfi{kj}\vec{d}_{\mu k}^*e^{-i\omega_0 t}
	\right)
	\cdot 
	\hat{B}(\vec{r},t),
	\\
	[\hat{H}_I(t), \trfi{j\mu}(\vec{r}, t)] &= -m_0
	\left(
		\sum_{\nu}\trfi{\nu\mu}(\vec{r},t)\vec{d}_{\nu j}e^{i \omega_0 t} -\sum_k \trfi{jk}\vec{d}_{\mu k}e^{i\omega_0 t}
	\right)
	\cdot 
	\hat{B}(\vec{r},t),
\end{align}
where $\mu,\nu$ index excited states, $j,k$ index ground states, and 
\begin{equation}\label{eq:pos-freq-rotating}
	\hat{B}(\vec{r},t) = \hat{B}_+(\vec{r},t)e^{-i \omega_0 t} + \hat{B}_-(\vec{r},t)e^{i\omega_0 t}
\end{equation}
is the total magnetic field in the rotating frame.
The nuclear Hamiltonian commutators give
\begin{align}
	[\hat{H}_N - H_{T,N}, \trfi{\mu \nu}(\vec{r}, t)] &= \hbar(\Delta_\mu - \Delta_\nu) \trfi{\mu \nu}(\vec{r}, t),
	\\
	[\hat{H}_N - H_{T,N}, \trfi{j k}(\vec{r}, t)] &= \hbar(\Delta_j - \Delta_k)\trfi{j k}(\vec{r},t),
	\\
	[\hat{H}_N - H_{T,N}, \trfi{\mu j}(\vec{r}, t)] &= \hbar(\Delta_\mu - \Delta_j)\trfi{\mu j}(\vec{r},t).
\end{align}
The Lindblad terms acting on excited state transition operators give
\begin{equation}
	L^H[\trfi{\mu\nu}(\vec{r},t)] = -\gamma \trfi{\mu\nu}(\vec{r},t),
\end{equation}
where $\gamma$ is the total line-width, summed over all multi-polarities.

For ground state transition operators they give
\begin{equation}
	\begin{aligned}
		L^H[\trfi{jk}(\vec{r},t)] = \delta_{jk}\sum_\mu \Gamma(\mu \to j)\trfi{\mu\mu}(\vec{r},t), 
	\end{aligned}
\end{equation}
where the partial decay rate is given by
\begin{equation}
	\Gamma(\mu \to j) =\sum_{\lambda=\mathcal{E},\mathcal{M}}\sum_{l}\Gamma(\lambda l, \mu \to j).
\end{equation}

Finally, for excited-ground state transition operators they give
\begin{equation}
	\begin{aligned}
		L^H[\trfi{\mu j}(\vec{r},t)] =& -\frac{\gamma}{2}\trfi{\mu j}(\vec{r},t).
	\end{aligned}
\end{equation}
Combined, we have the following nuclear Bloch equations, as required,
\begin{align}
	\label{eq:app-bloch-1}
	\partial_t \trfi{\mu\nu}(\vec{r},t) &= 
	\left(i(\Delta_\mu - \Delta_k)-\gamma\right)\trfi{\mu	\nu}(\vec{r},t)
	\\
	&\hspace{40pt}\nonumber 
	+\frac{i m_0}{\hbar}\sum_j \left(\trfi{\mu j}(\vec{r},t)\vec{d}_{\nu j}e^{i\omega_0 t}-\trfi{j \nu}(\vec{r},t)\vec{d}_{\mu j}^* e^{-i\omega_0 t}\right)\cdot \hat{B}(\vec{r},t),
	\\
	\label{eq:app-bloch-2}
	\partial_t\trfi{jk}(\vec{r},t) &=
	i(\Delta_j - \Delta_k)\trfi{j k}(\vec{r},t) + \delta_{jk}\sum_\mu \Gamma(\mu \to j)\trfi{\mu\mu}(\vec{r},t)
	\\
	&\hspace{40pt}\nonumber 
	-\frac{i m_0}{\hbar} \sum_\mu \left(\trfi{\mu k}(\vec{r},t)\vec{d}_{\mu j}e^{i\omega_0 t} - \trfi{j \mu}(\vec{r},t) \vec{d}_{\mu k}^* e^{-i\omega_0 t}\right)\cdot \hat{B}(\vec{r},t),
	\\
	\label{eq:app-bloch-3}
	\partial_t \trfi{\mu j}(\vec{r},t) &= 
	\left(i(\Delta_\mu - \Delta_j) - \frac{\gamma}{2}\right)\trfi{\mu j}(\vec{r},t)
	\\
	&\hspace{40pt}\nonumber 
	+\frac{i m_0}{\hbar}\left(\sum_{\nu}\trfi{\mu\nu}(\vec{r},t)\vec{d}_{\nu j}^*e^{-i\omega_0 t} -\sum_k \trfi{kj}\vec{d}_{\mu k}^*e^{-i\omega_0 t}\right)\cdot \hat{B}(\vec{r},t),
	\\
	\label{eq:app-bloch-4}
	\partial_t \trfi{j \mu}(\vec{r},t) &= 
	\left(-i(\Delta_\mu - \Delta_j) - \frac{\gamma}{2}\right)\trfi{j\mu}(\vec{r},t)
	\\
	&\hspace{40pt}\nonumber 
	-\frac{i m_0}{\hbar}\left(\sum_{\nu}\trfi{\nu\mu}(\vec{r},t)\vec{d}_{\nu j}e^{i\omega_0 t} -\sum_k \trfi{jk}\vec{d}_{\mu k}e^{i\omega_0 t}\right)\cdot \hat{B}(\vec{r},t).
\end{align}
Finally, let us derive the linear response susceptibility. In the linear response regime, we can approximate the ground state population as remaining fixed, and the excited state population as zero. At room temperature, the population is uniformly distributed amongst the available ground states, 
\begin{align}
	\trfi{\mu\nu}(\vec{r}) &= 0,
	\\
	\trfi{jk}(\vec{r}) &= \frac{\delta_{jk}}{2I_g+1}.
\end{align}
Using this approximation, and expanding the field into its positive and negative frequency components using \eqref{eq:pos-freq-rotating}, we have
\begin{align}
	\label{eq:app-bloch-5}
	\partial_t \trfi{\mu j}(\vec{r},t) &= 
	\left(i(\Delta_\mu - \Delta_j) - \frac{\gamma}{2}\right)\trfi{\mu j}(\vec{r},t)
	\\
	&\hspace{40pt}\nonumber 
	-\frac{i m_0}{\hbar(2I_g + 1)}\vec{d}_{\mu j}^*\cdot\left(\hat{B}_+(\vec{r},t)e^{-2i \omega_0 t} + \hat{B}_-(\vec{r},t)\right),
	\\
	\label{eq:app-bloch-6}
	\partial_t \trfi{j \mu}(\vec{r},t) &= 
	\left(-i(\Delta_\mu - \Delta_j) - \frac{\gamma}{2}\right)\trfi{j\mu}(\vec{r},t)
	\\
	&\hspace{40pt}\nonumber 
	+\frac{i m_0}{\hbar(2I_g+1)}\vec{d}_{\mu j}\cdot \left(\hat{B}_+(\vec{r},t)+ \hat{B}_-(\vec{r},t)e^{2i\omega_0 t}\right).
\end{align}
We now apply the rotating wave approximation: the $e^{\pm 2i\omega_0 t}$ factors oscillate very rapidly under integration, and can be neglected. We can then solve the response in Fourier space to obtain
\begin{align}
	\label{eq:app-bloch-7}
	\trfi{\mu j}(\vec{r},\omega) &= 
	\frac{m_0}{\hbar(2I_g + 1)}\frac{1}{\omega + \Delta_\mu - \Delta_j + i\gamma/2}\vec{d}_{\mu j}^*\cdot\hat{B}_-(\vec{r},\omega)
	\\
	\label{eq:app-bloch-8}
	\trfi{j \mu}(\vec{r},\omega) &= 
	-\frac{m_0}{\hbar(2I_g+1)}\frac{1}{\omega - \Delta_\mu +\Delta_j + i\gamma/2 }\vec{d}_{\mu j}\cdot \hat{B}_+(\vec{r},\omega).
\end{align}
Finally, using the definition 
\begin{equation}
	\hat{m}_+(\vec{r},t) = m_0 \sum_{\mu,j}\vec{d}_{\mu j}^*\trfi{j \mu}(\vec{r},t),
\end{equation}
we find that
\begin{equation}
	\hat{m}_+(\vec{r},\omega) = -\frac{m_0^2}{
		\hbar (2I_g + 1)
	}
	\sum_{\mu j}
	\frac{1}{\omega - \Delta_\mu +\Delta_j + i\gamma/2 }\vec{d}_{\mu j}^* \otimes \vec{d}_{\mu j}\cdot \hat{B}_+(\vec{r},\omega)
\end{equation}
Finally, we can express $m_0^2$ in terms of the decay rate as follows: the radiative decay rate is given in terms of the reduced transition probability by
\begin{equation}
	\Gamma_{rad}(\mathcal{M}1,I_e\to I_g) = \frac{\gamma}{1+\alpha}=\frac{k_0^3 \mu_0}{3\pi \hbar} \mathcal{B}(\mathcal{M}1,I_e\to I_g).
\end{equation}
Since we have
\begin{equation}
	m_0^2 = f_{LM}\mathcal{B}(\mathcal{M}1,I_e\to I_g),
\end{equation}
it follows that
\begin{equation}
	\hat{m}_+(\vec{r},\omega) = -\frac{6\pi }{\mu_0 k_0^3}\frac{1}{1+\alpha}\frac{1}{2I_g+1}
	\sum_{\mu, j}
	\frac{\gamma /2}{\omega - \Delta_\mu +\Delta_j + i\gamma/2 }\vec{d}_{\mu j}^* \otimes \vec{d}_{\mu j}\cdot \hat{B}_+(\vec{r},\omega).
\end{equation}
Noting that the cross-section of elastic resonant scattering is
\begin{equation}
	\sigma_{res} = \frac{2\pi}{k_0^2}\frac{f_{LM}}{1+\alpha} \frac{2I_e+1}{2I_g+1},
\end{equation}
we have
\begin{equation}
	\hat{m}_+(\vec{r},\omega) = -\frac{3}{2I_e+1}\frac{\sigma_{res}}{\mu_0 k_0}\sum_{\mu, j}
	\frac{\gamma /2}{\omega - \Delta_\mu +\Delta_j + i\gamma/2 }\vec{d}_{\mu j}^* \otimes \vec{d}_{\mu j}\cdot \hat{B}_+(\vec{r},\omega).
\end{equation}
Finally, defining the susceptibility $\dyarr{\chi}_m$, we have
\begin{align}
	\hat{m}_+(\vec{r},\omega) =& \frac{1}{\mu_0}\dyarr{\chi}_m \cdot \hat{B}_+(\vec{r},\omega),
	\\
	\dyarr{\chi}_m(\omega) =& -\frac{\sigma_{res}}{k_0}\dyarr{F}(\omega),
	\\
	\dyarr{F}(\omega) =& \frac{3}{2I_e+1}\sum_{\mu, j}
	\frac{\gamma /2}{\omega - \Delta_\mu +\Delta_j + i\gamma/2 }\vec{d}_{\mu j}^* \otimes \vec{d}_{\mu j},
\end{align}
as required. In the limit of vanishing hyperfine splitting, we have
\begin{equation}
	\dyarr{F}(\omega) = \frac{\gamma /2}{\omega + i\gamma/2 } \frac{3}{2I_e+1}\sum_{m_e, m_g}
	\vec{d}_{m_e m_g}^* \otimes \vec{d}_{m_e m_g},
\end{equation}
where we note that since all excited and ground states are degenerate in their respective subspaces, we are free to use the angular momentum eigenstates $\ket{m_e},\ket{m_g}$ as our basis. Using equations \eqref{eq:gen-clebsch} and \eqref{eq:definition-d-vector} we then have
\begin{equation}
	\sum_{m_e, m_g}
	\vec{d}_{m_e m_g}^* \otimes \vec{d}_{m_e m_g}
	= 
	(2I_e + 1)\sum_{q,q'}\sum_{m_e,m_g}
	\hat{e}_{q}^* 
	\otimes 
	\hat{e}_{q'}
	\begin{pmatrix}
		I_e & I_g & 1
		\\
		-m_e & m_g & q
	\end{pmatrix}
	\begin{pmatrix}
		I_e & I_g & 1
		\\
		-m_e & m_g & q'
	\end{pmatrix}
\end{equation}
Using the orthogonality relations of the Wigner 3j symbols, one can show that the sum evaluates to 
\begin{equation}
	(2I_e + 1)\sum_{q,q'}\sum_{m_e,m_g}
	\hat{e}_{q}^* 
	\otimes 
	\hat{e}_{q'}
	\begin{pmatrix}
		I_e & I_g & 1
		\\
		-m_e & m_g & q
	\end{pmatrix}
	\begin{pmatrix}
		I_e & I_g & 1
		\\
		-m_e & m_g & q'
	\end{pmatrix}
	=
	\frac{2I_e + 1}{3}\dyarr{1},
\end{equation}
giving us 
\begin{equation}
	\dyarr{\chi}_m(\omega) = -\frac{\sigma_{res}}{k_0}\frac{\gamma /2}{\omega + i \gamma/2}\dyarr{1}
\end{equation}
as expected.

\section{Explicit expressions for Dyson series}\label{app:dyson}
In this appendix, we will derive the explicit expressions for the Dyson series describing the scattered field.

The Laplace transformed Dyson series is given by
\begin{align}
	\tilde{B}(s,\omega)
	=& 
	\sum_{n-0}^{\infty}
	(-i\frac{\zeta}{2} F(\omega))^n \tilde{U}(s)^n
	\tilde{B}_{in}(s,\omega),
	\\
	\tilde{U}(s) 
	=& \sum_i \frac{\xi_i}{s - i q_i},
	\\
	\tilde{B}_{in}(s,\omega)
	=&
	\sum_i \frac{\beta_i}{s-iq_i}.
\end{align}
For each coefficient of the Dyson series, the Laplace transform is a rational function that only has poles located at each wave-vector $q_i$. Therefore, to Laplace invert the coefficient, we can use the following Bromwich contour,
\begin{align}
	b_n(x,\omega) =& \frac{1}{2\pi i}\int_{-i\infty}^{i\infty}\dd{s} e^{s x} \tilde{b}_n(s,\omega),
	\\
	\tilde{b}_n(s,\omega) =& \tilde{U}(s)^n \tilde{B}_{in}(s,\omega).
\end{align}
To begin, we will expand $\tilde{U}(s)$ using the multinomial series. Letting $N$ be the number of modes, we have
\begin{equation}
	\tilde{U}(s)^n = \sum_{
		k_1,k_2\ldots k_N \in p_N(n)
	}
	\begin{pmatrix}
		n
		\\
		k_1,k_2,\ldots k_N
	\end{pmatrix}
	\prod_{i=1}^N
	\xi_i^{k_i}
	\prod_{i=1}^{N}\frac{1}{(s- q_i)^{k_i}}
\end{equation}
where 
\begin{equation}
	\begin{pmatrix}
		n
		\\
		k_1,\ldots k_N
	\end{pmatrix}
	= \frac{n!}{k_1! k_2! \ldots k_N!}
\end{equation}
is a multinomial coefficient, and the sum is over all combinations of $N$ natural numbers $k_1 \ldots k_N$ that sum to $n$,
\begin{equation}
	p_N(n) = 
	\left\{
		(k_1, \ldots k_N)	:
		(k_i \geq 0,  \; i =1,\ldots N)
		\land
		(
		\sum_{i=1}^{N} k_i = n
		)
	\right\}
\end{equation}
Next, we must evaluate the residues at each pole. 
These can be given by the limit formula,
\begin{equation}
	R
	\left(\begin{matrix}
		q_1 & q_2 &\ldots & q_N
		\\
		k_1 & k_2 & \ldots & k_N
	\end{matrix}; x
	\right)= 
	\sum_{j=1}^{N} 
	\frac{1}{(k_j - 1)!} \lim_{s\to i q_j}
	\partial_{k_j - 1}
	\left(
	e^{s x}
	\prod_{i\neq j =1}^{N}
	\frac{1}{(s-iq_i)^{k_i}}
	\right),
\end{equation}
where
\begin{equation}
	\partial_m = \pdv[m]{}{s}.
\end{equation}
To evaluate the derivatives, we may use the generalized product rule,
\begin{equation}
	\partial_m \left(\prod_{i=1}^{n} f_i(s)\right)
	=
	\sum_{l_1\ldots l_n \in p_n(m)}
	\begin{pmatrix}
		m
		\\
		l_1,l_2,\ldots l_n
	\end{pmatrix}
	\prod_{i=1}^{n} \partial_{l_i} f_i(s).
\end{equation}
Additionally, the $m$th derivatives of the remaining poles can be expressed using the falling factorial,
\begin{align}
	\partial_m 
		\frac{1}{\left(s-iq_i\right)^{k_i}}
	=&
	(-k_i)_{(m)} \frac{1}{\left(s-iq_i\right)^{k_i+m}}
	\\
	=& 
	m!
	\begin{pmatrix}
	-k_i
	\\
	m
	\end{pmatrix}\frac{1}{\left(s-iq_i\right)^{k_i+m}}
	\\
	=&
	m!
	(-1)^m
	\begin{pmatrix}
		k_i + m - 1
		\\
		m
	\end{pmatrix}\frac{1}{\left(s-iq_i\right)^{k_i+m}},
\end{align}
while for the exponential we have
\begin{equation}
	\partial_m e^{sx} = x^m e^{sx}.
\end{equation}
We then have
\begin{equation}
	R
	\left(\begin{matrix}
		q_1 & q_2 &\ldots & q_N
		\\
		k_1 & k_2 & \ldots & k_N
	\end{matrix}; x
	\right)
	=
	\sum_{j=1}^{N}
	e^{i q_j x}
	\sum_{l_1,l_2 \ldots l_N : p_{N}(k_j - 1)}
	\frac{x^{l_j}}{l_j!} 
	\prod_{i\neq j=1}^{N}
	\begin{pmatrix}
		-k_i
		\\
		l_i
	\end{pmatrix}
	\frac{1}{(iq_j - i q_i)^{k_i + l_i}}.
\end{equation}
The inclusion of the incident field can be done in a similar fashion,
giving us
\begin{equation}
	B(x,\omega)
	= \sum_{n=0}^{\infty}
	n!\left(-i\frac{\zeta}{2} F(\omega)\right)^n
	\sum_{i=1}^{N}
	\beta_i
	\sum_{k_1,\ldots k_N 
	\in p_N(n)}
	\prod_{j=1}^{N}
	\frac{\xi_j^{k_j}}{k_j!}
	R
	\left(\begin{matrix}
		q_1 &\ldots & q_i & \ldots & q_N
		\\
		k_1 & \ldots & k_i + 1 & \ldots & k_N
	\end{matrix}; x
	\right).
\end{equation}
Finally, we perform the Fourier inversion of the frequency dependence,
\begin{equation}
	\int_{-\infty}^{\infty}\frac{\dd{\omega}}{2\pi}\left(\frac{\gamma/2}{\omega + i\gamma/2}\right)^{n}e^{-i\omega t}
	= 
	(\gamma/2)^n \frac{(-i)^{n} t^{n-1}}{(n-1)!}\Theta(t)e^{-\gamma t/2},
	\quad n > 0.
\end{equation}
This gives us the final expression
\begin{equation}
	B(x,t)
	=
	\frac{B_{in}(x)}{\Gamma_{in}}\delta(t)
	-\frac{\zeta \gamma}{4\Gamma_{in}} \Theta(t)e^{-\gamma t/2}
	\sum_{n=1}^{\infty}
	n\left(-\frac{\zeta \gamma t}{4}\right)^{n-1}
	\sum_{i=1}^{N}
	\beta_i 
	\sum_{k_1,\ldots k_N 
	\in p_N(n)}
	\prod_{j=1}^{N}
	\frac{\xi_j^{k_j}}{k_j!}
	R
	\left(\begin{matrix}
		q_1 &\ldots & q_i & \ldots & q_N
		\\
		k_1 & \ldots & k_i + 1 & \ldots & k_N
	\end{matrix}; x
	\right).
\end{equation}
\subsection*{Recovery of single mode expression}
For the single mode limit, we note that 
\begin{equation}
	R\left(\begin{matrix}
		q_1 
		\\
		k_1
	\end{matrix};
	x\right)
	= \frac{1}{(k_1-1)!}
	\lim_{s \to iq_1}
	\partial_{k_1 - 1}e^{s x}
	= 
	\frac{x^{k_1 - 1}}{(k_1 -1)!}e^{iq_1 x}.
\end{equation}
We then have
\begin{align}
	B(x,t)
	=&
	\frac{B_{in}(x)}{\Gamma_{in}}\delta(t)
	-\frac{\zeta \gamma}{4\Gamma_{in}} \Theta(t)e^{-\gamma t/2}
	\sum_{n=1}^{\infty}
	n\left(-\frac{\zeta \gamma t}{4}\right)^{n-1}
	\beta_1
	\frac{\xi_1^{n}}{n!}
	R
	\left(\begin{matrix}
		q_1
		\\
		n+1
	\end{matrix}; x
	\right)
	\\
	=&
	\frac{B_{in}(x)}{\Gamma_{in}}\delta(t)
	-\frac{B_{in}(x)}{\Gamma_{in}}\frac{\xi_1 \zeta \gamma x}{4} \Theta(t)e^{-\gamma t/2}
	\sum_{n=1}^{\infty}
	\left(-\frac{\xi_1 \zeta \gamma x t}{4}\right)^{n-1}
	\frac{1}{n!(n-1)!}.
\end{align}
Here, we note that for a single mode,
\begin{equation}
	B_{in}(x) = e^{iq_1 x}B_{in}(0).
\end{equation}
This is the expected expression from single mode scattering, and with some manipulation gives the usual Bessel function solution.
\section{Temporal response for constructively interfering micro-strips}\label{app:temporal}
In this section, we derive the temporal response of the case of constructively interfering micro-strips from section~\ref{sec:two-mode-micro}.

The frequency domain response of the scattered field is given by
\begin{equation}
	R(\omega) = (1+\chi(\omega))^n -1 = \frac{1}{(1 + i \frac{\tau}{2} F(\omega)\tr{\Lambda})^n} - 1,
\end{equation}
where we have subtracted off the prompt input field response.
For a Lorentzian line-shape, we have $F(\omega) = \frac{\gamma/2}{\omega + i \gamma/2}$. We therefore have
\begin{equation}
	R(\omega) = \frac{(\omega + i\gamma/2)^n}{(\omega + i \gamma/2  + i\gamma \tau \tr\{\Lambda\}/4)^n} - 1.
\end{equation}
Let us define $\nu = \omega + i \gamma/2 + \nu_0$, where $\nu_0 = i\gamma \tau \tr\{\Lambda\}/4$. We then have
\begin{equation}
	R(\nu) = \frac{(\nu - \nu_0)^n}{\nu^n} - 1 = \sum_{m=0}^{n-1}\binom{n}{m}
	\left(-\frac{\nu_0}{\nu}\right)^{n-m}.
\end{equation}
The time domain response is then given by
\begin{equation}
	R(t) = e^{-\gamma t/2 + i \nu_0 t}\sum_{m=0}^{n-1}\binom{n}{m}\int_{i\gamma /2+ \nu_0 - \infty}^{i\gamma /2+ \nu_0 + \infty}\frac{\dd\nu}{2\pi}e^{-i\nu t}\left(-\frac{\nu_0}{\nu}\right)^{n-m}.
\end{equation}
The Fourier inversion gives
\begin{equation}
	\int_{i\gamma/2 + \nu_0 - \infty}^{i\gamma /2+ \nu_0 + \infty}\frac{\dd\nu}{2\pi}e^{-i\nu t} \frac{1}{\nu^l} = (-i)^l\frac{t^{l-1}}{(l-1)!}\Theta(t).
\end{equation}
We then have
\begin{align}
	R(t) =& e^{-\gamma t/2 + i \nu_0 t}\sum_{m=0}^{n-1}\binom{n}{m}
	\frac{1}{(n-m-1)!}(i\nu_0 t)^{n-m-1}\Theta(t)'
	\\
	=& i\nu_0 e^{-\gamma t/2 + i \nu_0 t}\sum_{m=0}^{n-1}\binom{n}{n-m-1}
	\frac{1}{m!}(i\nu_0 t)^{m}\Theta(t)'.
\end{align}
In particular, we identify the sum with a generalized Laguerre polynomial,
\begin{equation}
	\sum_{m=0}^{n-1}\binom{n}{n-m-1}
	\frac{1}{m!}(i\nu_0 t)^{m} 
	= L_{n-1}^{(1)}(-i\nu_0 t).
\end{equation}
We therefore have the compact expression for the temporal response,
\begin{equation}
	R(t) = i\nu_0 e^{-\gamma t /2+ i \nu_0 t} L^{(1)}_{n-1}(-i\nu_0 t).
\end{equation}


\end{document}